\documentclass[conference]{IEEEtran}
\usepackage{tikz}
\usepackage{amsmath}

\usepackage{filecontents}

\usepackage{graphicx}
\usepackage{subfigure}
\usepackage{amsmath}
\usepackage{amsthm}
\usepackage{mathrsfs}
\usepackage[vlined,ruled,linesnumbered,algo2e]{algorithm2e}
\usepackage{xcolor}
\usepackage{multicol}
\usepackage{multirow}
\usepackage{diagbox}
\usepackage{bbm}
\usepackage{makecell}
\usepackage{booktabs}
\usepackage{algorithm}
\usepackage{algorithmic}
\usepackage{hyperref}
\usepackage{amsfonts}

\newtheorem{thm}{Theorem}

\newtheorem{lem}{Lemma}

\ifCLASSINFOpdf
\else
\fi
\begin{document}
%
\title{Reinforcement Unlearning}




\author{Dayong Ye$^*$, Tianqing Zhu$^{\dagger}$, Congcong Zhu$^{\dagger}$, Derui Wang$^{\ddagger}$, Kun Gao$^*$, \\Zewei Shi$^{\ddagger}$, Sheng Shen$^{\P}$, Wanlei Zhou$^{\dagger}$, and Minhui Xue$^{\ddagger}$\\
$^*$University of Technology Sydney, $^{\dagger}$City University of Macau, \\$^{\ddagger}$CSIRO's Data61, $^{\P}$Torrens University Australia}


%


\IEEEoverridecommandlockouts
\makeatletter\def\@IEEEpubidpullup{6.5\baselineskip}\makeatother
\IEEEpubid{\parbox{\columnwidth}{
    Network and Distributed System Security (NDSS) Symposium 2025\\
    23 - 28 February 2025, San Diego, CA, USA\\
    ISBN 979-8-9894372-8-3\\
    https://dx.doi.org/10.14722/ndss.2025.23080\\
    www.ndss-symposium.org
}
\hspace{\columnsep}\makebox[\columnwidth]{}}

\maketitle

\begin{abstract}
Machine unlearning refers to the process of mitigating the influence of specific training data on machine learning models based on removal requests from data owners. 
However, one important area that has been largely overlooked in the research of unlearning is reinforcement learning. 
Reinforcement learning focuses on training an agent to make optimal decisions within an environment to maximize its cumulative rewards. During the training, the agent tends to memorize the features of the environment, which raises a significant concern about privacy. 
As per data protection regulations, the owner of the environment holds the right to revoke access to the agent's training data, thus necessitating the development of a novel research field, termed \emph{reinforcement unlearning}.
Reinforcement unlearning focuses on revoking entire environments rather than individual data samples. This unique characteristic presents three distinct challenges: 1) how to propose unlearning schemes for environments; 
2) how to avoid degrading the agent's performance in remaining environments; and 3) how to evaluate the effectiveness of unlearning. 
To tackle these challenges, we propose two reinforcement unlearning methods. The first method is based on decremental reinforcement learning, which aims to erase the agent's previously acquired knowledge gradually. 
The second method leverages environment poisoning attacks, which encourage the agent to learn new, albeit incorrect, knowledge to remove the unlearning environment. 
Particularly, to tackle the third challenge, we introduce the concept of ``environment inference'' to evaluate the unlearning outcomes. 
The source code is available at \url{https://github.com/cp-lab-uts/Reinforcement-Unlearning}.
\end{abstract}


%

\section{Introduction}
Machine learning relies on the acquisition of vast amounts of data. 
To safeguard the data privacy of individual users, data protection regulations, e.g., the General Data Protection Regulation \cite{GDPR}, empower users to request the removal of their data. 
It is imperative for model owners to adhere to users' requests by removing revoked data from their datasets and ensuring that any influence these revoked data may have on the model is eliminated. 
This process is referred to as machine unlearning \cite{Cao15,Bourtoule21}.


While significant progress has been made in conventional machine unlearning \cite{Bourtoule21,Guo20,Zhang22}, one area that remains an unfilled gap for unlearning is reinforcement learning (RL).  
RL is an essential research field in machine learning due to its ability to address complex decision-making problems in dynamic environments \cite{Wang22}. 
In RL, the primary objective is to train an intelligent entity, known as an agent, to interact with the environment through a specific policy. This policy guides its actions based on the current state. 
With each action taken, the agent receives a reward and updates its state, creating an experience sample used to update its policy. 
The aim of the agent is to learn an optimal policy that maximizes its cumulative rewards over time. 


However, in the course of RL, agents tend to memorize features of their environments, raising security concerns. 
Consider an agent designed for providing navigation guidance through real-time data from Google Maps. During its training, the agent learns from a dynamic environment using Google Street Views for its photographic content \cite{Mirowski18}. However, privacy issues may arise when the agent inadvertently learns sensitive information, e.g., the locations of restricted areas. 

Another example arises from RL-based recommendation systems. Major platforms like YouTube \cite{Chen19WSDM}, Netflix \cite{Tang23RecSys}, and Amazon \cite{Nunez22} have successfully implemented RL to enhance the quality of their recommendations, providing users with more personalized services. In these systems, each user's interaction history and preferences represent a unique environment. The system, i.e., the agent, learns to make personalized recommendations by adapting its strategy to each user’s specifics, which often contain sensitive information, such as purchasing habits. When users opt out or request data deletion, the recommendation system must comply by forgetting users' specifics, giving rise to the novel field of \emph{reinforcement unlearning}.

Conventional machine unlearning methods are not directly applicable to reinforcement unlearning due to fundamental differences in their learning paradigms. In machine learning, unlearning involves removing specific data samples from the static training set, where data samples are independently and identically distributed. In contrast, RL is a dynamic and sequential decision-making process, where agents interact with an environment in a series of actions, and agents' experience samples are temporally dependent. 

Reinforcement unlearning is also distinct from privacy-preserving RL \cite{Vietri20, Garcelon21}. 
Reinforcement unlearning aims to selectively erase learned knowledge from the agent's memory, ensuring the privacy of environment owners, while privacy-preserving RL focuses on preserving the agent's personal information. 
In essence, reinforcement unlearning presents three specific challenges. 
\begin{itemize}
    \item \textbf{How can we unlearn an environment from the agent's policy?} 
    In machine unlearning, a data owner can specify which data samples should be removed. However, in reinforcement unlearning, the environment owner cannot access the experience samples, as these samples are managed by the agent. 
    Thus, the key challenge lies in effectively associating the environment that needs to be unlearned with the corresponding experience samples. 


    \item \textbf{How can we prevent a degradation in the agent's performance after unlearning?} 
    In conventional machine unlearning, removing samples may lead to a decrease of performance. It is more challenging in reinforcement unlearning as unlearning an environment requires forgetting a significant number of experience samples. 
    
    \item \textbf{How can we evaluate the effectiveness of reinforcement unlearning?} 
    In machine unlearning, one commonly used evaluation is using membership inference ~\cite{Shokri17} to assess if the model has forgotten the revoked data. However, this methodology cannot be directly applied to reinforcement unlearning, as the environment owner cannot specify which samples should be unlearned. 

\end{itemize}

To address these challenges, we propose two distinct unlearning methods: \emph{decremental reinforcement learning} and \emph{environment poisoning}.
Decremental reinforcement learning involves deliberately erasing an agent's learned knowledge about a specific environment. 
This method can be applied to scenarios where certain environments become obsolete or need to be forgotten due to privacy concerns. 
Environment poisoning-based method aims to create poisoning experience samples by modifying the unlearning environment. This method ensures that the agent's performance in other environments remains unaffected, eliminating any negative impact on its overall capabilities. This method finds application in situations where attacks or misinformation may be present. 
Both methods enable an agent to unlearn specific environments while maintaining its performance in others, thereby tackling the first two challenges. 
To tackle the third challenge, we utilize environment inference to infer an agent's training environment by observing its behavior. 
If the inference result after unlearning shows a substantial degradation compared to the result before unlearning, it is indicative that the agent has effectively unlearned that environment.

In summary, we make three main contributions:
\begin{itemize}
\item We provide a valuable step forward in machine unlearning by pioneering the research of reinforcement unlearning. 
The concept of reinforcement unlearning that selectively forgets learned knowledge of the training environment from the agent's memory offers novel insights and lays a foundation for future research in this emerging domain. %

\item Reinforcement unlearning exposes an impactful vulnerability of RL -- the risk of exposing the privacy of the environment owner. This vulnerability can disclose sensitive information about the environment owner's preferences and intentions. We introduce two innovative reinforcement unlearning methods: decremental RL-based and environment poisoning-based approaches to tackle this vulnerability. 

\item 
To confirm the unlearning results, we introduce a novel evaluation approach, ``environment inference''. By visualizing the unlearning results, this approach provides an intuitive means of measuring the efficacy of unlearning techniques. 
\end{itemize}

\section{Preliminaries}

The notations used in this paper are listed in Table~\ref{tab:notation}.

\begin{table}[!ht]\scriptsize
	\centering
	\caption{Summary of Notations}
\begin{tabular} {cl}
\toprule
{\bf Notations} & \hspace{8em}{\bf Description} \\
\midrule
$\mathcal{M}$ & A learning environment/task  \\
$\mathcal{S}$ & A state set, $\mathcal{S}=\{s_1,...,s_n\}$  \\
$\mathcal{A}$ & An action set, $\mathcal{A}=\{a_1,...,a_m\}$  \\
$\mathcal{T}$ & A transition function \\
$r$ & A reward function \\
$\gamma$ & The discount factor \\
$\pi$ & A policy learned by an agent \\
$Q(s,a)$ & The value of a $Q$-function by taking action $a$ in state $s$  \\
$e$ & An experience sample  \\
$B$ & A batch of experience samples for agent learning  \\
$\mathcal{B}$ & A batch of experience samples for poisoning strategy update  \\
$\tau$ & A trajectory consisting of a sequence of state-action pairs  \\
$||\mathbf{x}||_{\infty}$ & The maximum dimension in a vector $\mathbf{x}$ \\
\bottomrule
\end{tabular}
	\label{tab:notation}
\end{table}

\noindent\textbf{Reinforcement Learning.~}
In RL, a learning environment is formulated by the tuple $\mathcal{M}=\langle\mathcal{S},\mathcal{A},\mathcal{T},r\rangle$ \cite{Mnih15Nature,Sutton18}.
$\mathcal{S}$ and $\mathcal{A}$ denote the state and action sets, respectively, while $\mathcal{T}$ represents the transition function, and $r$ represents the reward function.
At each time step $t$, the agent, given the current environmental state $s_t\in\mathcal{S}$, selects an action $a_t\in\mathcal{A}$ based on its policy $\pi(s_t,a_t)$.
This action causes a transition in the environment from state $s_t$ to $s_{t+1}$ according to the transition function: $\mathcal{T}(s_{t+1}|s_t,a_t)$.
The agent then receives a reward $r_t(s_t,a_t)$, along with the next state $s_{t+1}$. 
This tuple of information, denoted as $(s_t,a_t,r_t(s_t,a_t),s_{t+1})$, is collected by the agent as an experience sample utilized to update its policy $\pi$.
Typically, the policy $\pi$ is implemented using a Q-function: $Q(s,a)$, estimating the accumulated reward the agent will attain in state $s$ by taking action $a$.
Formally, the Q-function is defined as: 
\begin{equation}\label{eq:Q}
    Q_{\pi}(s,a)=\mathbb{E}_{\pi}[\sum^{\infty}_{i=1}\gamma^i\cdot r(s_i,a_i)|s_i=s,a_i=a],
\end{equation}
where $\gamma$ represents the discount factor. 

In deep RL, a neural network is employed to approximate the Q-function \cite{Mnih13}, denoted as $Q(s,a;\theta)$, where $\theta$ represents the weights of the neural network.
The neural network takes the state $s$ as input and produces a vector of Q-values as output, with each Q-value corresponding to an action $a$.
To learn the optimal values of $Q(s,a;\theta)$, the weights $\theta$ are updated using a mean squared error loss function $\mathcal{L}(\theta)$. 
\begin{equation}\label{eq:loss}
\begin{aligned}
    \mathcal{L}=\frac{1}{|B|}\sum_{e\in B}[(r(s_t,a_t)+\gamma\max_{a_{t+1}}Q(s_{t+1},a_{t+1};\theta)\\-Q(s_t,a_t;\theta))^2],
\end{aligned}
\end{equation}
where $e=(s_t,a_t,r(s_t,a_t),s_{t+1})$ is an experience sample showing a state transition, 
and $B$ consists of multiple experience samples used to train the neural network.

\vspace{2mm}
\noindent\textbf{Machine Unlearning.~}
Machine unlearning aims to erase the impact of certain data samples on a trained model's behavior \cite{Cao15}. 
A straightforward unlearning method is removing the revoked data and retraining the model from scratch. 
Formally, given a dataset $D$, an unlearning set $D_u$ and a learning algorithm $A$, the objective of machine unlearning is to guarantee that the unlearned model $A(D/D_u)$ performs as if it had never seen the unlearning set $D_u$. 




\section{Reinforcement Unlearning}

\subsection{Problem Statement and Threat Model}\label{sec:security model}
\noindent\textbf{Problem Definition.~}
The definition of ``forgetting'' is application-dependent \cite{Kurmanji23}. 
For example, in a privacy-centric application, the main goal of unlearning user data is to ensure that the unlearned model has no exposure to the data, and a successful membership inference would reveal that the data is not in the training set for the unlearned model. Conversely, in a bias-removing application, the aim of unlearning is to prevent the unlearned model from predicting the assigned labels of the forgotten data, as these labels may indicate unintended and biased behavior. 

The objective of reinforcement unlearning is to eliminate the influence of a specific environment on the agent, i.e., ``forgetting an environment''. We define ``forgetting an environment'' as equivalent to ``performing deterioratively in that environment''. This definition aligns with common sense. For example, when we have thoroughly explored a place and are highly familiar with it, we can efficiently find things within it, resulting in high performance. Conversely, when we have forgotten a place, our ability to locate things diminishes, leading to deteriorative performance. 
This definition may not align with the conventional machine unlearning approach, which usually focuses on ensuring the model's performance matches that of one retrained from scratch. This difference arises because reinforcement unlearning operates within a distinct learning paradigm, characterized by sequential decision-making and dynamic learning. Our definition is motivated by the need to address privacy concerns and mitigate the impact of sensitive or incorrect information within the environment. 



Formally, let us consider a set of $n$ learning environments: \\$\{\mathcal{M}_1,\ldots,\mathcal{M}_n\}$. Each environment $\mathcal{M}_i$ has the same state and action spaces but differs in state transition and reward functions. 
Consider the target environment to be unlearned as $\mathcal{M}_u = \langle \mathcal{S}_u, \mathcal{A}_u, \mathcal{T}_u, r \rangle$, denoted as the `unlearning environment'. 
The set of remaining environments, denoted as $\{\mathcal{M}_1, \ldots, \mathcal{M}_{u-1}, \mathcal{M}_{u+1}, \ldots, \mathcal{M}_n\}$, will be referred to as the `retaining environments'.  
Given a learned policy $\pi$, the goal is to update the policy $\pi$ to $\pi'$ such that the accumulated reward obtained in $\mathcal{M}_u$ is minimized: 
\begin{equation}\label{eq:aim}
    \min_{\pi'}||Q_{\pi'}(s)||_{\infty},
\end{equation}
where $s\in\mathcal{S}_u$, 
while the accumulated reward received in the retaining environments remains the same: 
\begin{equation}\label{eq:constraint}
    \min_{\pi'}||Q_{\pi'}(s)-Q_{\pi}(s)||_{\infty}, 
\end{equation} 
where $s\notin\mathcal{S}_u$. 
Note that if the agent needs to regain knowledge of previously unlearned environments, it can treat them as new environments. This process involves employing an incremental learning approach, where the model is fine-tuned by focusing specifically on the distinct features of these environments.

We assume that the owner of the trained RL model can access $\mathcal{M}_u$ and gather trajectories within $\mathcal{M}_u$. A trajectory $\tau$ is denoted as a sequence of state-action pairs: $\tau = ((s_1, a_1), \ldots, (s_k, a_k))$, where $k$ is the length of the trajectory.
In scenarios where environments become inaccessible after training, the agent stores the experience samples collected during its training and labels them based on their respective environments. These labeled samples can be used to implement the proposed unlearning methods by separating the samples into those from the unlearning environment and those from the remaining environments. 

Note that our work can also be applied to unlearning in a single environment, where the agent is trained within one environment and aims to selectively forget specific aspects of that environment. Detailed discussion and experimental results regarding this adaptation are included in Appendix \ref{appendix: single environment}.



\vspace{2mm}
\noindent\textbf{Threat Model.~} 
Reinforcement unlearning primarily focuses on mitigating the influence of a designated unlearning environment on the trained agent. This essentially involves safeguarding the distinctive features of that environment by thwarting a particular type of attack, namely environment inference attacks. In these attacks, adversaries seek to infer a learning environment by closely observing the actions of the agent within that specific environment. Formally, consider the unlearning environment as $\mathcal{M}_u = \langle \mathcal{S}_u, \mathcal{A}_u, \mathcal{T}_u, r \rangle$ and the unlearned policy as $\pi'$. The adversary's objective is to infer the transition function $\mathcal{T}_u$ by accessing $\mathcal{S}_u$, $\mathcal{A}_u$, $r$, and $\pi'$. 

Environment inference attack differs significantly from conventional membership inference attacks in machine learning. Membership inference attacks typically involve point-level inference, where the focus is on deducing information about an individual sample. In contrast, the environment inference attack operates at the object level, concentrating on the inference of features characterizing an entire environment that encompasses a substantial number of samples. 


\vspace{2mm}
\noindent\textbf{Methods Overview.~} 
Both decremental RL-based and the poisoning-based methods share the common aim of intentionally degrading the agent's performance within the unlearning environment while preserving its performance in other environments. However, they employ distinct strategies to achieve this outcome.
The decremental RL-based method involves updating the agent by minimizing its reward specifically in the unlearning environment. This is achieved through iterative adjustments to the agent's policy, aiming to reduce its effectiveness within the unlearning environment. 
In contrast, the environment poisoning-based method focuses on modifying the unlearning environment itself. This method involves introducing deliberate changes to the state transition function of the environment and subsequently updating the agent in this modified environment. The intention is to disrupt the agent's learned behavior in the unlearning environment.

\subsection{Decremental RL-based Method}

\begin{figure}[ht]
\centering
	\includegraphics[scale=0.5]{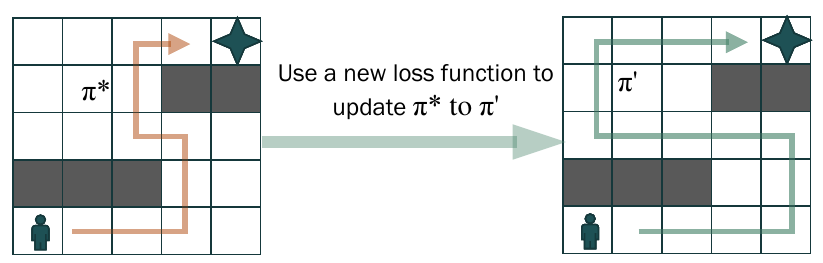}
	\caption{The schematic diagram of the decremental reinforcement learning-based method} 
	\label{fig:overviewDRL}
\end{figure}

The implementation of this method involves two main steps. The first one is the exploration of the unlearning environment $\mathcal{M}_u$. Initially, the agent is allowed to explore the unlearning environment, collecting experience samples specific to that environment. The nature of this exploration depends on the scenario. For instance, in the grid-world setting, the agent might traverse the unlearning grid using a random walk for a predefined number of steps. 
The second step is fine-tuning the agent using the collected experience samples. This fine-tuning process employs a newly defined loss function (Eq. \ref{eq:lossu}) to update the policy $\pi^*$ with the experience samples accumulated in the first step. This loss function is carefully designed to ensure that the agent's performance within the unlearning environment $\mathcal{M}_u$ degrades while preserving its performance in other environments. Essentially, it guides the agent to unlearn the knowledge associated with the unlearning environment.
The schematic diagram of this method is depicted in Figure \ref{fig:overviewDRL}, where the agent's policy is updated from the optimal policy $\pi^*$ to a deteriorative policy $\pi'$.


To accomplish the aim of unlearning, we establish an optimization objective to guide the unlearning process (Eqs. \ref{eq:aim} and \ref{eq:constraint}).
We also introduce a new loss function (Eq. \ref{eq:lossu}) that will be used to update the agent. This loss function is designed to minimize the influence of the previously learned knowledge and encourage the agent to modify its existing policy. With this loss function, we can steer the agent's learning process towards unlearning the knowledge from the given environment.
\begin{equation}\label{eq:lossu}
    \mathcal{L}_u=\mathbb{E}_{s\sim\mathcal{S}_u}[||Q_{\pi'}(s)||_{\infty}]+\mathbb{E}_{s\not\sim\mathcal{S}_u}[||Q_{\pi'}(s)-Q_{\pi}(s)||_{\infty}].
\end{equation}

In Eq. \ref{eq:lossu}, the first term encourages the new policy $\pi'$ to work deficiently in the unlearning environment $\mathcal{M}_u$, while the second term drives the new policy $\pi'$ to have the same performance as the current policy $\pi$ in other environments. 
Notably, the two terms in Eq. \ref{eq:lossu} have favorable properties. 
The first term directs an agent to search and attempt different policies to sufficiently explore the state space of environment $\mathcal{M}_u$. 
Thus, $\mathcal{M}_u$ can be adequately unlearned. 
This property is particularly useful when $\mathcal{M}_u$ is a sparse reward setting, i.e., the reward is $0$ in most of the states in $\mathcal{S}_u$. 
The second term motivates an agent to strategically modify policies. 
This property ensures that the agent performs consistently in those states which are not in $\mathcal{S}_u$. 
Note that the accurate computation of the second term is infeasible due to its involvement with all states except $\mathcal{S}_u$. 
Thus, during implementation, we uniformly select a consistent set of states across all environments, excluding $\mathcal{M}_u$. This approach is taken to mitigate computational burden and ensure a balanced impact on performance preservation across the remaining environments.

The decremental RL-based method is formally described as follows. First, the agent explores the unlearning environment using a random policy. 
Here, a random policy, defined by the agent owner, refers to a policy where the agent selects actions uniformly at random from the set of all possible actions available in any given state. This means that the probability of selecting any action $a$ from the set of possible actions $\mathcal{A}$ is $1/(|\mathcal{A}|)$, where $|\mathcal{A}|$ is the number of available actions in that state.
For instance, in the grid world setting, where the agent has four possible actions (moving up, down, left, and right), the random policy dictates that, in each grid (i.e., state), the agent has an equal likelihood of selecting any of the four directions.
This policy ensures that the agent does not leverage any previously learned strategies specific to the unlearning environment.
If the agent were to use a well-established policy from its training, there is a risk of swift achievement of the target, resulting in insufficient collection of experience samples within the unlearning environment. 
During its exploration, in each time step $t$, after taking an action $a_t$, the agent receives a corresponding reward $r_t$. Coupled with the current state $s_t$ and the subsequent state $s_{t+1}$, the agent effectively creates an experience sample in the form of $(s_t,a_t,r_t,s_{t+1})$. When the agent is directed to explore the unlearning environment for a specified number of steps, denoted as $m$, the agent accumulates a total of $m$ experience samples: $(s_1,a_1,r_1,s_2 ),\ldots,(s_m,a_m,r_m,s_{m+1})$. In the second step, the agent employs these collected experience samples to fine-tune its current optimal policy $\pi^*$ to a new policy $\pi'$ by minimizing the custom loss function defined in Eq. \ref{eq:lossu}.


\vspace{2mm}
\noindent\textbf{Convergence Analysis of the Method.~}
We proceed with examining the convergence of the method by conducting a separate analysis for each term in Eq. \ref{eq:lossu}. 

Let $y_{\pi}=r(s,\pi(s))+\gamma max_{\pi(s')}Q(s',\pi(s'))$ and $\delta_{\pi}=y_{\pi}-Q(s,\pi(s))$. 
Then, Eq. \ref{eq:loss} can be rewritten as: 
\begin{equation}\label{eq:newloss}
    \mathcal{L}=\mathbb{E}_{s\sim\mathcal{S}}[\delta_{\pi}].
\end{equation}
Similarly, the first term in Eq. \ref{eq:lossu} can also be rewritten as: $\mathbb{E}_{s\sim\mathcal{S}_u}[|y_{\pi'}-\delta_{\pi'}|]$. 
Based on the triangle inequality, we have: 
\begin{equation}\label{eq:inequality}
    \mathbb{E}_{s\sim\mathcal{S}_u}[|y_{\pi'}-\delta_{\pi'}|]\leq\mathbb{E}_{s\sim\mathcal{S}_u}[|y_{\pi'}|]+\mathbb{E}_{s\sim\mathcal{S}_u}[|\delta_{\pi'}|].
\end{equation}
As the convergence of the learning on the loss function in Eq. \ref{eq:newloss} has been both theoretically and empirically proven \cite{Mnih13,Tosatto17,Fan20}, we can also conclude the convergence of $\mathbb{E}_{s\sim\mathcal{S}_u}[|\delta_{\pi'}|]$ in Eq. \ref{eq:inequality}. 
For the term $\mathbb{E}_{s\sim\mathcal{S}_u}[|y_{\pi'}|]$ of Eq. \ref{eq:inequality}, as it is computed by accumulating the previously collected discounted rewards (Eq. \ref{eq:Q} and \ref{eq:loss}), the term converges if the rewards are bounded. 
The reward bound can be acquired by proper definition, i.e., $r\in[-R_{max},R_{max}]$. 
Thus, as both $\mathbb{E}_{s\sim\mathcal{S}_u}[|y_{\pi'}|]$ and $\mathbb{E}_{s\sim\mathcal{S}_u}[|\delta_{\pi'}|]$ converge, $\mathbb{E}_{s\sim\mathcal{S}_u}[|y_{\pi'}-\delta_{\pi'}|]$ also converges, i.e., $\mathbb{E}_{s\sim\mathcal{S}_u}[||Q_{\pi'}(s)||_{\infty}]$ converges.


For the second term in Eq. \ref{eq:lossu}, to analyze its convergence, we need the following theorem.

\begin{thm}[Error Propagation \cite{Tosatto17}]\label{thm:error}
Let $(\pi_i)^K_{i=0}$ be a sequence of policies with regard to the sequence $(Q_i)^K_{i=0}$ of Q-functions learned using a fitted Q-iteration. 
Then, the following inequality holds.
\begin{equation}\nonumber
    ||Q_i-Q^*||_{\infty}\leq||\xi_{i-1}||_{\infty}+\gamma||Q_{i-1}-Q^*||_{\infty}+\zeta||Q_{i-1}||_{\infty},
\end{equation}
where $Q^*$ is the optimal value function, $\xi_i$ denotes the approximation error: $\xi_i=T^{\pi_i}Q_i-Q_{i+1}$ which is also bounded, $T$ is the Bellman operator, and $\zeta$ is a constant.
\end{thm}
Theorem \ref{thm:error} provides evidence that the disparity between the learned Q-function and the optimal Q-function diminishes as the learning process advances. This reduction signifies convergence, given that the Q-function, denoted as $Q$, remains uniformly bounded by $\frac{R_{max}}{1-\gamma}$ for any policy $\pi$ \cite{Tosatto17}. Consequently, if the number of learning iterations is sufficiently large, the method converges. 

In our problem, the second term in Eq. \ref{eq:lossu}, $\mathbb{E}_{s\not\sim\mathcal{S}_u}[||Q_{\pi'(s)}(s)-Q_{\pi(s)}(s)||_{\infty}]$, is intended to minimize the performance discrepancy between the unlearned policy $\pi'$ and the well-trained policy $\pi$ across all environments except $\mathcal{M}_u$. In this context, the well-trained policy $\pi$ can be considered as the optimal policy, while the unlearned policy $\pi'$ represents the policy we aim to learn. Notably, this learning process is analogous to the one described in Theorem \ref{thm:error}, implying that the second term also exhibits convergence.

\begin{figure}[ht]
\centering
	\includegraphics[scale=0.25]{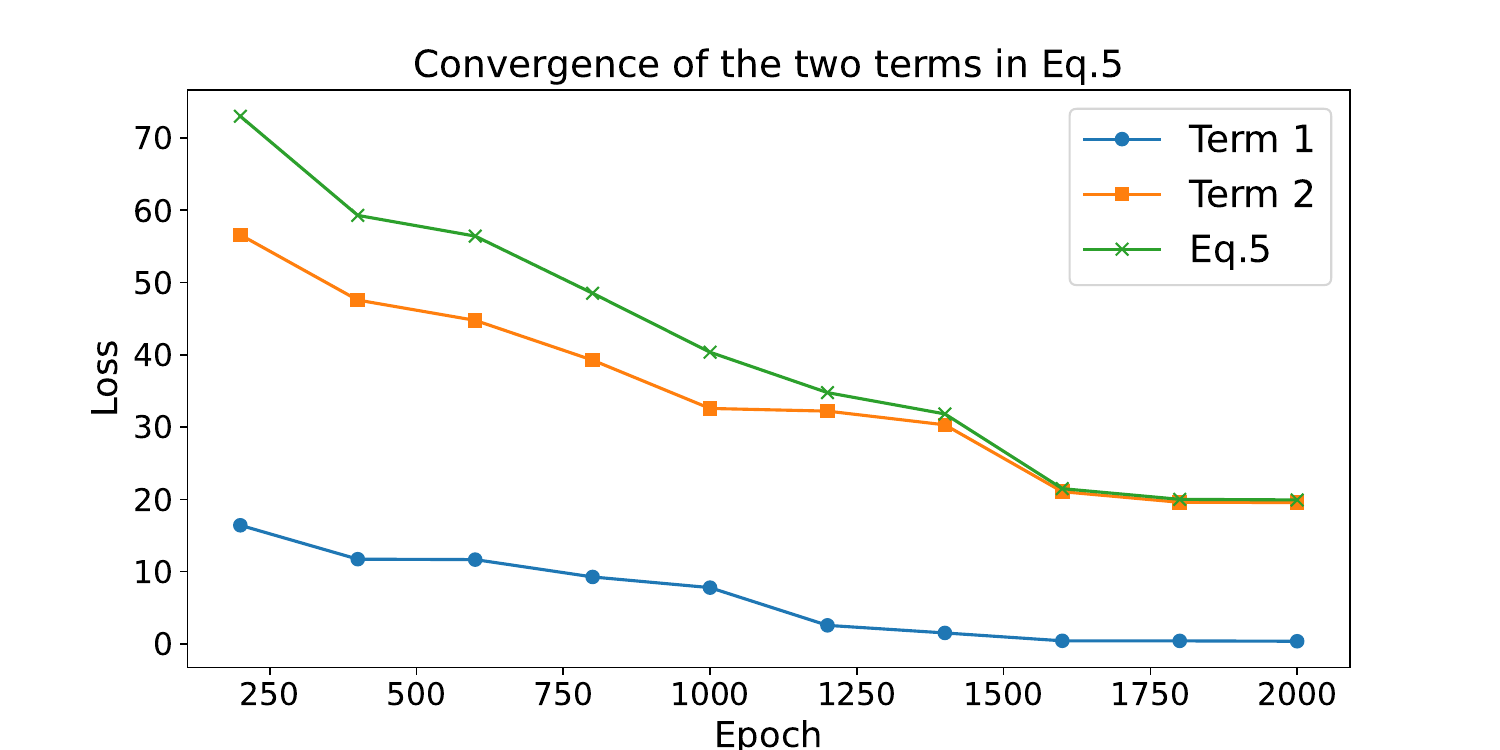}
	\caption{Convergence of Eq. \ref{eq:lossu} and its two terms} 
	\label{fig:Convergence_dqn}
\end{figure}

Note that the first and second terms in Eq. \ref{eq:lossu} may not converge simultaneously due to their optimization for different objectives. However, their convergence contributes to the overall convergence of Eq. \ref{eq:lossu}. When both terms converge, it indicates that the agent's policy is gradually adjusting to achieve the desired objectives outlined by each term. As the agent updates its policy over time, these adjustments lead to a convergence of the Q-values towards their respective targets. This convergence implies that the agent's policy becomes increasingly optimized to fulfill both objectives. 
Our experimental results, depicted in Figure \ref{fig:Convergence_dqn}, reveal that both terms converge gradually over time. Although they exhibit different rates of convergence, they still contribute to the overall convergence of Eq. \ref{eq:lossu}.


\subsection{Environment Poisoning-based Method}
This method is implemented by modifying the unlearning environment itself. This modification can include various changes, i.e., the poisoning actions, such as altering the layout in the grid world scenario by adding or removing obstacles and repositioning targets within the environment. After these changes are introduced, the agent is updated in this modified environment. This method aims to influence the agent's policy learning by creating a situation where its previously learned knowledge becomes less effective. 

The poisoning process operates as a Markov decision process, resembling an RL problem. Here, the agent's policies serve as states, its modifications to the unlearning environment act as actions, and the disparity between the current policy's performance and that of the updated policy within the unlearning environment functions as the reward. Consequently, the poisoning strategy is updated using Eq. \ref{eq:loss} as the loss function.

\begin{figure}[ht]
\centering
	\includegraphics[scale=0.5]{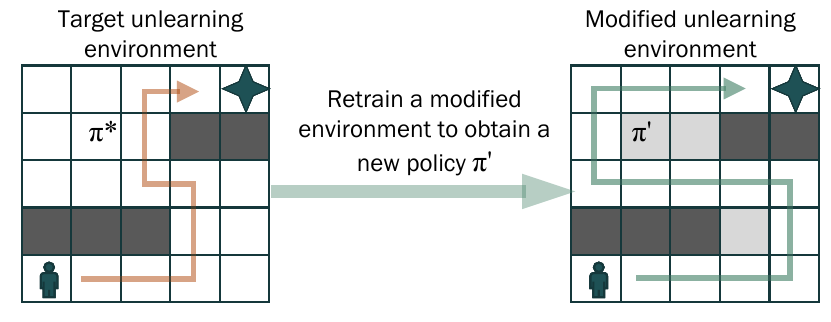}
	\caption{The schematic diagram of the environment poisoning-based method} 
	\label{fig:overviewPoison}
\end{figure}

Specifically, this method consists of three steps.
Firstly, we apply a random poisoning strategy to alter the transition function of the unlearning environment.
Secondly, the agent learns a new policy in this modified environment. 
Lastly, based on the agent's learned policy, we update the poisoning strategy and re-poison the unlearning environment.
These three steps are iteratively repeated until a predetermined number of poisoning epochs is reached. 
The schematic diagram is presented in Figure \ref{fig:overviewPoison}, illustrating that the unlearning environment $\mathcal{M}_u$ is altered to $\mathcal{M}'_u$ with strategically introduced perturbations, e.g., adding fake obstacles, and the agent is retrained in this poisoned environment to learn a new policy $\pi'$. 

Let the learned policy be $\pi^*$, which is regarded as the optimal policy. To refine the agent's policy, we manipulate the given environment $\mathcal{M}_u$, involving poisoning the transition function $\mathcal{T}_u(s'|s,a)$ \cite{Xu22}, where $\mathcal{M}_u = \langle\mathcal{S}_u, \mathcal{A}_u, \mathcal{T}_u, r\rangle$.
We introduce a poisoned transition function denoted as $\hat{\mathcal{T}}_u(\hat{s}'|s,a)$. After the agent takes action $a$ in state $s$, instead of observing the intended state $s'$, it will observe the manipulated state $\hat{s}'$. The challenge now lies in determining the appropriate state $\hat{s}'$, which can mislead the agent's learning process.
To address this, we define a new learning environment for poisoning, denoted as $\mathcal{M}_p = \langle\mathcal{O}, \mathcal{G}, \mathcal{P}, \mathcal{R}\rangle$. 

\begin{itemize}
    \item $\mathcal{O}$ is the set of poisoning states. Each state $\pi_i\in\mathcal{O}$ is the policy used by the agent during the $i$-th poisoning epoch. 
    \item $\mathcal{G}$ is the set of poisoning actions. A poisoning action $g\in\mathcal{G}$ is a modification made to the transition function of the unlearning environment. This modification signifies which state should be presented to the agent as the new state. 
    \item $\mathcal{P}:\mathcal{O}\times\mathcal{G}\times\mathcal{O}\rightarrow[0,1]$ defines the poisoning state transition. It describes how the agent adjusts its policy in response to poisoning actions. Here, $\mathcal{P}(\pi'|\pi,g)$ is the probability of the agent transitioning from policy $\pi$ to $\pi'$ when the unlearning environment's transition function is modified by $g$.
    \item $\mathcal{R}:\mathcal{O}\times\mathcal{G}\times\mathcal{O}\rightarrow\mathbb{R}$ represents the reward function, which serves two purposes. Firstly, it quantifies the disparity between the current policy $\pi_i$ and the updated policy $\pi'$ in the unlearning environment $\mathcal{M}_u$. Secondly, it incorporates the rewards obtained by the agent in other environments while utilizing $\pi_i$. 
    Specifically, the reward function is defined as 
    \begin{equation}\label{eq:reward}
        \mathcal{R}_i:=\lambda_1\Delta(\pi_i(s_i)||\pi'(s_i))+\lambda_2\sum_{s\not\sim\mathcal{S}_u}\sum_a\pi_i(s,a)r(s,a).
    \end{equation}
    $\mathcal{R}_i$ represents the reward received during the $i$-th poisoning epoch.
    The term $\Delta(\pi_i(s_i)||\pi'(s_i))$ indicates the difference between $\pi_i(s_i)$ and $\pi'(s_i)$, which can be measured using KL-divergence. Here, $\pi_i(s_i)$ is the probability distribution over the available actions in state $s_i$ under policy $\pi_i$, and $\lambda_1$ and $\lambda_2$ are introduced to balance the two terms.
    Precisely computing the second term is computationally infeasible due to the involvement of states from all environments except $\mathcal{M}_u$. 
    Thus, similar to the decremental RL-based method, in the implementation of the poisoning-based method, a uniform selection of states across all the environments, except $\mathcal{M}_u$, is performed in each poisoning epoch. 
\end{itemize}

\begin{algorithm}\small
\caption{The poisoning-based method}
\label{alg:poison}
\begin{algorithmic}[1]
\REQUIRE{The learned policy $\pi^*$, the learning tasks $\mathcal{M}_1,...,\mathcal{M}_n$, and the unlearning task $\mathcal{M}_u$;}
\ENSURE{A refined policy $\hat{\pi}$;}
\FOR{poisoning epoch $i=1,...,m$}
\STATE Select a poisoning action $g_i$ from $\mathcal{G}$;
\STATE Alter the transition function of the unlearning task from $\hat{\mathcal{T}}_{u,{i-1}}$ to $\hat{\mathcal{T}}_{u,i}$ based on $g_i$;
\STATE The agent learns a policy $\pi_i$ according to $\hat{\mathcal{T}}_{u,i}$;
\STATE Receive reward $\mathcal{R}_i$;
\STATE Update poisoning strategy using samples $(\pi_{i-1},g_i,\pi_i,\mathcal{R}_i)\in\mathcal{B}$ by optimizing loss function Eq. \ref{eq:loss};
\ENDFOR
\RETURN $\hat{\pi}\leftarrow\pi_m$; 
\end{algorithmic}
\end{algorithm}

The poisoning-based method is outlined in Algorithm \ref{alg:poison}. In each poisoning epoch $i$, we take the first step by choosing a poisoning action $g_i$ to modify the transition function of the unlearning environment (Lines 1-3). The selection process can be implemented using an $\epsilon$-greedy strategy, where the best action, which results in the highest Q-value, is chosen with a probability of $1-\epsilon$, and the remaining actions are chosen uniformly with a probability of $\frac{\epsilon}{|\mathcal{G}|-1}$.
Here, $\epsilon$ is typically set within the range of $0.1$ to $0.2$ to achieve a balance between exploration and exploitation. 

Next, the agent performs the second step by learning a new policy $\pi_i$ in the altered environment (Line~4). This learning phase can be performed using any RL algorithms, such as deep Q-learning \cite{Mnih13}. Once $\pi_i$ is learned, we execute the third step by evaluating the reward using Eq. \ref{eq:reward} and updating the poisoning strategy using samples $(\pi_{i-1},g_i,\pi_i,\mathcal{R}_i)$ from batch $\mathcal{B}$ (Lines 5 and 6). The update process is carried out using the DDPG algorithm~\cite{DDPG}.
After all poisoning epochs are completed, we obtain a refined policy $\hat{\pi}$ (Line 8), which allows the agent to perform poorly in the unlearning environment $\mathcal{M}_u$, while maintaining satisfactory performance in other environments.

\begin{figure}[ht]
\centering
	\includegraphics[scale=0.45]{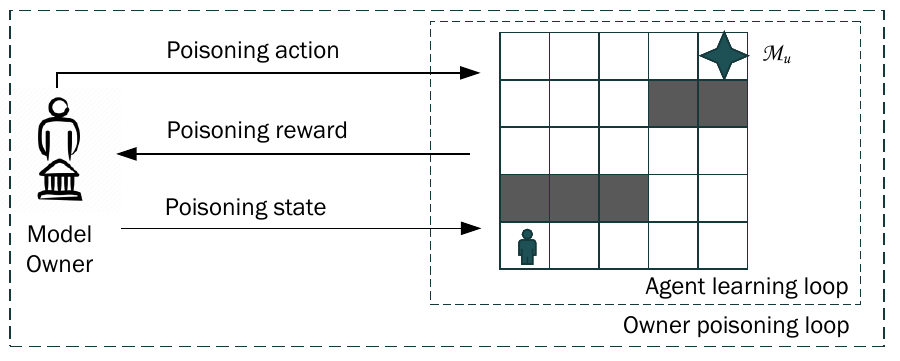}
	\caption{The interaction between agent learning and model owner poisoning. The owner observes the poisoning states about the unlearning environment $\mathcal{M}_u$, takes poisoning actions against $\mathcal{M}_u$, and receives poisoning rewards. The agent, on the other hand, is learning in the poisoned environment.}
	\label{fig:poisoning}
\end{figure}

\vspace{2mm}
\noindent\textbf{Convergence Analysis of the Method.~}
Algorithm \ref{alg:poison} represents a deep RL algorithm employed by the model owner. 
The interaction between the algorithm and the agent's learning process is illustrated in Figure \ref{fig:poisoning}. In this figure, the model owner is engaged in learning how to poison the unlearning environment $\mathcal{M}_u$, while the agent is concurrently learning within the poisoned environment $\mathcal{M}'_u$.

Given that our primary concern lies in the performance of the agent's policies, our analysis primarily revolves around these policies. Each policy $\pi$ is associated with a state distribution denoted as $\mu_{\pi}$, which can be defined as:
\begin{equation}\nonumber
    \mu_{\pi}:=(1-\gamma)\sum^{\infty}_{t=0}\gamma^t\mathbb{P}[s_t=s|s_0\sim d_0,\pi],
\end{equation}
where $d_0$ is the initial state distribution and $\mu_{\pi}>0$ for each state $s$. 
Here, $\mu_{\pi}$ satisfies the following Bellman flow constraints \cite{Rakhsha21}:
\begin{equation}\label{eq:mu}
    \mu_{\pi}=(1-\gamma)d_0+\gamma\sum_{s'}\mathcal{T}(s'|\pi(s'),s)\mu_{\pi(s')}.
\end{equation}
Then, the score of policy $\pi$ can be defined as: 
\begin{equation}\nonumber
    \rho_{\pi}(\mathcal{M},d_0):=\sum_s\mu_{\pi}(s)r(s,\pi(s)).
\end{equation}
The policy score $\rho_{\pi}$ quantifies the quality of a policy $\pi$, with a higher score indicating a better policy. 
Specifically, $\rho_{\pi}$ has the following property.
\begin{lem}[\cite{Even04}]\label{lem:twoPolicies}
For two policies $\pi$ and $\pi'$, the following equation holds: 
$\rho_{\pi}-\rho_{\pi'}=\sum_{s\in\mathcal{S}}\mu_{\pi'}(s)(Q_{\pi}(s,\pi(s))-Q_{\pi}(s,\pi'(s)))$.
\end{lem}

Let us examine the expression $Q_{\pi}(s,\pi(s))-Q_{\pi}(s,\pi'(s))$. To simplify the analysis, we introduce a seminorm called the span. The span of $Q$ is defined as $sp(Q) = \max_i Q(s_i,a_i) - \min_j Q(s_j,a_j)$. This seminorm measures the maximum difference between the highest and lowest values of the function $Q$ across different states and actions. 
Certainly, we have: $Q_{\pi}(s,\pi(s))-Q_{\pi}(s,\pi'(s))\leq sp(Q_{\pi})$. 
Then, we have: $\rho_{\pi}-\rho_{\pi'}\leq sp(Q_{\pi})|\mathcal{S}|\sum_{s\in\mathcal{S}}\mu_{\pi'}(s)$. 

Based on Eq. \ref{eq:Q} and \ref{eq:mu}, it can be inferred that the span $sp(Q_{\pi})$ is limited by the cumulative reward, while $\mu_{\pi'}$ is bounded by the transition function. The reward is predefined by users, and the initial state distribution $d_0$ remains fixed for a given environment. Thus, the only variable that influences the difference in policy scores, $\rho_{\pi}-\rho_{\pi'}$, is the transition function dictated by the environment. This rationale underscores why we opt for environment-poisoning as our unlearning method.

\subsection{Discussion of the Two Methods}


\noindent\textbf{Over-unlearning.} 
The decremental RL-based method may inadvertently suffer from over-unlearning, which occurs when the model is fine-tuned to degrade the agent's learning performance in $\mathcal{M}_u$. 
Even if efforts are made to restrict the deterioration to $\mathcal{M}_u$, it may still affect other environments due to the shared distribution among them. However, the poisoning-based method inherently avoids this issue by focusing on enabling the agent to learn new knowledge rather than intentionally forgetting existing knowledge. Thus, the poisoning-based method has the potential to achieve superior performance in non-unlearning environments compared to the decremental RL-based method. 

Another factor contributing to the over-unlearning issue in the decremental RL-based method is the delayed influence of the second term in Eq. \ref{eq:lossu}. This term is responsible for fine-tuning the policy to sustain performance in the retained environments. However, its impact may not manifest immediately. This delayed interference implies that the policy is tuned to initially focus solely on minimizing performance in the unlearning environment, without considering the impact on performance in the retained environments. As a result, the policy may prioritize forgetting the unlearning environment over preserving knowledge from the retained environments.

\vspace{2mm}
\noindent\textbf{Catastrophic Forgetting.} A valid concern regarding the poisoning-based method is the potential occurrence of catastrophic forgetting, which arises when the continual updating of the model results in the overwriting of previously acquired knowledge. However, this issue does not arise in the context of the poisoning-based method.
The primary cause of catastrophic forgetting is a shift in the input distribution across different environments \cite{Kirk17,Shi21}. In our scenario, the modified environment $\mathcal{M}'_u$ retains the same distribution as the other environments. This is because the modification is solely applied to the transition function, while the state and action spaces, as well as the reward function, remain unchanged. 
Specifically, the transition function dictates the evolution of states based on the actions taken by the agent, governing how states change over time. In contrast, the state and action spaces, as well as the reward function, are pre-defined by the model owner during the training of the agent. These foundational elements remain relatively unaffected by modifications to the transition function.
Thus, there is no distribution shift across environments in our problem, thereby mitigating the risk of catastrophic forgetting.

\vspace{2mm}
\textbf{Offline RL Settings.} Although our methods are designed to online RL settings, they can be adapted to offline settings, where the agent cannot interact with the environment in real-time but relies on previously collected experience samples. The decremental RL-based method can utilize these samples by adjusting the policy to minimize performance in the unlearning environment while maintaining performance in retained environments. The poisoning-based method can modify the collected experience samples, e.g., states and rewards, to simulate the effect of altering the transition function.

\vspace{2mm}
\textbf{Direct Reward Inversion.} As both unlearning methods focus on reducing the agent's received reward in the unlearning environment, a seemingly straightforward approach involves directly inverting the received reward: changing a real reward $r$ to $-r$. However, this method risks deteriorating the agent's performance not only in the unlearning environment but also in remaining ones. This occurs because such inversion disrupts the entire reward structure foundational to the agent’s learning and decision-making processes, potentially leading to unintended behavior across all environments.

\textbf{Mitigating Security Issues.} The potential security issues associated with our unlearning methods can be effectively mitigated, particularly concerning adversarial attempts to reconstruct the unlearning environment by observing the agent's behavior and leveraging knowledge that this behavior results from minimizing rewards. This is because both methods employ randomness in their action selection processes that goes beyond merely minimizing rewards. In the decremental RL-based method, while rewards are minimized, the policy also includes a degree of randomness in action selection that are not directly tied to reward signals. This randomness ensures that the behavior observed by an adversary does not consistently align with the most minimized rewards, thereby obscuring the true characteristics of the unlearning environment. Likewise, the poisoning-based method not only reduces reward, but also alters the transition dynamics between states. This means that even if an adversary can observe that certain actions are less rewarded, the underlying state transitions are altered. This alteration further obscures the patterns in the unlearning environment, complicating any adversary attempts to reconstruct the environment based on minimized rewards alone.

\subsection{Environment Inference}

One of our contributions is to propose a new evaluation methodology named environment inference. This kind of inference aims to infer an agent's training environments by observing agent's behavior~\cite{Pan19}. By using this approach, we can assess if the agent has successfully erased the knowledge of the unlearning environment. If the removal of the unlearning environment's knowledge is executed correctly, the agent's behavior in that environment should be random rather than purposeful. Thus, by observing the agent's behavior, an adversary can only infer a randomized environment. 

A notable environment inference \cite{Pan19} utilizes a genetic algorithm to identify a transition function that not only satisfies specific constraints but also provides the best possible explanation for the observed policy. Inspired by this approach, our research also employs the same genetic algorithm to infer the unlearning environments. 
Initially, we randomly generate $L$ transition functions, representing the population. These transition functions undergo iterative updates through crossover and mutation operations.
During the crossover process, pairs of transition functions are selected based on their fitness, determined by their similarity to the learned optimal policy. New transition functions are then created by combining elements of the selected pairs. This promotes the exchange of information between transition functions, leading to improved performance.
In the mutation stage, individual transition functions undergo small, random changes to introduce diversity and explore new regions of the solution space. The changes are made by randomly adjusting the resulting next states for some state-action pairs. This helps prevent premature convergence and allows for the discovery of novel and effective transition strategies.
After completing a predefined number of iterations, the transition function with the highest fitness score is chosen as the optimal representation. 


\section{Experimental Evaluation}\label{sec:experiments}




\subsection{Experimental Setup}\label{sub:setup} 

Evaluation metrics in machine unlearning \cite{Xu23} are not applicable to reinforcement unlearning. 
For instance, in reinforcement unlearning, there are no specific datasets to be forgotten, rendering metrics like ``accuracy on forget set'' irrelevant. 
Thus, it is necessary to propose new metrics. 

\vspace{2mm}
\noindent\textbf{Cumulative Reward} 
quantifies the total sum of rewards accumulated by an agent while utilizing the acquired policy.

\noindent\textbf{The Number of Steps} 
quantifies the total number of steps taken by an agent to reach its goal or complete a task. 

\noindent\textbf{Environment Similarity} quantifies the resemblance between the inferred environment and the original one. It is evaluated by doing an environment inference and calculating the percentage of agreement between the inferred and original environments.

\subsubsection{Tasks}
The experiments were primarily conducted across three learning tasks: grid world, virtual home and maze explorer. The grid world was developed by us, while the virtual home and maze explorer tasks were sourced from the VirtualHome \cite{VirtualHome} and MazeExplorer \cite{MazeExplorer}, respectively. 
Although there are well-known RL tasks available, e.g., Gym~\cite{Gym} and Atari \cite{Artari}, they were deemed unsuitable for our research 
as those RL tasks are designed for single-environment and do not support multiple environments. 

\noindent\textbf{Grid World.~} 
The objective for the agent in this task is to navigate towards the predetermined destination. 
This task mirrors numerous challenges present in real-world autonomous driving and navigation scenarios, with each square akin to a section of road, intersection, or obstacle. 

\noindent\textbf{Virtual Home.~} It is a multi-agent platform designed to simulate various activities within a household setting. 
This task mirrors real-world smart home challenges, where homeowners can request the agent to forget certain features of their home. 

\noindent\textbf{Maze Explorer.~} It is a customizable 3D platform. The objective is to guide an agent through a procedurally generated maze to collect a predetermined number of keys. This task represents scenarios where agents need to learn from visual information, making it relevant to applications such as robotic exploration in intricate environments.

In the three tasks, four available actions include moving up (forward), down (backward), left, and right. Environments are instantiated with predetermined sizes. The instantiation process involves two steps. Firstly, each environment is randomly generated, introducing variability in the placement of obstacles. Subsequently, a manual inspection is carried out to eliminate any instances of ``dead locations'', which are inaccessible due to being entirely surrounded by obstacles. 

We have also expanded our study to include two additional tasks: recommendation systems and aircraft landing. Recommendation systems are utilized to evaluate privacy risks associated with unlearning, while aircraft landing scenarios serve to assess the safety-critical aspects of our methods. 

\noindent\textbf{Recommendation Systems.} We employ the MovieLens dataset \cite{MovieLens} to simulate a recommendation system using reinforcement learning. In this setup, each user represents a unique training environment. Within this environment, a state is defined by the movies the user has watched and their corresponding ratings. An action corresponds to recommending a specific movie to the user. The reward is determined by the user’s rating of the recommended movie.

\noindent\textbf{Aircraft Landing.~} This task simulates an aircraft landing on the ground by avoiding the obstacles with four available actions: moving up, down, left, and right.

While our unlearning methods are evaluated in tasks with discrete state and action spaces, they can also be applied to tasks with continuous state and action spaces. Our approaches are independent of the underlying RL algorithms. To address continuous spaces, one can simply integrate our unlearning techniques with suitable RL algorithms, e.g., \cite{Lee18}, designed for such environments.



\subsubsection{Comparison Methods}
As there are no closely related existing works, to establish a benchmark, we propose two baseline methods. 

\vspace{2mm}
\noindent\textbf{Learning From Scratch (LFS).~} 
This method entails removing the unlearning environment and then retraining the agent from scratch using the remaining environments. 
However, this method is not a desirable criterion for reinforcement unlearning. 
We will experimentally show that this approach fails to meet the objectives of reinforcement unlearning as defined in Section \ref{sec:security model}.

\noindent\textbf{Non-transferable Learning From Scratch (Non-transfer LFS).} To align with the objectives of reinforcement unlearning, we introduce a non-transferable learning-from-scratch approach. This approach is similar to the previously mentioned learning-from-scratch approach. However, a crucial distinction lies in the non-transferable version, which incorporates the non-transferable learning technique \cite{Wang22bICLR} to restrict the approach's generalization ability within the unlearning environments. In this approach, while training an agent, the model owner meticulously stores experience samples acquired from all learning environments, labeling them according to their source environment. When an unlearning request is initiated, the model owner engages in an offline retraining process. Specifically, all the collected experience samples are utilized in retraining the agent. If a sample originates from the unlearning environment, an inverse loss function is applied to minimize the agent's cumulative reward. Conversely, for samples from other environments, the standard loss function is used to maximize the agent's overall reward. Denoting the unlearning environment as $\mathcal{M}_u=(\mathcal{S}_u,\mathcal{A}_u,\mathcal{T}_u,r)$, the loss functions are defined in Eq. \ref{eq:non-transfer}.

\begin{equation}\label{eq:non-transfer}
    \mathcal{L}=\begin{cases}
    -\frac{1}{|B|}\sum_{e\in B}[(r(s_t,a_t)+\gamma\max\limits_{a_{t+1}}Q(s_{t+1},a_{t+1};\theta)\\-Q(s_t,a_t;\theta))^2], & \hspace{-10mm} \text{if $s_t\in\mathcal{S}_u$}, \\
    \frac{1}{|B|}\sum_{e\in B}[(r(s_t,a_t)+\gamma\max\limits_{a_{t+1}}Q(s_{t+1},a_{t+1};\theta)\\-Q(s_t,a_t;\theta))^2], & \hspace{-10mm} \text{otherwise}.
    \end{cases}
\end{equation}



\subsubsection{Sample Complexity of Unlearning Methods}
In the learning-from-scratch method (LFS), the retraining process involves leveraging all the experience samples collected from various environments, excluding the unlearning environment. This extensive dataset is used for the comprehensive retraining of the agent. Similarly, in the Non-transfer LFS, retraining utilizes all experience samples, encompassing those from the unlearning environment. 
In contrast, when evaluating the performance of the decremental RL-based and the poisoning-based methods, only a small subset of these samples, approximately one-tenth, is employed to fine-tune the agent to generate the final experimental results. 

There might be a concern about our proposed methods, as they allow the agent to have additional interactions with the unlearning environment. In contrast, both LFS and Non-transfer LFS do not involve further interactions with any environments. However, this additional interaction in our methods does not bring any extra advantages. The purpose of engaging with the unlearning environment is solely to collect experience samples.
These experience samples are not required for LFS, and for Non-transfer LFS, they have already been collected. Therefore, the lack of such interactions, does not impact the performance of both LFS and Non-transfer LFS.

\subsubsection{Underlying RL Algorithms}
The primary RL algorithm employed in this study is the Deep Q-Network (DQN) \cite{Mnih13}, a well-established value-based method. We also evaluated other RL algorithms, such as Proximal Policy Optimization (PPO) \cite{PPO} and Deep Deterministic Policy Gradient (DDPG) \cite{DDPG}. However, their performance was found to be inferior to that of DQN (refer to Appendix \ref{app:overall performance}). Thus, PPO and DDPG were not included in our experiments.

\subsection{Overall Performance}\label{sub:overall}
The presented experimental results were derived by averaging the outcomes across $100$ rounds of repeated experiments, and a $95\%$ confidence interval of $±3\%$ was calculated. The variances of the average reward and steps are both below 7 and 10, respectively. However, for clarity, they are not visually presented in the figures.


To illustrate the effectiveness of unlearning, we introduce the concept of ``forget quality'' as a quantitative measure of the strength of forgetting, aligning with the latest machine unlearning evaluation criteria \cite{Maini24} and tailoring it to reinforcement unlearning. We define the ``cumulative truth ratio'' to gauge this strength, serving as the foundation for the computation of forget quality. Specifically, the cumulative truth ratio $R_{truth}$ can be written as follows.
\begin{equation}\label{eq:TruthRatio}
   R_{truth}=\frac{\sum_{s\in\tau}\frac{1}{|\mathcal{A}_w|}\sum_{\hat{a}\in\mathcal{A}_w}\pi(\hat{a}|s)}{\sum_{s\in\tau}\pi(\tilde{a}|s)},
\end{equation}
where $\hat{a}$ represents a wrong action from the wrong action set $\mathcal{A}_w$, $\tilde{a}$ is the correct action, and $\tau=((s_1,a_1), \ldots, (s_N,a_N))$ signifies a trajectory in $\mathcal{M}_u$. Here, the correct action is defined as the one capable of moving the agent closer to the target compared to other available actions in a given state, while the remaining actions are considered wrong. In particular, the cumulative truth ratio $R_{truth}$ quantifies the ratio between the average probability of selecting wrong actions and the probability of taking the correct action. Thus, a to-be-unlearned agent is expected to achieve a low $R_{truth}$, while an effectively unlearned agent in the unlearning environment should exhibit a high $R_{truth}$, resembling an agent that has never seen $\mathcal{M}_u$.
For each agent, we obtain $N$ values of $R_{truth}$ each from the first $i$ ($i\in[1,N]$) step(s), and normalize them to serve as an empirical cumulative distribution function (ECDF). 
A two-sample Kolmogorov-Smirnov (KS) test~\cite{massey1951kolmogorov} is conducted between the ECDFs of the unlearned and retained agent to generate a $p$-value to quantify forget quality. 
A high $p$-value indicates a strong forgetting, implying that the $R_{truth}$ distributions of the unlearned and retained agents are identical.

\begin{figure}[ht]
\centering
    \includegraphics[scale=0.21]{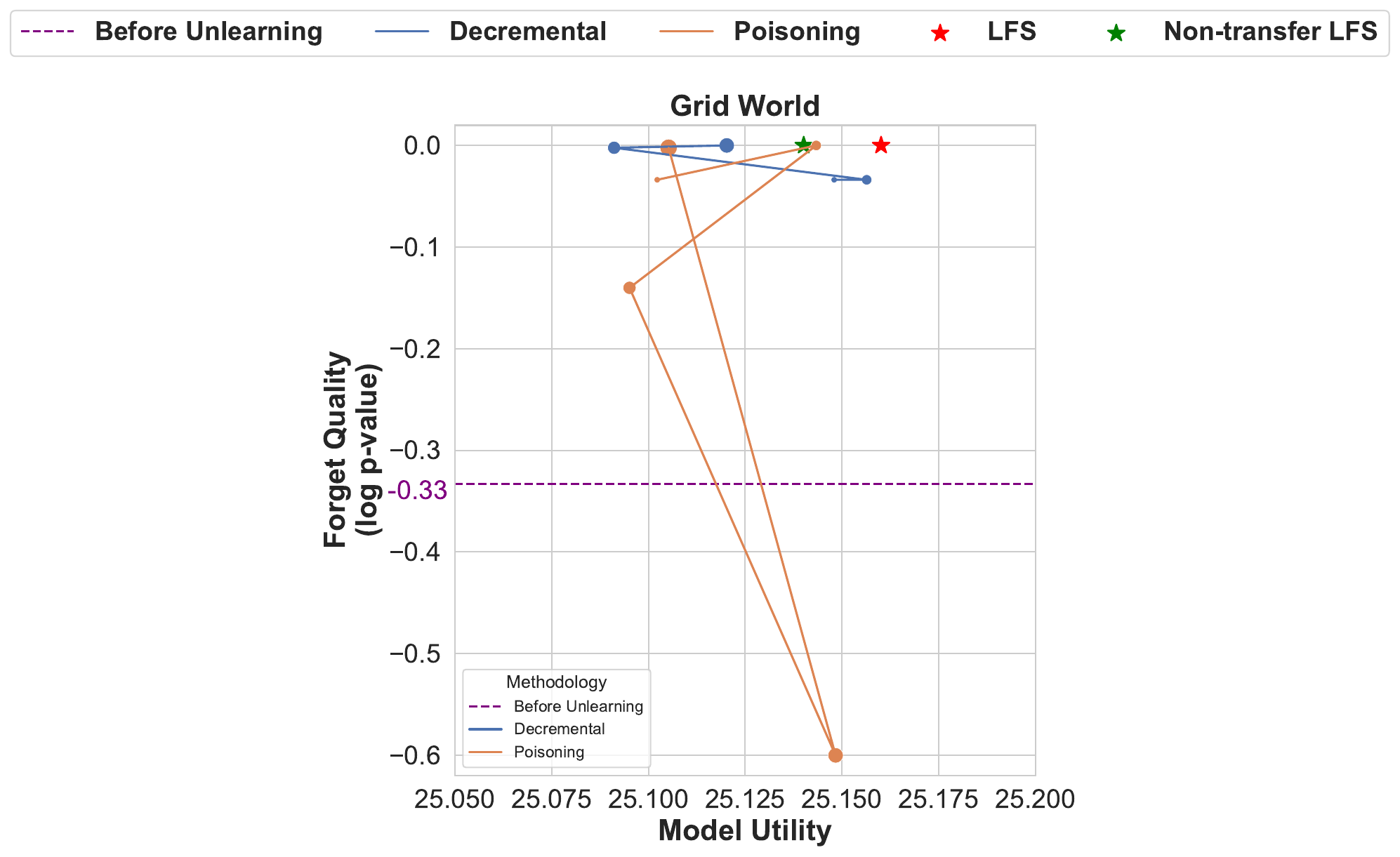}
	\caption{Forget Quality vs. Model Utility. Larger marker represents more unlearning epochs.}
	\label{fig:ForgetQuality}
\end{figure}

We compute the utility of the unlearned/retained model as the average cumulative rewards of it in the retaining environments. 
Figure~\ref{fig:ForgetQuality} illustrates the trade-off between model utility and forget quality during unlearning in grid world, where a larger marker denotes more unlearning epochs. The decremental RL-based method exhibits consistently high forget quality, while the poisoning-based method displays some fluctuations in forget quality but eventually attains a commendable level. This variance may be attributed to the randomness introduced by the RL algorithm in the poisoning-based approach. During the learning of the poisoning strategy, the method gradually converges, ultimately achieving a favorable result. 
Furthermore, both methods attain a higher forget quality than \emph{before unlearning}, indicating the success of the unlearning process.

For model utility, both methods demonstrate robust performance, fluctuating in a narrow range of $(25.075, 25.175)$. 
This performance is comparable to that of the retained models trained by the two baseline methods, i.e., LFS and Non-transfer LFS. 
A closer examination of model utility reveals a dynamic shift in utility in both methods as the unleanring progresses. This oscillation may arise from two factors. First, the decremental RL-based method aims to erase the agent's knowledge, potentially leading to over-forgetting~\cite{hu2023duty}. Second, the poisoning-based method involves an intricate interplay between the RL dynamics and the strategic introduction of poison. This interplay can introduce randomness.
\begin{figure}[ht]
\centering
	\begin{minipage}{1\textwidth}
    \subfigure[\scriptsize{Average steps of the four methods}]{
    \includegraphics[scale=0.2]{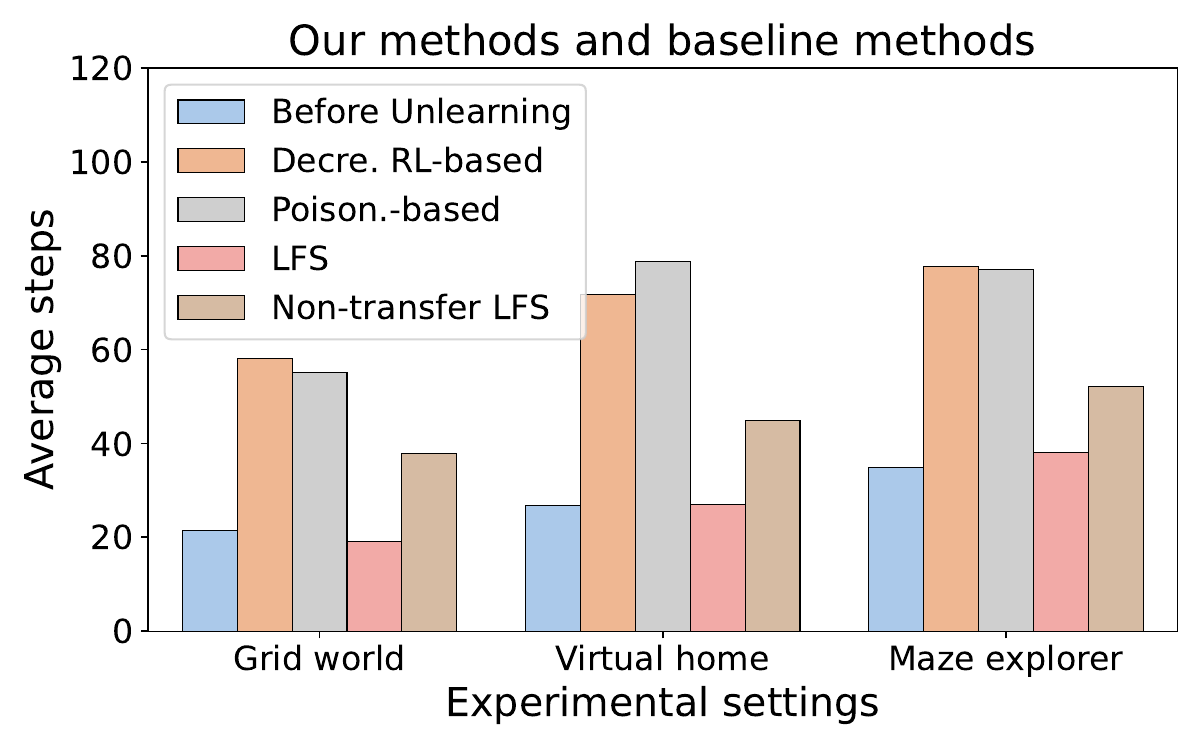}
			\label{fig:BaselineSteps}}
	\subfigure[\scriptsize{Rewards of the four methods}]{
    \includegraphics[scale=0.2]{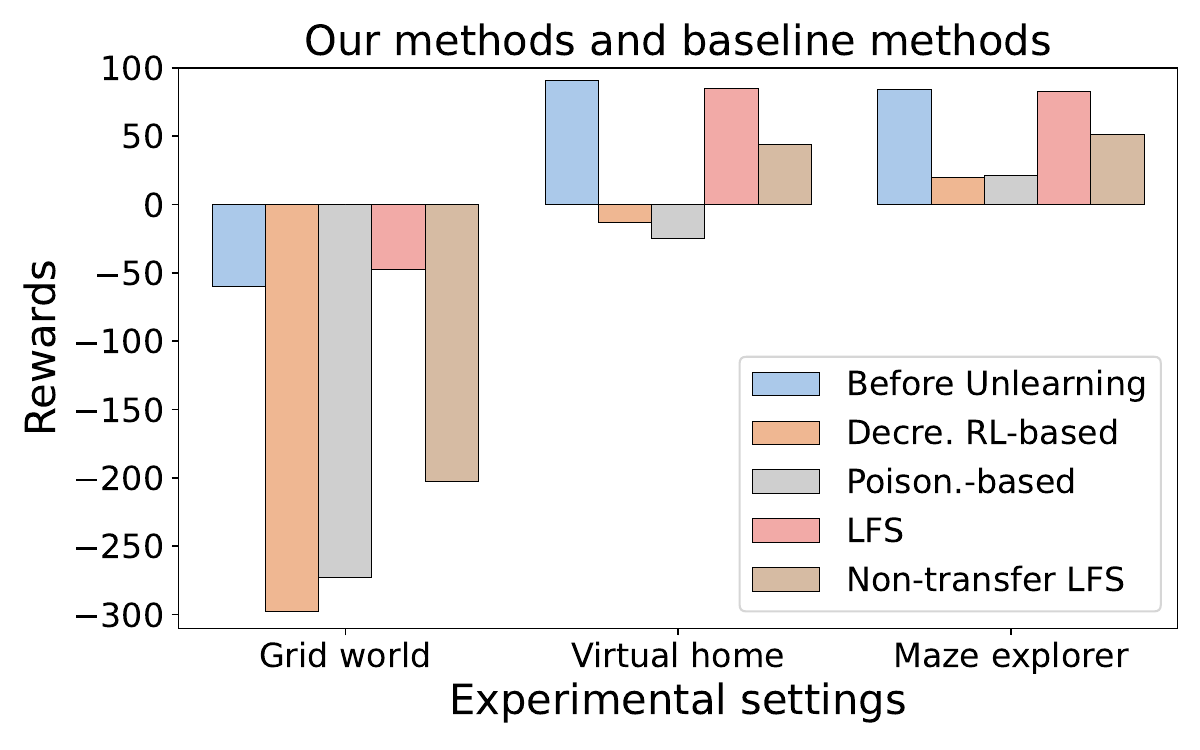}
			\label{fig:BaselineRewards}}\\[2ex]
    \end{minipage}
	\caption{Four methods in the unlearning environment.}
	\vspace{-0mm}
	\label{fig:Baseline}
\end{figure}

Next, we make detailed comparisons between our methods and the baselines.
Figure \ref{fig:Baseline} shows that the unlearning results of the LFS baseline method in all three experimental settings are subpar. The agent's performance in the unlearning environment remains nearly unchanged before and after unlearning. The reason for this result lies in the agent's ability to generalize knowledge from other environments and apply it to the unlearning environment, despite never having encountered it before.
During training, the agent learns underlying rules and strategies from various environments. For instance, in the grid world setting, the agent acquires knowledge that obstacles should be avoided while collecting the target as quickly as possible. This learned knowledge, even if it was acquired in different environments, enables the agent to still perform well in unseen environments, including the unlearning environment. 

In contrast, the unlearning results of the Non-transfer LFS method surpass those of the regular LFS due to the limitation on its generalizability. Notably, Non-transfer LFS shows a considerable performance deterioration in the unlearning environment while maintaining effectiveness in other environments. These outcomes underscore the effectiveness of using an inverse loss function to minimize the agent's cumulative reward in the unlearning environment.

\begin{figure}[ht]
\centering
	\begin{minipage}[c]{1\textwidth}
    \subfigure[\scriptsize{The average number of steps before and after unlearning}]{
    \centering\includegraphics[scale=0.115]{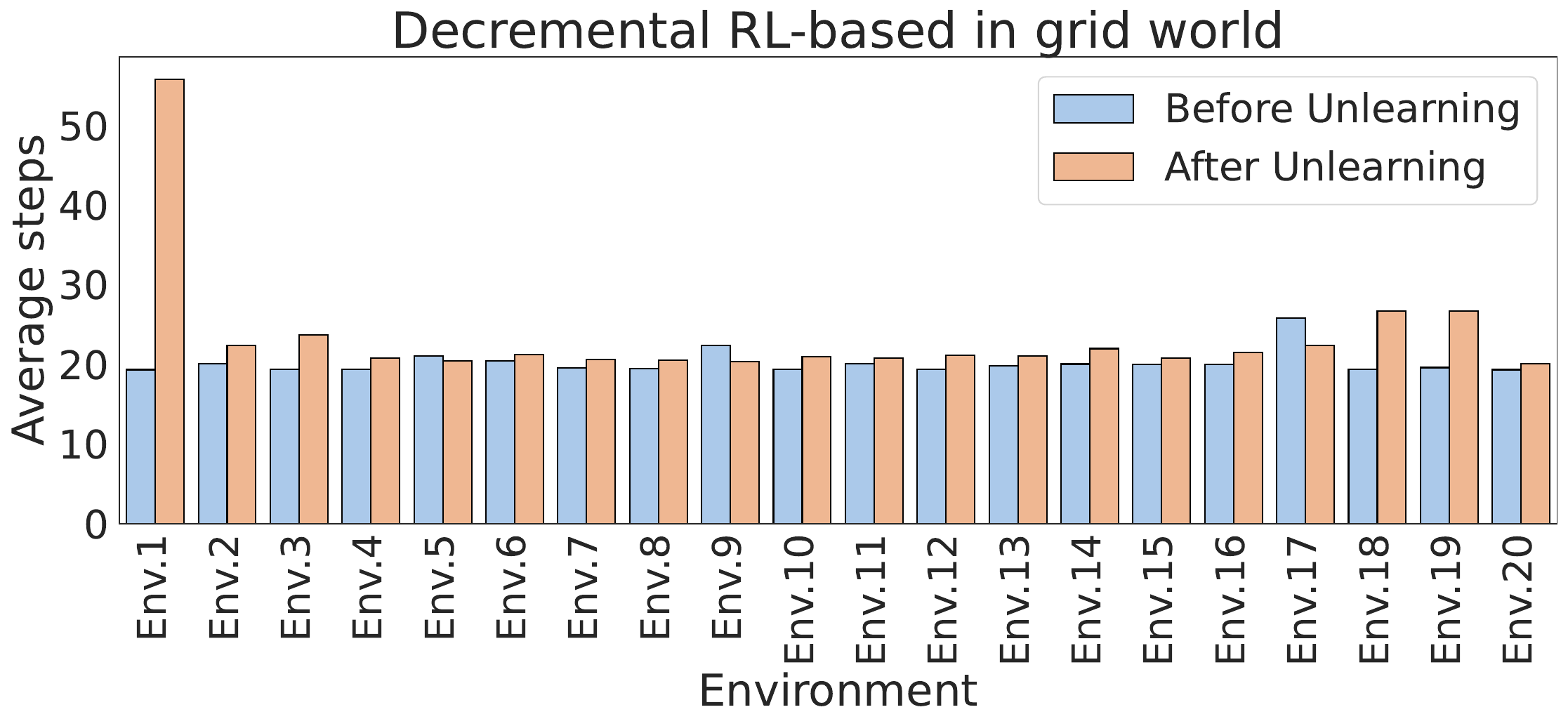}
			\label{fig:GridSize10StepsMethod2}}
	\subfigure[\scriptsize{The average rewards before and after unlearning}]{
    \centering\includegraphics[scale=0.115]{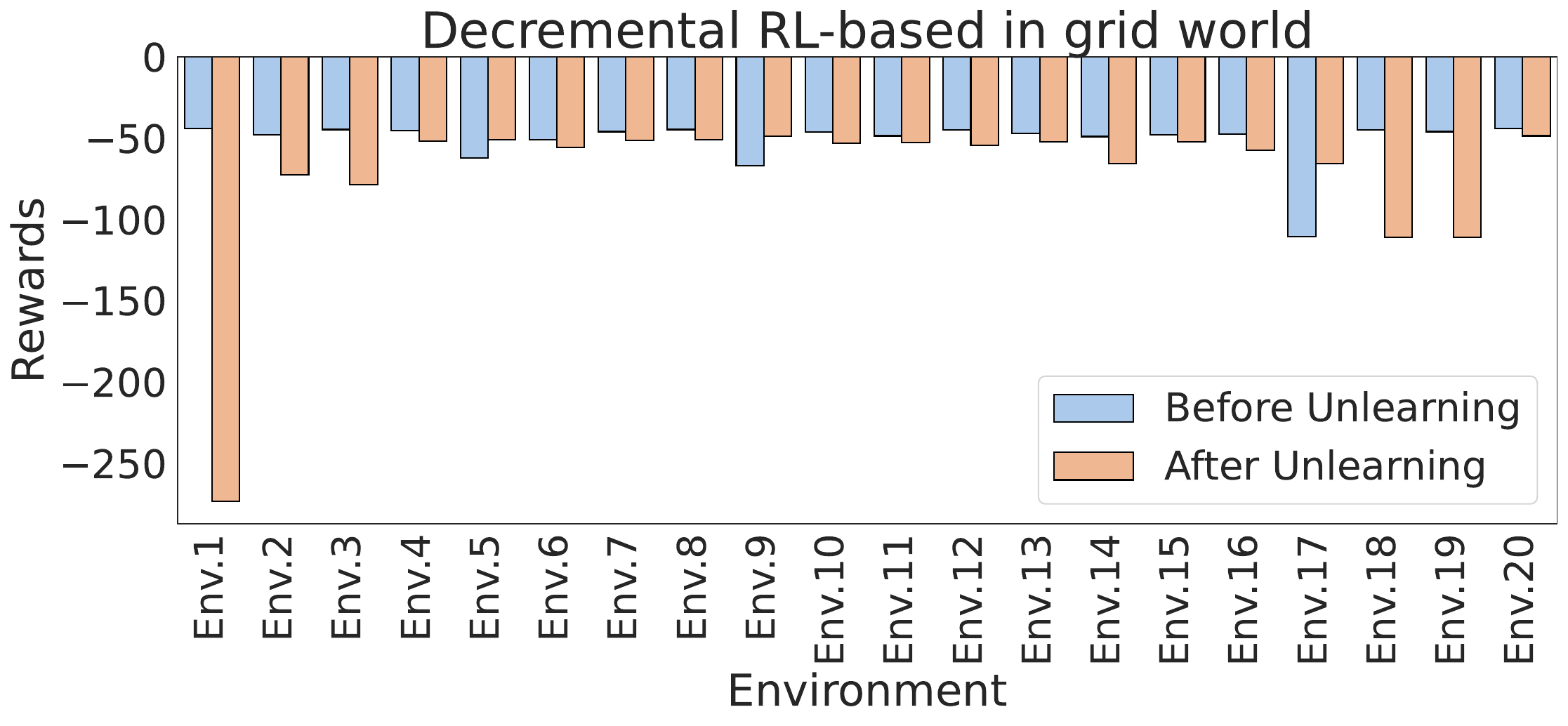}
			\label{fig:GridSize10RewardsMethod2}}
    \end{minipage}
	\caption{The decremental RL-based method in Grid World, where Environment 1 is the unlearning environment.}
	\label{fig:GridOverall}
\end{figure}

Figure \ref{fig:GridOverall} presents the overall performance of the decremental RL-based method in the grid world setting. The obtained results provide compelling evidence of the profound impact of the unlearning process on the agent's performance, as evident from the average number of steps taken and the average received rewards metrics. 
Following unlearning, the agent demonstrates a substantial increase in the average number of steps taken and a notable reduction in the average received rewards compared to the pre-unlearning stage. 

For example, in Figure \ref{fig:GridSize10StepsMethod2}, after unlearning, it is evident that the average number of steps taken by the agent in the unlearning environment substantially increases from $19.34$ to $55.8$, while in Figure \ref{fig:GridSize10RewardsMethod2}, its reward decreases from $-44$ to $-273.5$. These findings indicate a significant performance reduction in the unlearning environment, which can be interpreted as a successful unlearning outcome.
Conversely, in the retained environments, we observe minimal changes in the agent's steps and rewards. This implies a successful preservation of performance in these environments. 

\begin{figure}[ht]
\centering
	\begin{minipage}{1\textwidth}
    \subfigure[c][\scriptsize{The average number of steps before and after unlearning}]{
    \centering\includegraphics[scale=0.115]{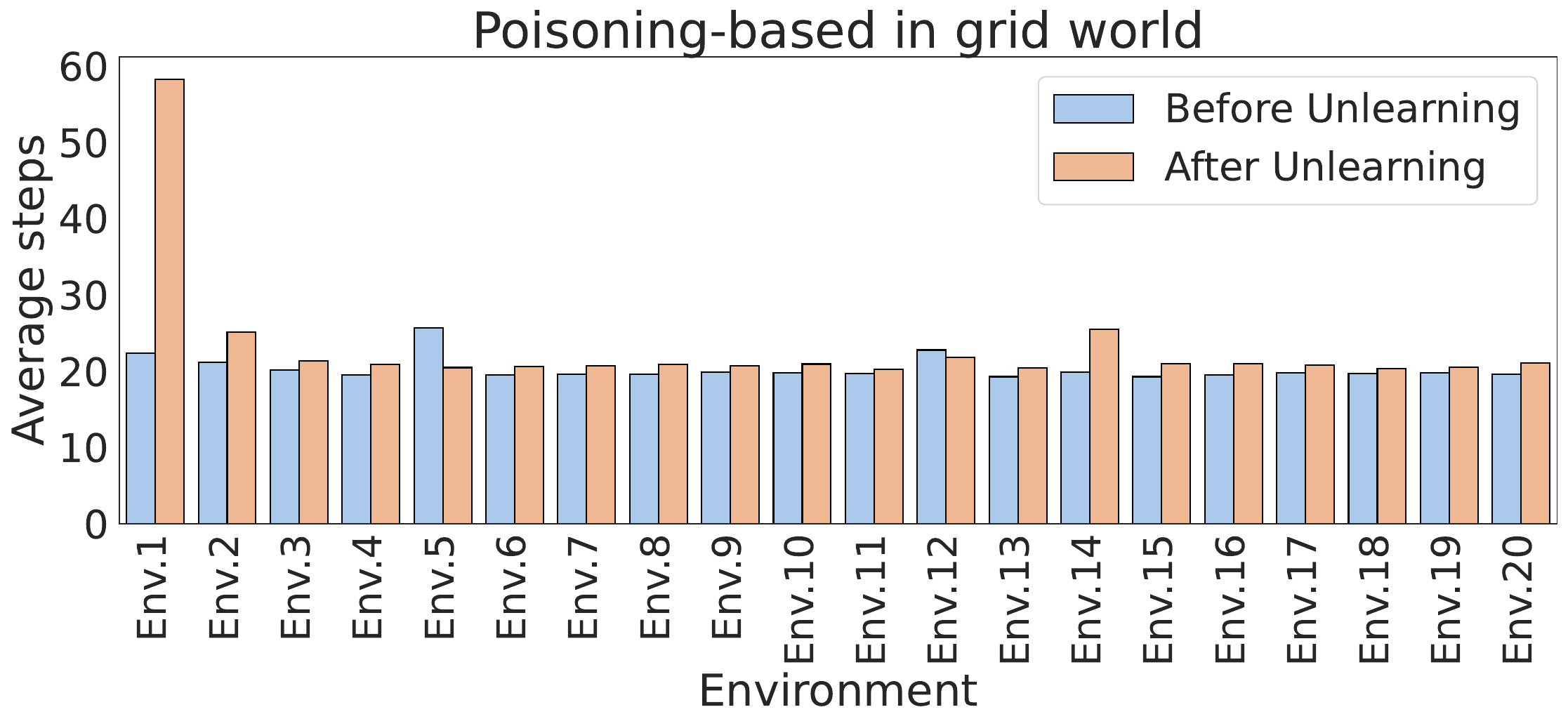}
			\label{fig:GridSize10Steps}}
	\subfigure[c][\scriptsize{The average rewards before and after unlearning}]{
    \centering\includegraphics[scale=0.115]{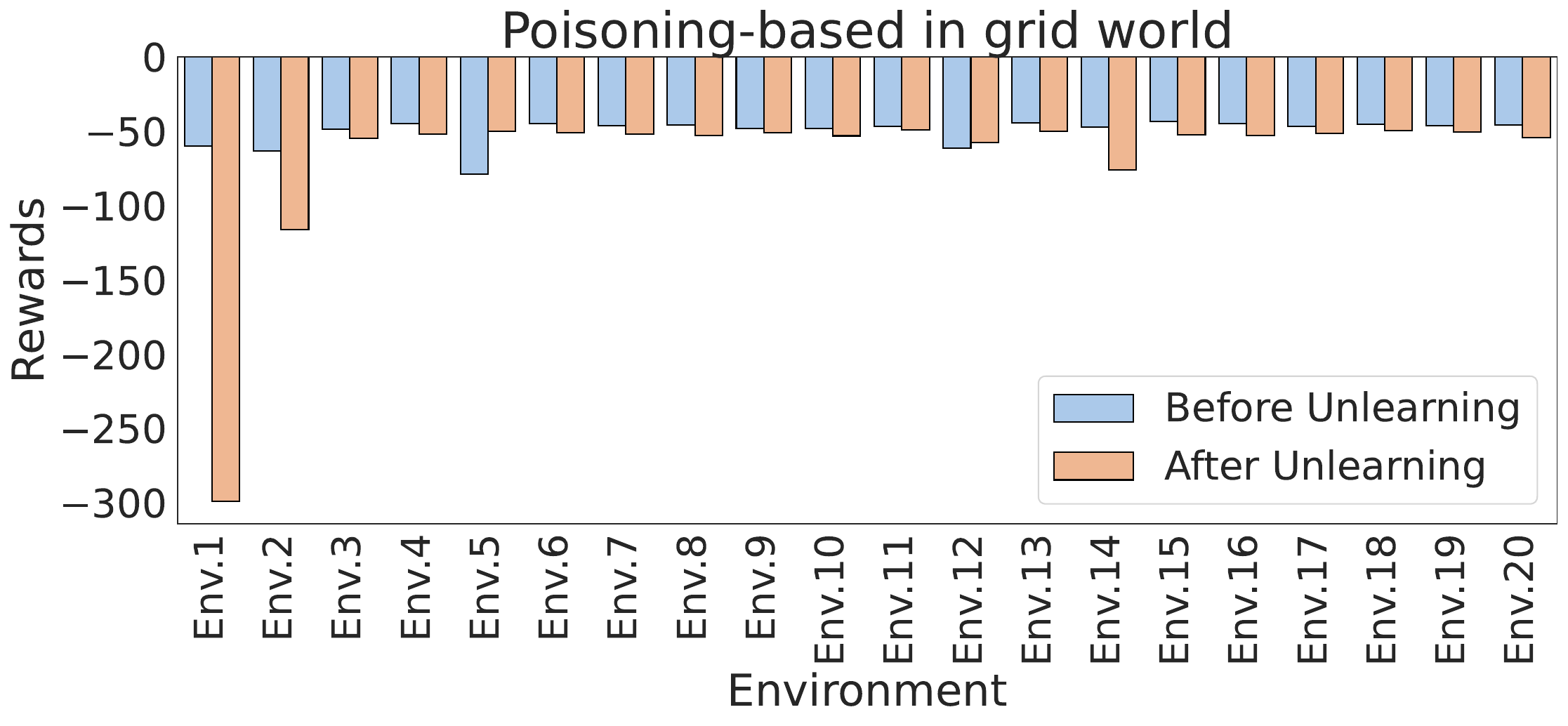}
			\label{fig:GridSize10Rewards}}
    \end{minipage}
	\caption{The poisoning-based method in Grid World, where Environment 1 is the unlearning environment.}
	\label{fig:GridOverall2}
\end{figure}

Figure \ref{fig:GridOverall2} presents the overall performance of the poisoning-based method in the grid world setting. It exhibits a similar trend to the decremental RL-based method. The reason for this similarity lies in the shared objective of both methods, which is to degrade the agent's performance within the targeted unlearning environment while maintaining its performance in other environments. As a result, both methods effectively achieve the goal of reinforcement unlearning by selectively modifying the agent's behavior within the specified context.

However, upon closer comparison between Figures \ref{fig:GridOverall} and \ref{fig:GridOverall2}, we can observe slight differences in the performance of the two methods in some remaining environments, such as Environments 18 and 19. In these environments, the poisoning-based method maintains almost unchanged steps and rewards between the pre-unlearning and post-unlearning stages, while the decremental RL-based method does not achieve this. This result suggests that the decremental RL-based method can potentially suffer from the over-unlearning issue to some extent, while the poisoning-based method demonstrates its ability to overcome this issue and retain better performance in the remaining environments after unlearning. These findings highlight the different characteristics and strengths of the two unlearning methods.

\begin{figure}[hbt]
\centering
	\begin{minipage}{.7\textwidth}
    \subfigure[c][\scriptsize{The average number of steps before and after unlearning}]{
    \centering\includegraphics[scale=0.115]{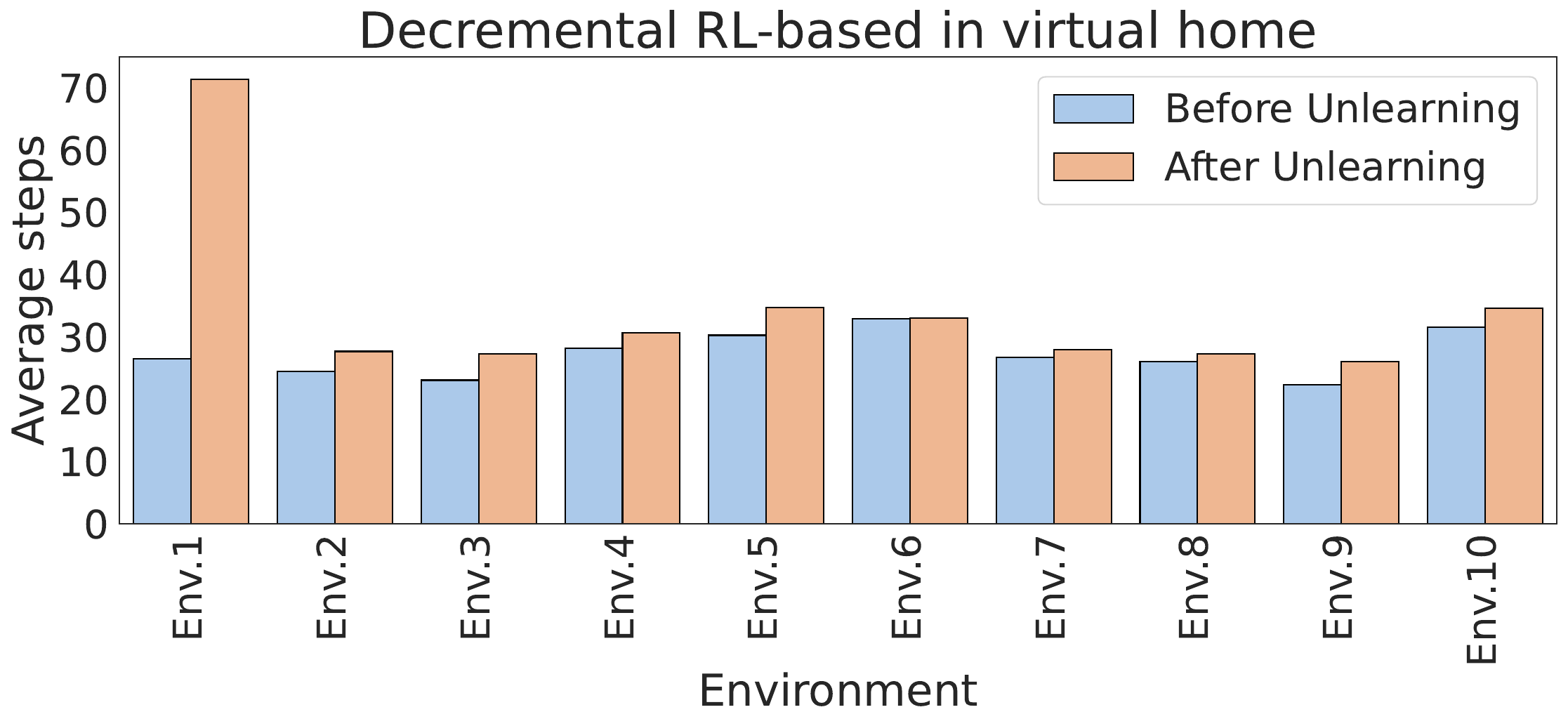}
			\label{fig:VHomeRobustSteps}}
	\subfigure[c][\scriptsize{The average rewards received by the agent before and after unlearning}]{
    \centering\includegraphics[scale=0.115]{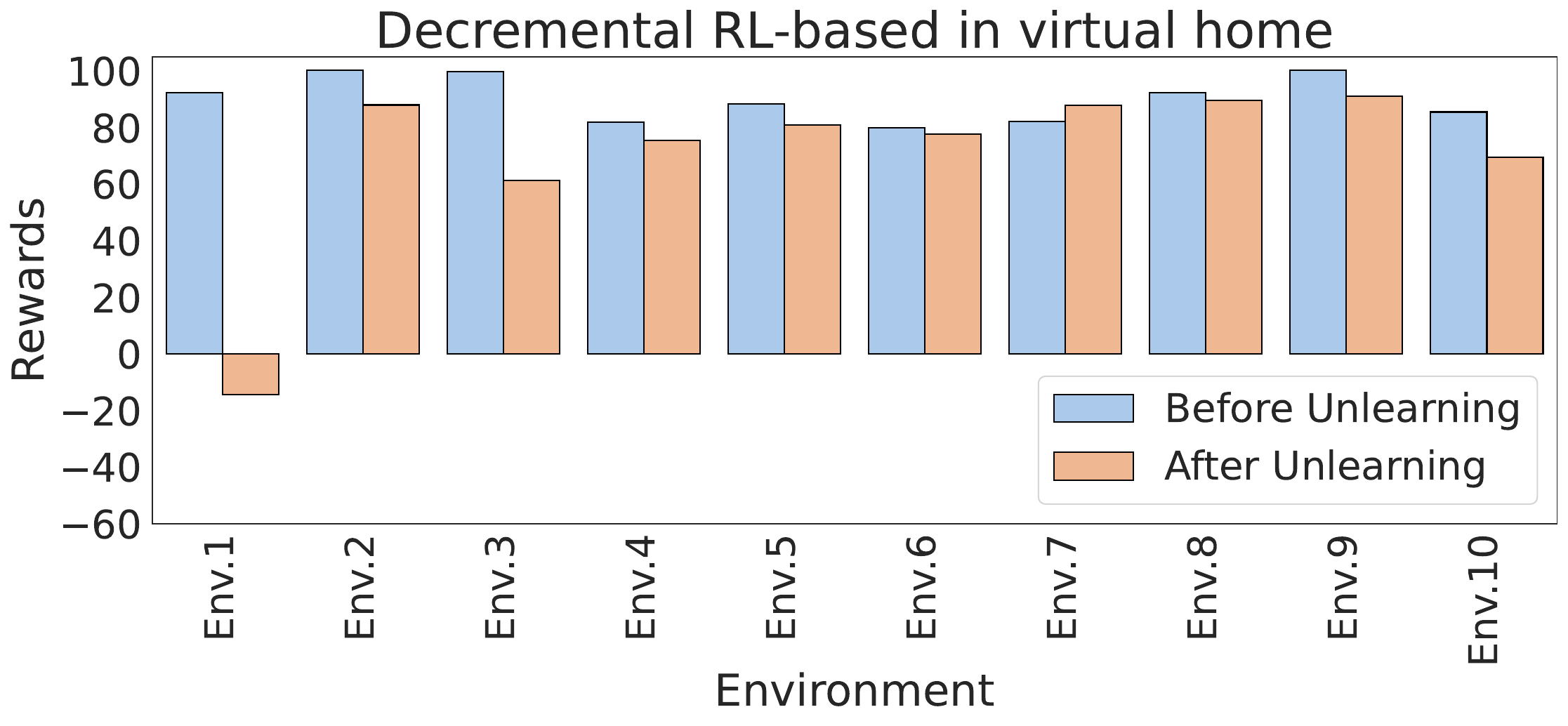}
			\label{fig:VHomeRobustRewards}}
    \end{minipage}
	\caption{The decremental RL-based method in Virtual Home, where Environment 1 is the unlearning environment.}
	\label{fig:VHomeOverall}
\end{figure}

\begin{figure}[hbt]
\centering
	\begin{minipage}{.7\textwidth}
    \subfigure[c][\scriptsize{The average number of steps before and after unlearning}]{
    \centering\includegraphics[scale=0.115]{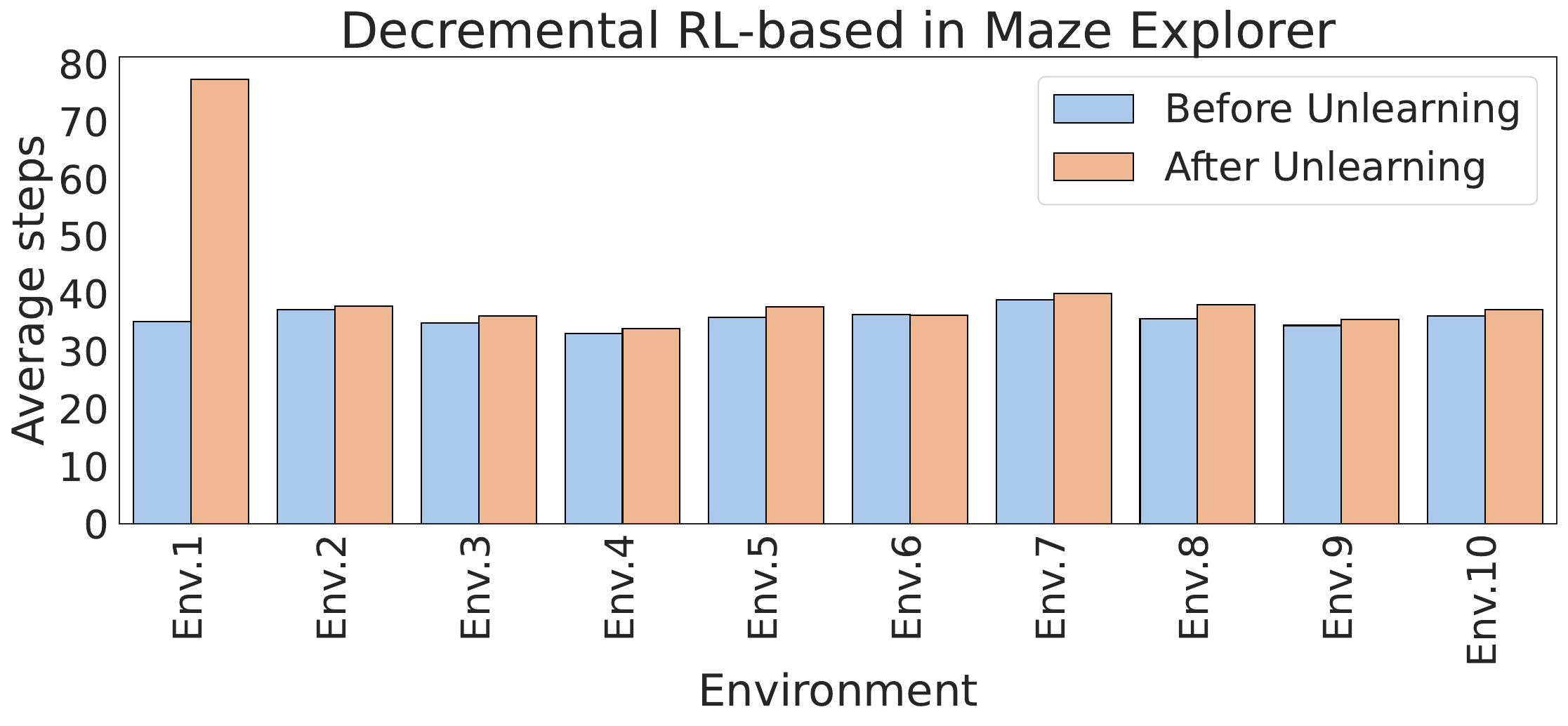}
			\label{fig:MazeSteps}}
	\subfigure[c][\scriptsize{The average rewards received by the agent before and after unlearning}]{
    \centering\includegraphics[scale=0.115]{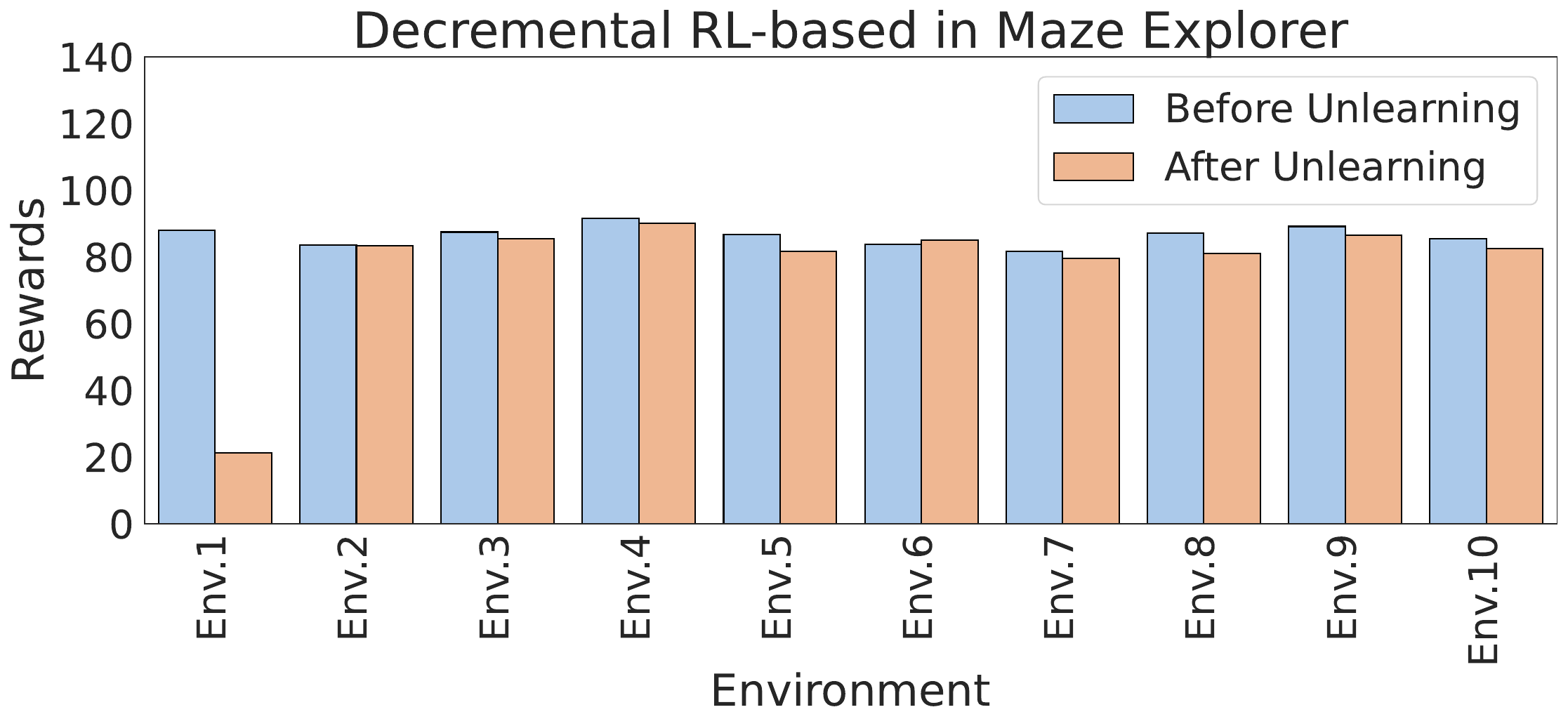}
			\label{fig:MazeRewards}}
    \end{minipage}
	\caption{The decremental RL-based method in Maze Explorer, where Environment 1 is the unlearning environment.}
	\label{fig:MazeOverall}
\end{figure}

Figures \ref{fig:VHomeOverall} and \ref{fig:MazeOverall} demonstrate the overall performance of the decremental RL-based method in the context of virtual home and maze explorer, respectively. In the two scenarios, the agent's behavior exhibits a notable increase in steps taken and a significant decrease in rewards achieved after the unlearning process. The reason for these trends in the two scenarios is rooted in the fundamental nature of reinforcement unlearning. The unlearning process seeks to selectively modify the agent's behavior to forget specific environments or aspects of its learning history. Thus, the agent must re-explore and adapt to new circumstances, leading to fluctuations in its performance. 


\begin{figure}[hbt]
\centering
	\begin{minipage}{.7\textwidth}
    \subfigure[c][\scriptsize{The average number of steps before and after unlearning}]{
    \centering\includegraphics[scale=0.115]{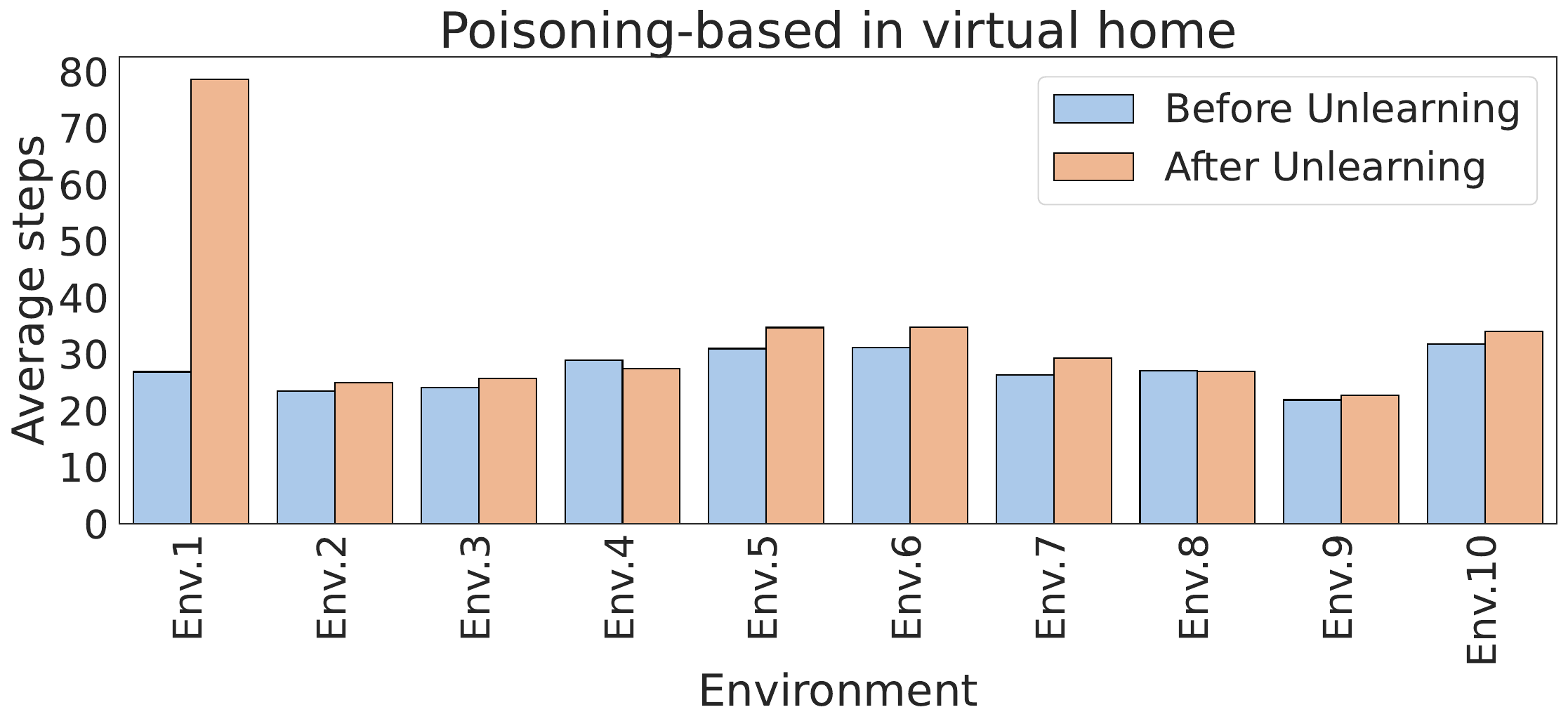}
			\label{fig:VHomeRobustStepsMethod2}}
	\subfigure[c][\scriptsize{The average rewards received by the agent before and after unlearning}]{
    \centering\includegraphics[scale=0.115]{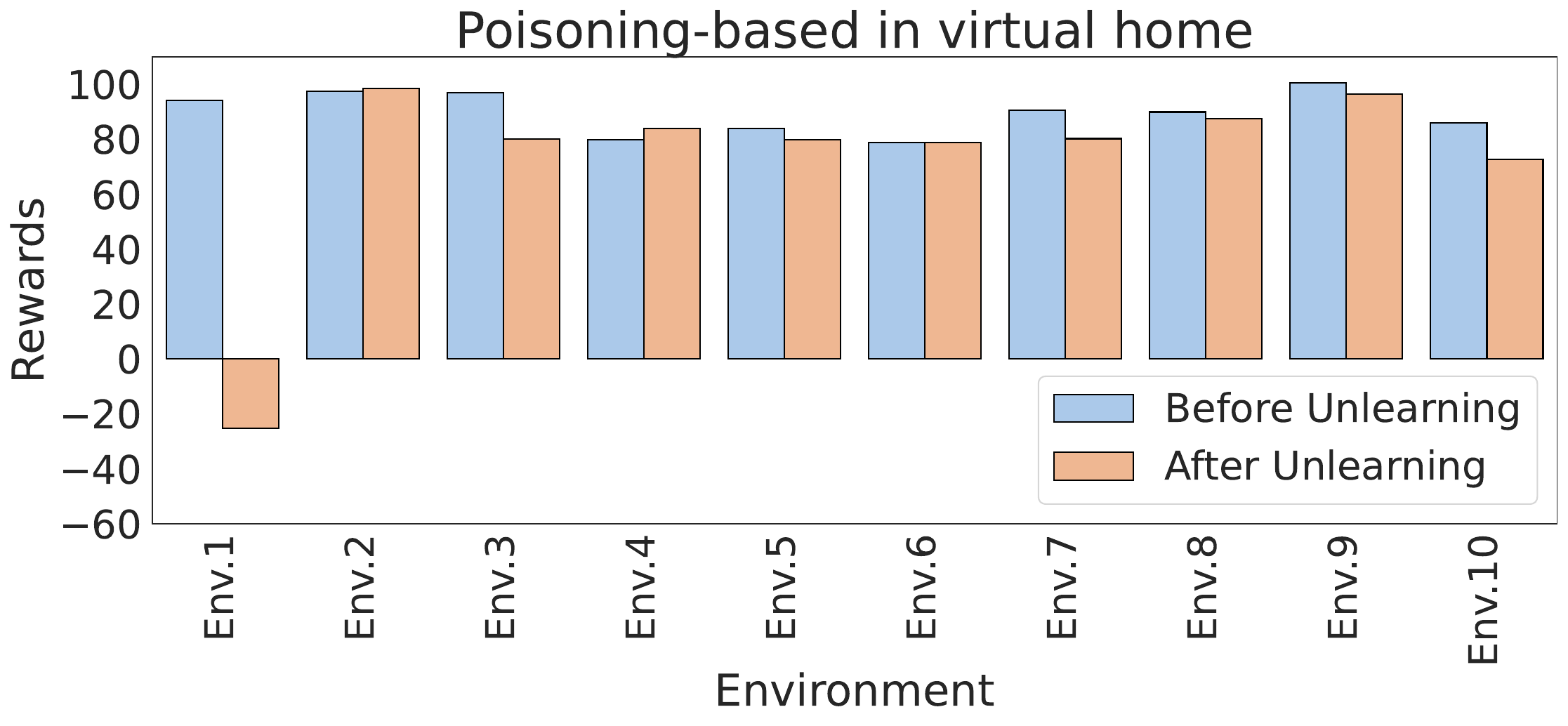}
			\label{fig:VHomeRobustRewardsMethod2}}
    \end{minipage}
	\caption{The poisoning-based method in Virtual Home, where Environment 1 is the unlearning environment.}
	\label{fig:VHomeOverall2}
\end{figure}

\begin{figure}[hbt]
\centering
	\begin{minipage}{.7\textwidth}
    \subfigure[c][\scriptsize{The average number of steps before and after unlearning}]{
    \centering\includegraphics[scale=0.115]{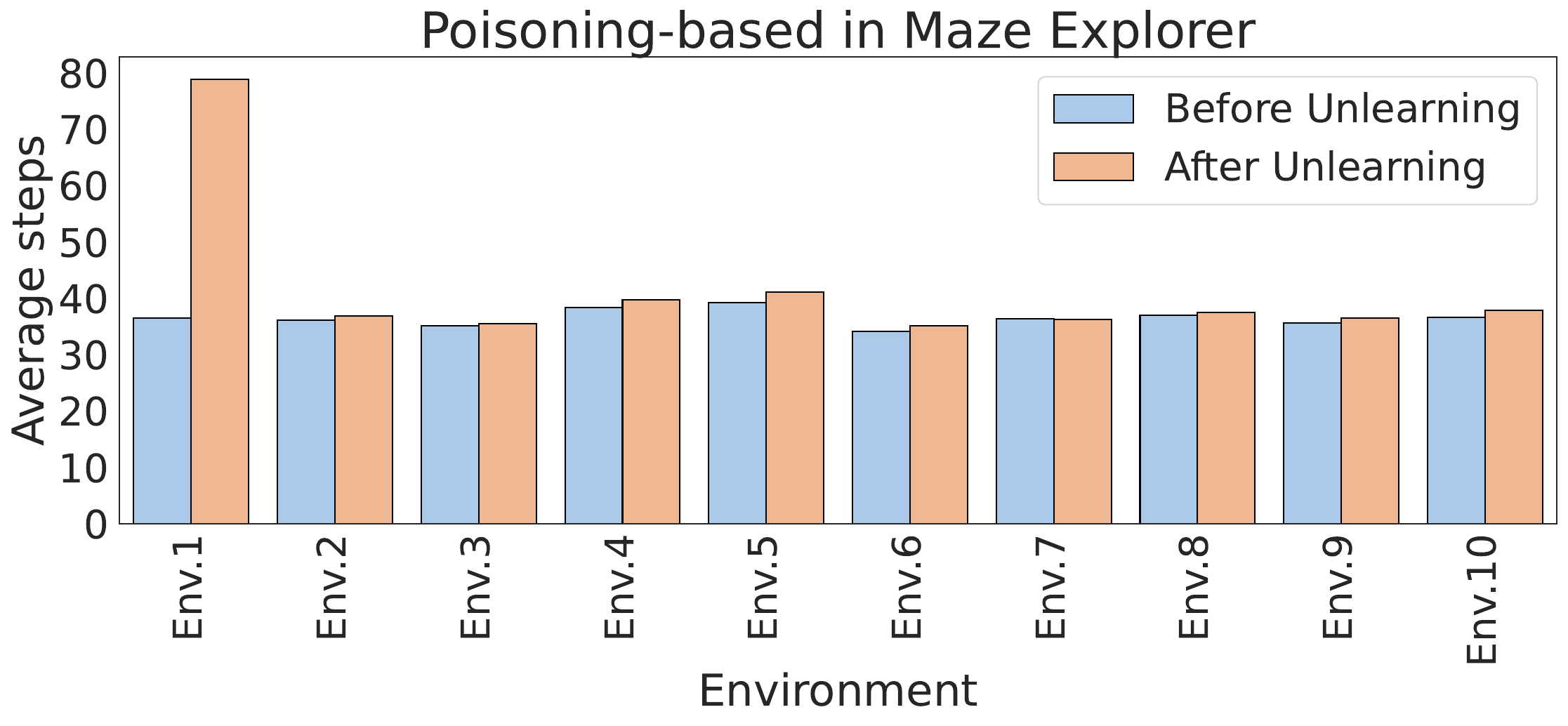}
			\label{fig:MazeStepsMethod2}}
	\subfigure[c][\scriptsize{The average rewards received by the agent before and after unlearning}]{
    \centering\includegraphics[scale=0.115]{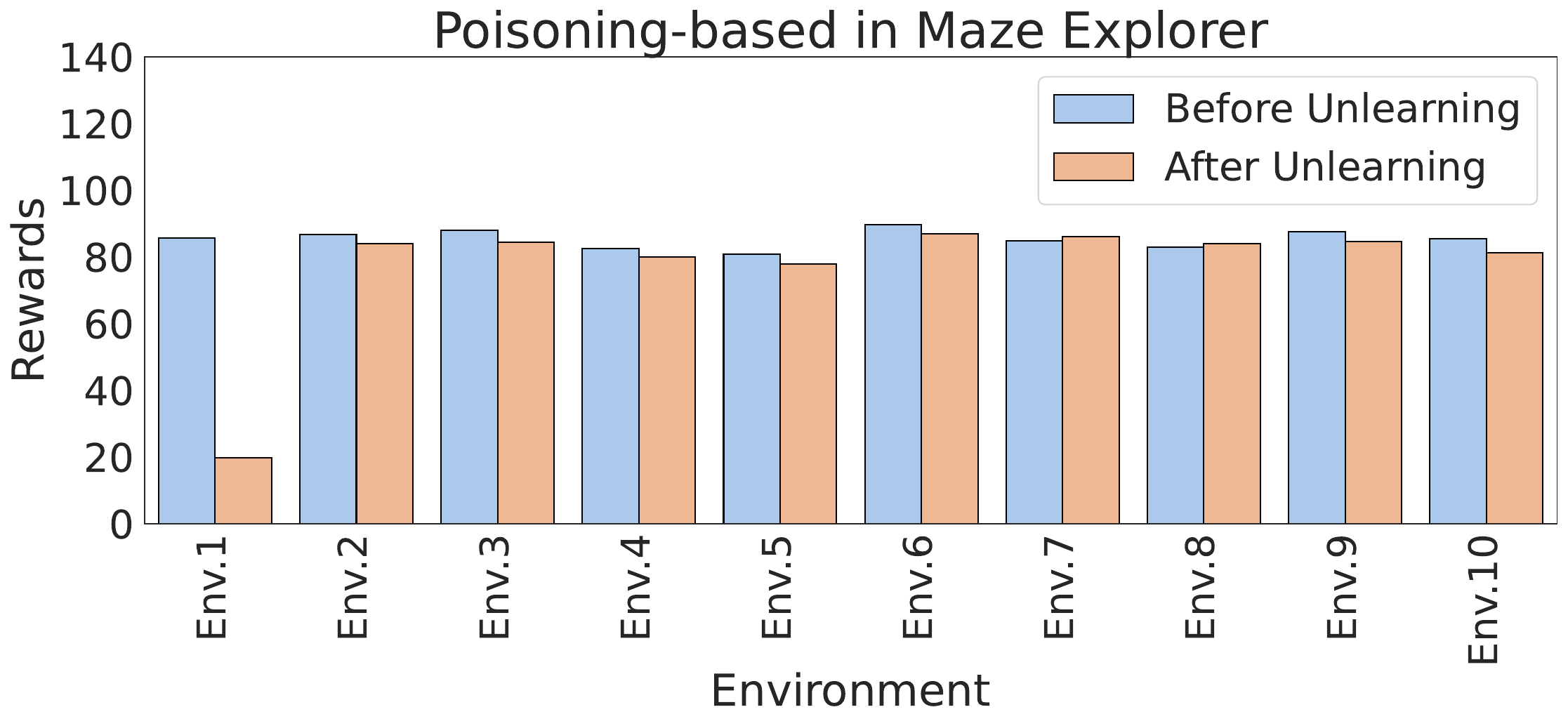}
			\label{fig:MazeRewardsMethod2}}
    \end{minipage}
	\caption{The poisoning-based method in Maze Explorer, where Environment 1 is the unlearning environment.}
	\label{fig:MazeOverall2}
\end{figure}

Figures \ref{fig:VHomeOverall2} and \ref{fig:MazeOverall2} provide a comprehensive view of the poisoning-based method's performance in the virtual home and maze explorer settings, respectively. Remarkably, the performance trend in the two scenarios is similar to that of the decremental RL-based method. This observation reinforces the effectiveness of the poisoning-based approach in reinforcement unlearning, as it consistently achieves the objective of degrading the agent's performance in the targeted unlearning environment while preserving its capabilities in other environments. The consistent performance trend across different settings shows the method's versatility and potential applicability in various RL scenarios.


\subsection{Hyperparameter Study}\label{sub:hyperparameter}

\noindent\textbf{Impact of Environment Size.~} The alteration in the environment size allows us to evaluate how well the unlearning methods adapt and perform across different scales.

\begin{figure}[ht]
\centering
	\begin{minipage}{1\textwidth}
    \subfigure[\scriptsize{Steps of decremental RL-based}]{
    \includegraphics[scale=0.15]{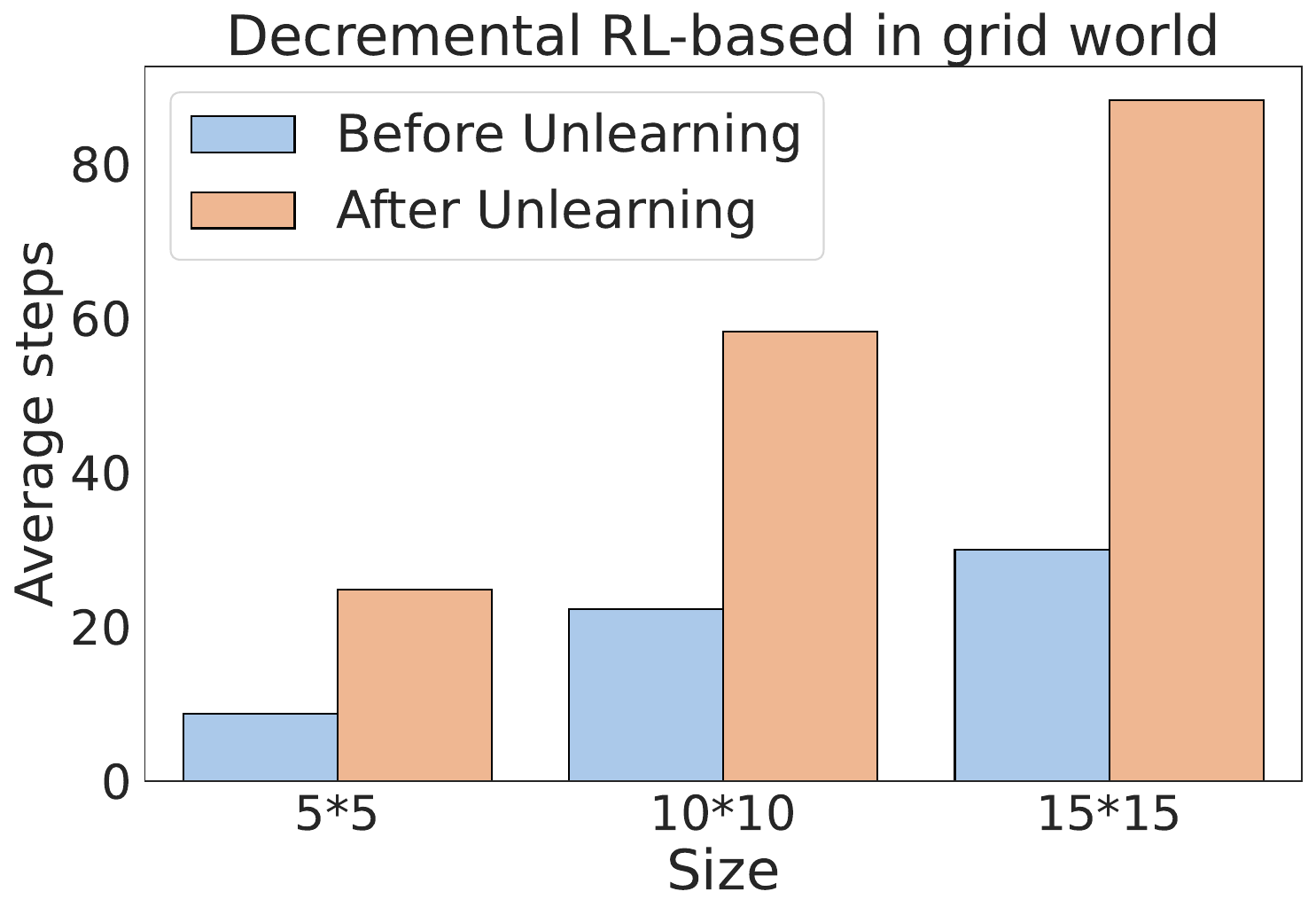}
			\label{fig:GridSizeStepsMethod1}}\vspace{-3mm}
	\subfigure[\scriptsize{Rewards of decremental RL-based}]{
    \includegraphics[scale=0.15]{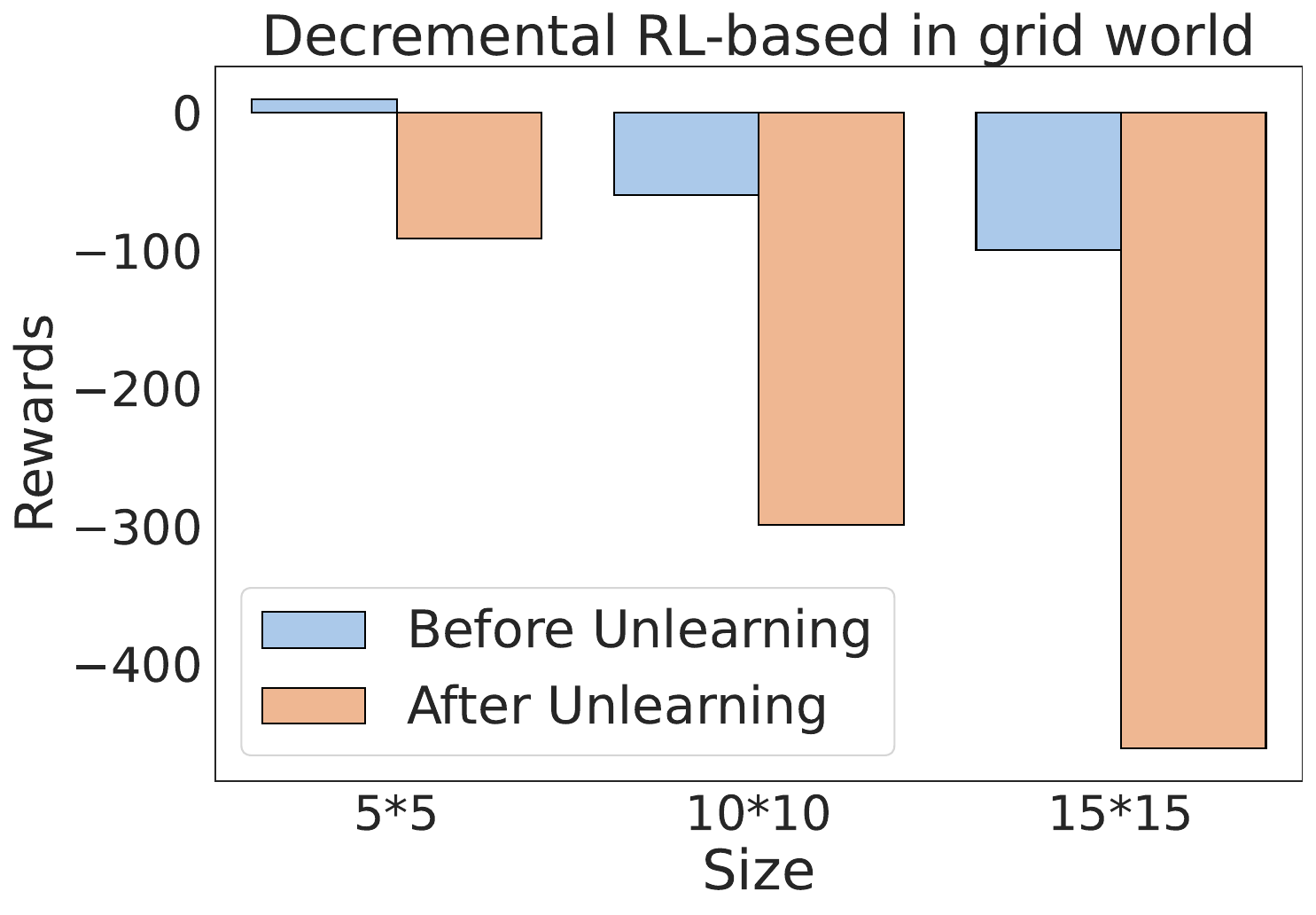}
			\label{fig:GridSizeRewardsMethod1}}\\[2ex]
    \subfigure[\scriptsize{Steps of poisoning-based}]{
    \includegraphics[scale=0.15]{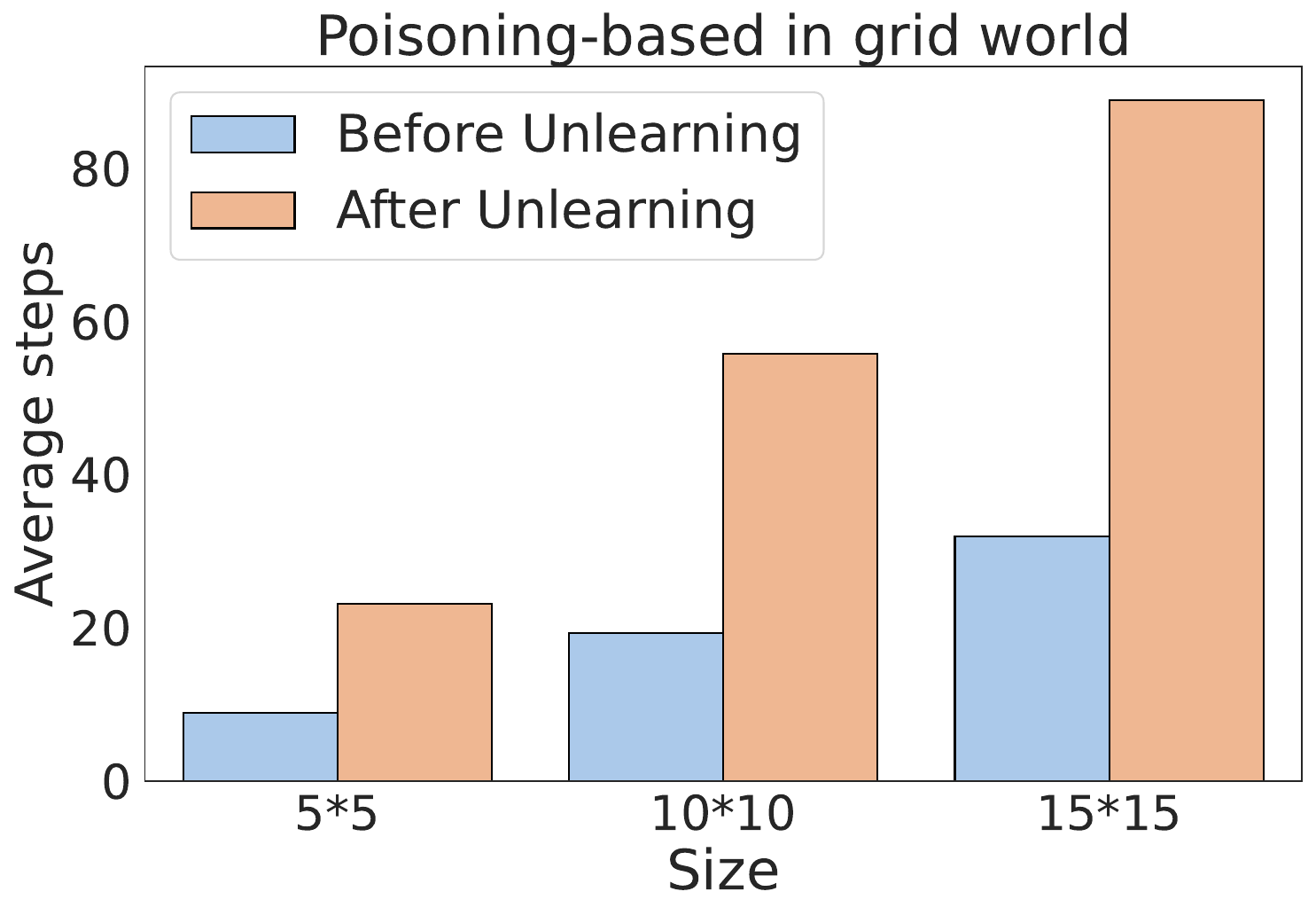}
			\label{fig:GridSizeStepsMethod2}}
	\subfigure[\scriptsize{Rewards of poisoning-based}]{
    \includegraphics[scale=0.15]{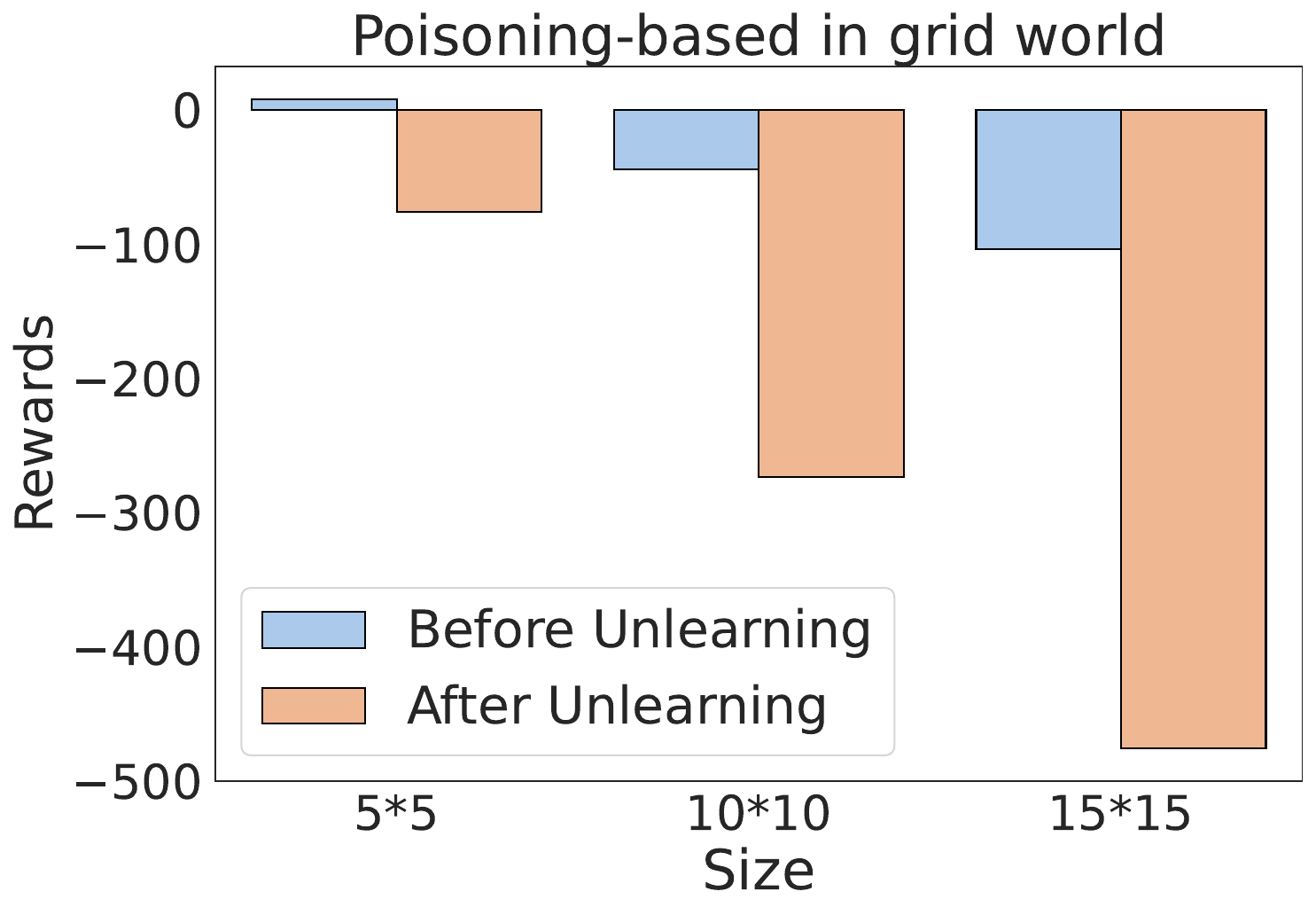}
			\label{fig:GridSizeRewardsMethod2}}\\[2ex]
    \end{minipage}
	\caption{The decremental RL-based and poisoning-based methods in Grid World with different sizes.}
	\label{fig:GridSize}
\end{figure}

In grid world, we extend the size of the environment from $5\times 5$ to $15\times 15$, resulting in a larger grid. Figure \ref{fig:GridSize} visually depicts the impact of this increased environment size on both methods.
As illustrated in the figure, we observe that with the expansion of the environment, the discrepancy in rewards between the pre-unlearning and post-unlearning stages is magnified for both methods. The reason is that the larger grid size introduces a greater number of states for the agent to navigate. Thus, unlearning becomes a more challenging task as the agent must modify its learned behavior to adapt to the enlarged environment.
Also, this magnification effect can be attributed to the increased number of possible trajectories and interactions in the expanded grid world. 
After unlearning, the behavior of the agent in the unlearning environment becomes randomized. Thus, a wider range of possible trajectories and interactions often leads to longer paths taken by the agent, thereby resulting in lower rewards attained.
Hence, the discrepancy in rewards between the pre-unlearning and post-unlearning stages becomes more pronounced.

\begin{figure}[ht]
\centering
	\begin{minipage}{1\textwidth}
    \subfigure[\scriptsize{Steps of decremental RL-based}]{
    \includegraphics[scale=0.16]{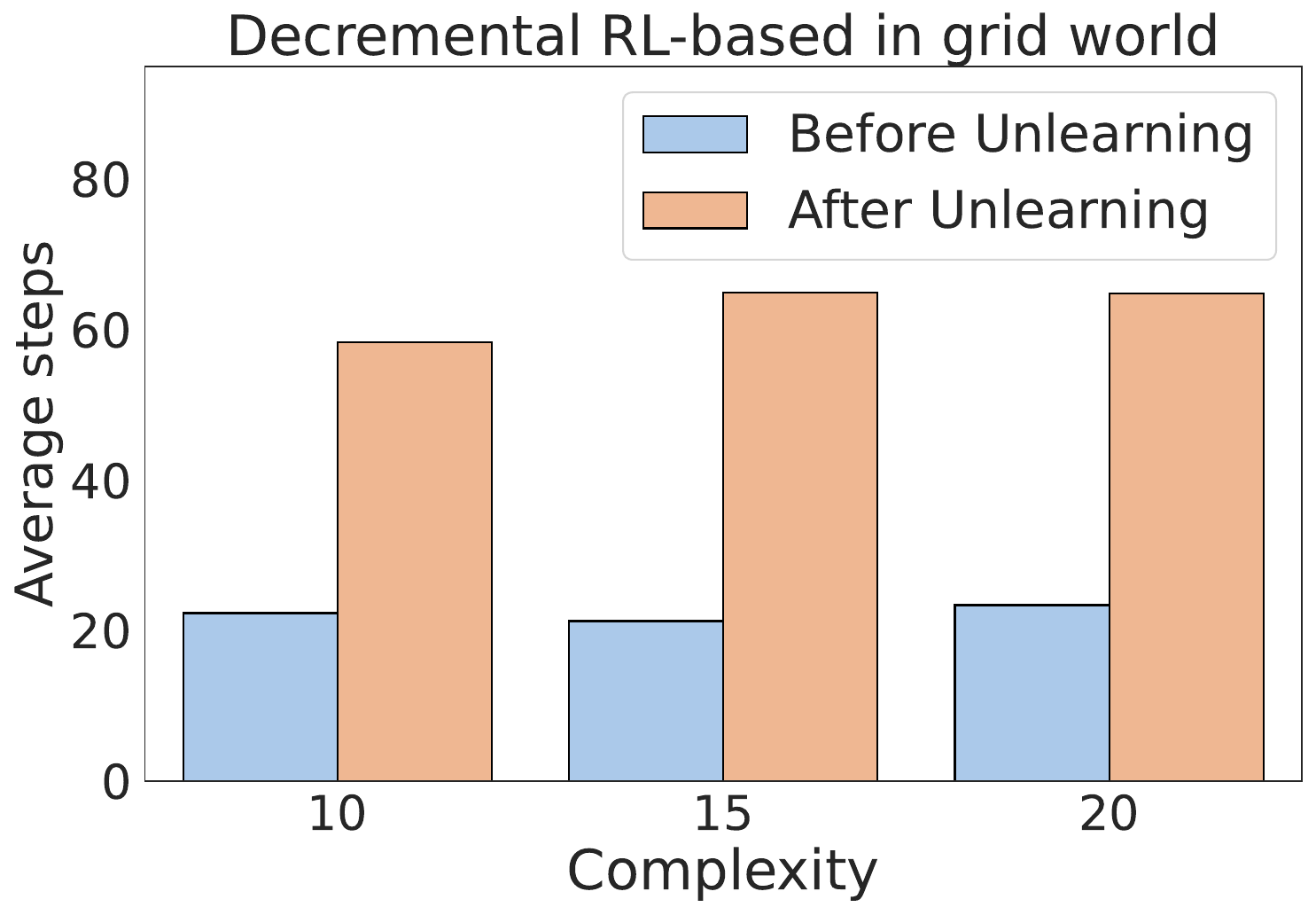}
			\label{fig:GridComplexityStepsMethod1}}
	\subfigure[\scriptsize{Rewards of decremental RL-based}]{
    \includegraphics[scale=0.16]{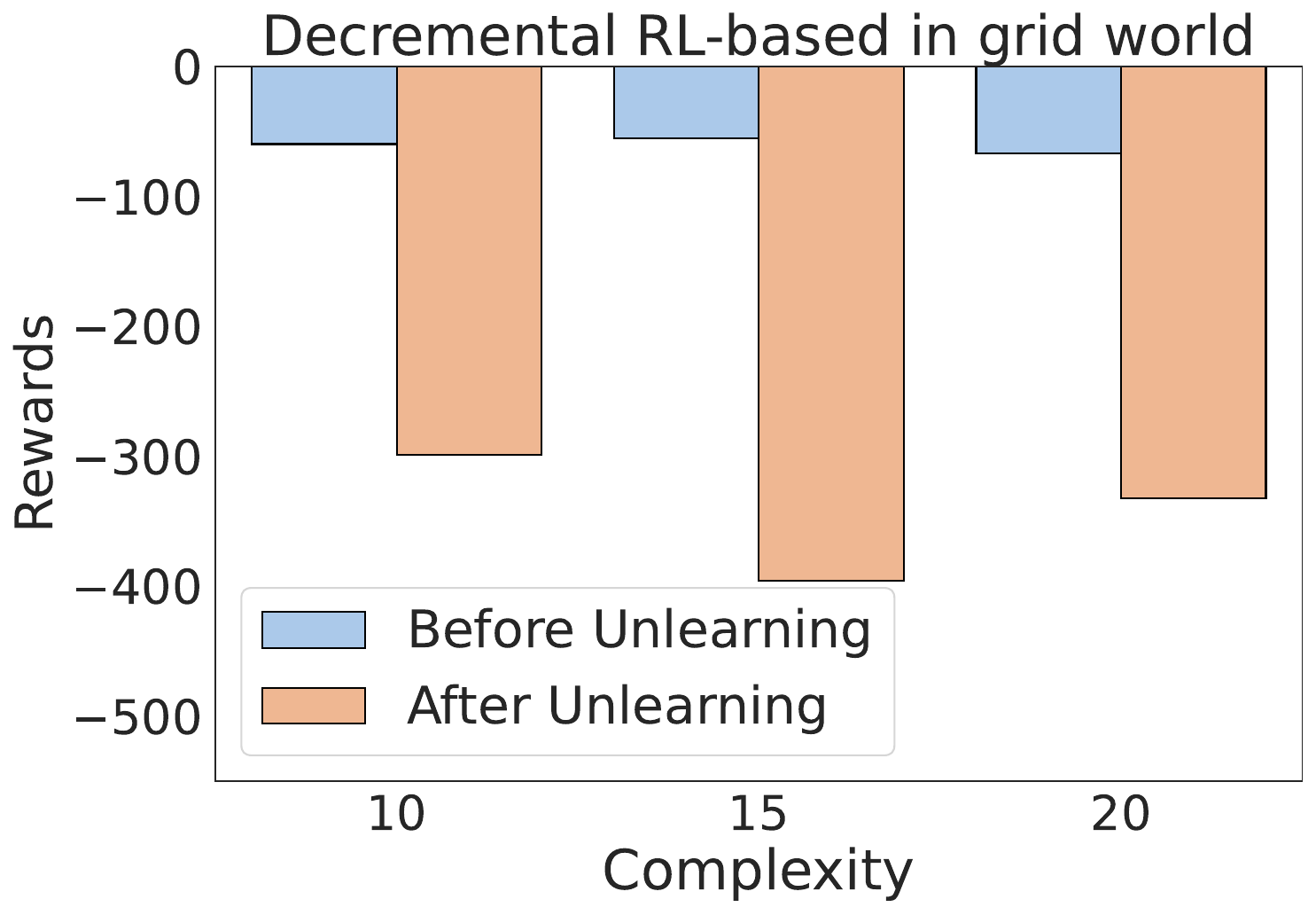}
			\label{fig:GridComplexityRewardsMethod1}}\\[2ex]
    \subfigure[\scriptsize{Steps of poisoning-based}]{
    \includegraphics[scale=0.16]{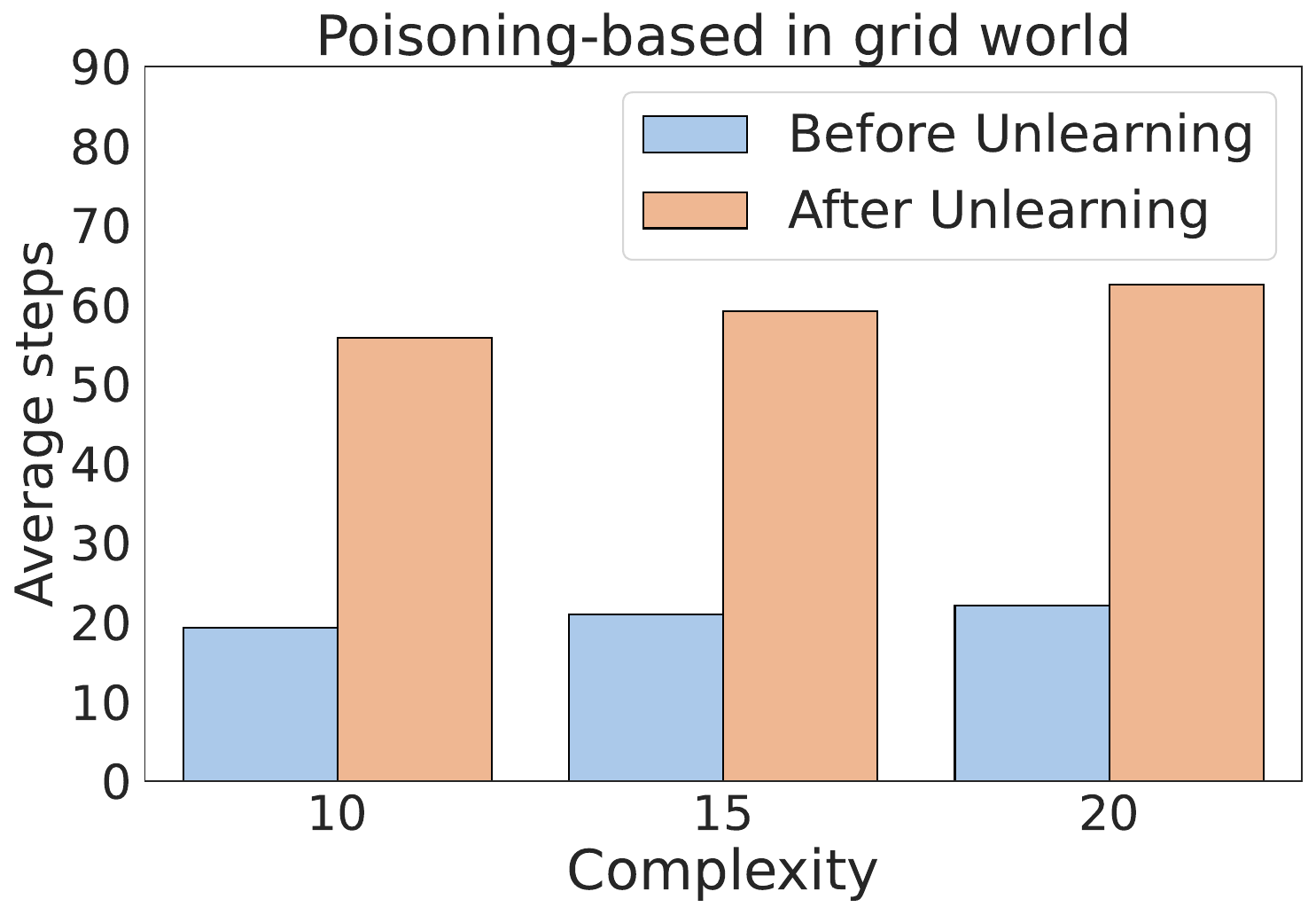}
			\label{fig:GridComplexityStepsMethod2}}
	\subfigure[\scriptsize{Rewards of poisoning-based}]{
    \includegraphics[scale=0.16]{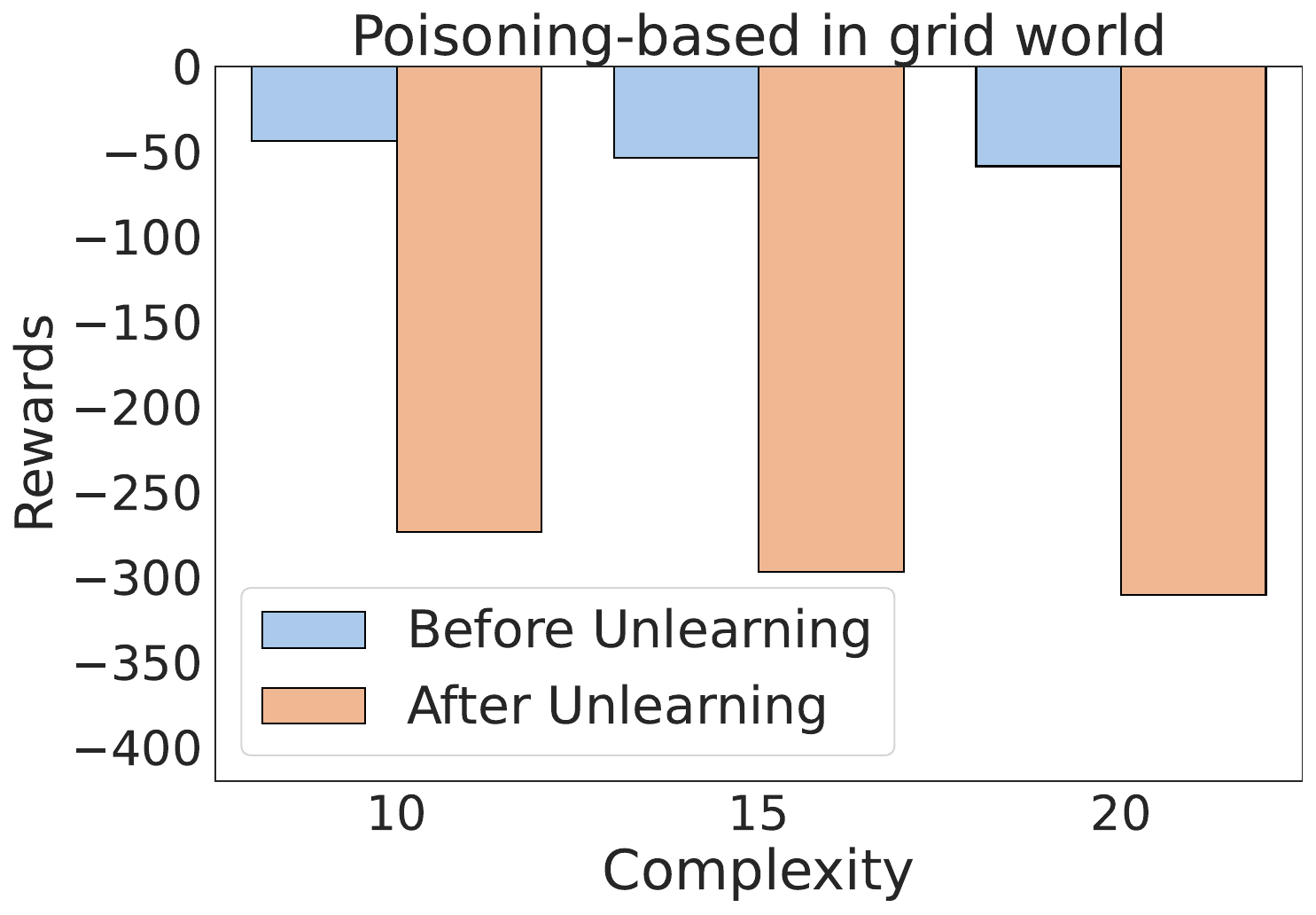}
			\label{fig:GridComplexityRewardsMethod2}}\\[2ex]
    \end{minipage}
	\caption{The decremental RL-based and poisoning-based methods in Grid World with different complexity}
	\label{fig:GridComplexity}
\end{figure}

\vspace{2mm}
\noindent\textbf{Impact of Environment Complexity.} The complexity of an environment can be characterized by the presence and arrangement of obstacles within it. By modifying the complexity of the environment, we can assess the adaptability of the proposed methods.
In the grid world setting, we examine the impact of increasing the environment complexity by introducing more obstacles, with the environment size maintained at $10\times 10$. Figures \ref{fig:GridComplexityStepsMethod1} and \ref{fig:GridComplexityRewardsMethod1} show the outcomes of the decremental RL-based method. 

We observe that as the number of obstacles is raised from $10$ to $15$, there is a notable increase in both the number of steps taken by the agent and the disparity in rewards between the pre-unlearning and post-unlearning stages. This observation suggests that with a moderate increase in complexity, the unlearning process becomes more challenging, resulting in a substantial alteration in the agent's behavior, leading to changes in both step count and rewards.
However, intriguingly, as we further augment the number of obstacles from $15$ to $20$, the difference in both steps and rewards between before and after unlearning seems to stabilize or vary less significantly. 
The reason behind this behavior lies in the agent's learning adaptability. When the environment complexity rises from $10$ to $15$ obstacles, the agent faces substantial alterations in the optimal path and must undertake considerable unlearning to adjust its behavior accordingly. As a result, we observe a noticeable increase in step count and disparity in rewards.
Conversely, when the number of obstacles increases from $15$ to $20$, the agent has already adapted its behavior to accommodate the increased complexity. As the agent's policy has already been modified, further increases in obstacle count have a diminishing impact on step count and reward disparity.

However, when employing the poisoning-based method (depicted in Figures \ref{fig:GridComplexityStepsMethod2} and \ref{fig:GridComplexityRewardsMethod2}), an interesting observation emerges. Unlike the decremental RL-based method, the difference in both steps and rewards between the pre-unlearning and post-unlearning stages remains relatively stable even as the number of obstacles increases.
The reason behind this intriguing behavior lies in the nature of the poisoning-based approach. When we introduce additional obstacles to the environment, the poisoning-based method operates differently compared to the decremental RL-based approach. Instead of modifying the agent's learned policy gradually, the poisoning-based method incorporates an element of targeted perturbation.
As the number of obstacles increases, the poisoning-based method strategically poisons the agent's policy by introducing deceptive information during the unlearning process. This targeted perturbation causes the agent's behavior to deviate from the optimal path more significantly, leading to relatively constant differences in both step count and reward between the pre-unlearning and post-unlearning phases.

By undertaking a comparative analysis, it becomes evident that the poisoning-based method introduces a higher level of stability in performance compared to the decremental RL-based method. This enhanced stability is of significant interest and has several underlying reasons.
Firstly, the poisoning-based approach leverages targeted perturbations to strategically poison the agent's policy during the unlearning process. By introducing adversarial elements in a controlled manner, this method consistently influences the agent's behavior, leading to more predictable changes in its performance.

Secondly, the poisoning-based method's targeted perturbations are designed to cause deviations from the optimal path. Thus, the agent's policy becomes consistently misled in the presence of additional obstacles, leading to a stable performance difference between pre- and post-unlearning.

Moreover, the consistent impact of the poisoning-based method can be advantageous in certain scenarios. For instance, in safety-critical environments, e.g., autonomous driving, where stability and predictability are crucial, the poisoning-based approach offers a more controlled and reliable means of unlearning unwanted behaviors.
On the other hand, the decremental RL-based method, gradually modifying the agent's policy, leads to more varied and less predictable changes in behavior as the environment complexity increases. This approach makes it challenging to precisely anticipate the agent's performance changes in response to additional obstacles.

\vspace{2mm}
\noindent\textbf{Impact of Poisoning Level.~} The hyperparameter, poisoning level, serves as a pivotal factor in evaluating and testing the poisoning-based method exclusively. This parameter governs the quantity of poison introduced to the agent during the unlearning process, enabling us to investigate how the method performs under varying levels of poisoning. 
Specifically, the poisoning level is measured by the difference between the intended state $s'$ and the manipulated state $\hat{s}'$. To illustrate, in the grid world context where an agent's state comprises eight dimensions, a poisoning level of $3$ indicates that the two states differ in three dimensions. 
The poisoning level is consistently set to $3$ throughout the experiments, unless otherwise specified.

\begin{figure}[ht]
\centering
	\begin{minipage}{1\textwidth}
    \subfigure[\scriptsize{Steps of poisoning-based}]{
    \includegraphics[scale=0.15]{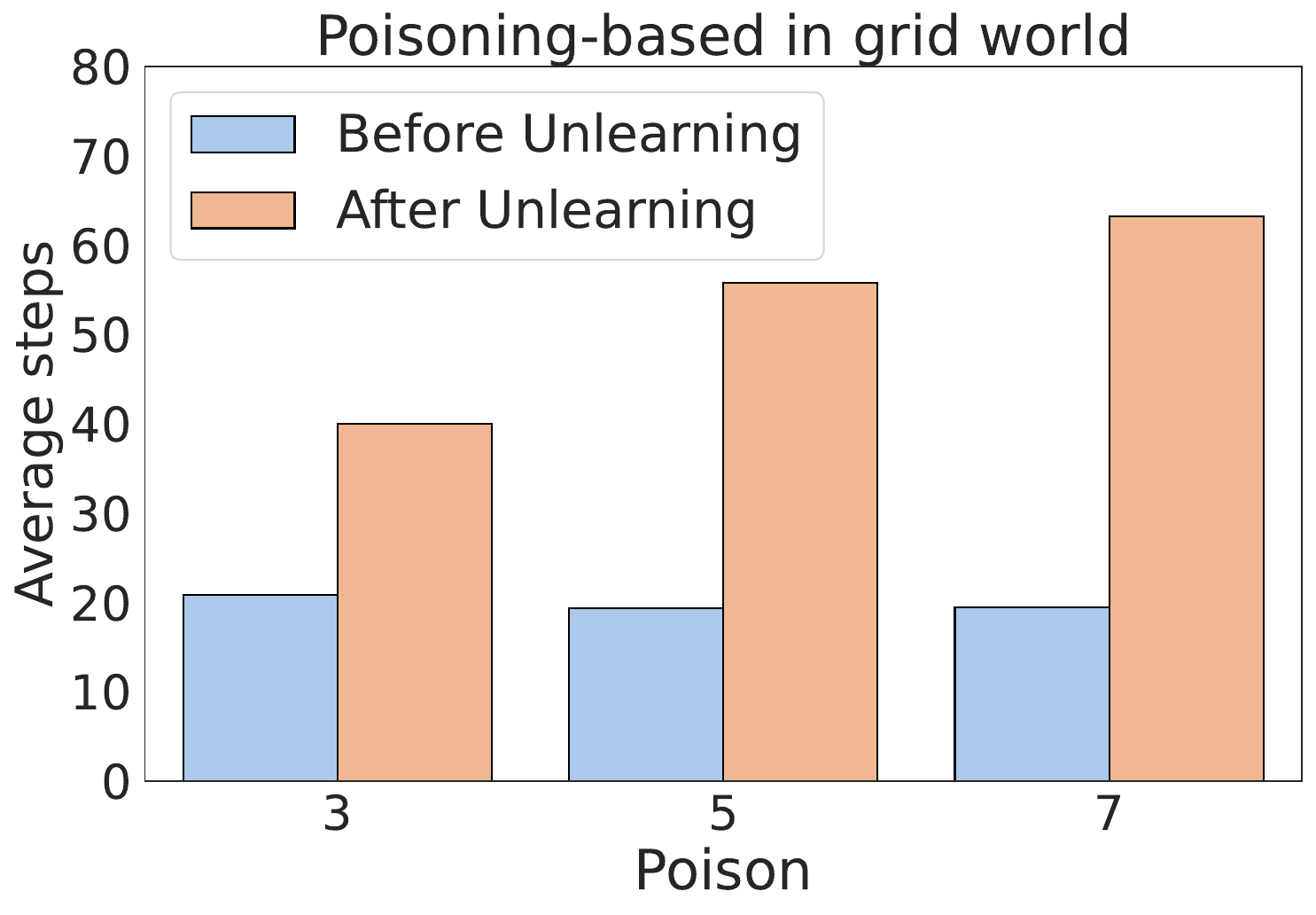}
			\label{fig:GridPoisonSteps}}
	\subfigure[\scriptsize{Rewards of poisoning-based}]{
    \includegraphics[scale=0.15]{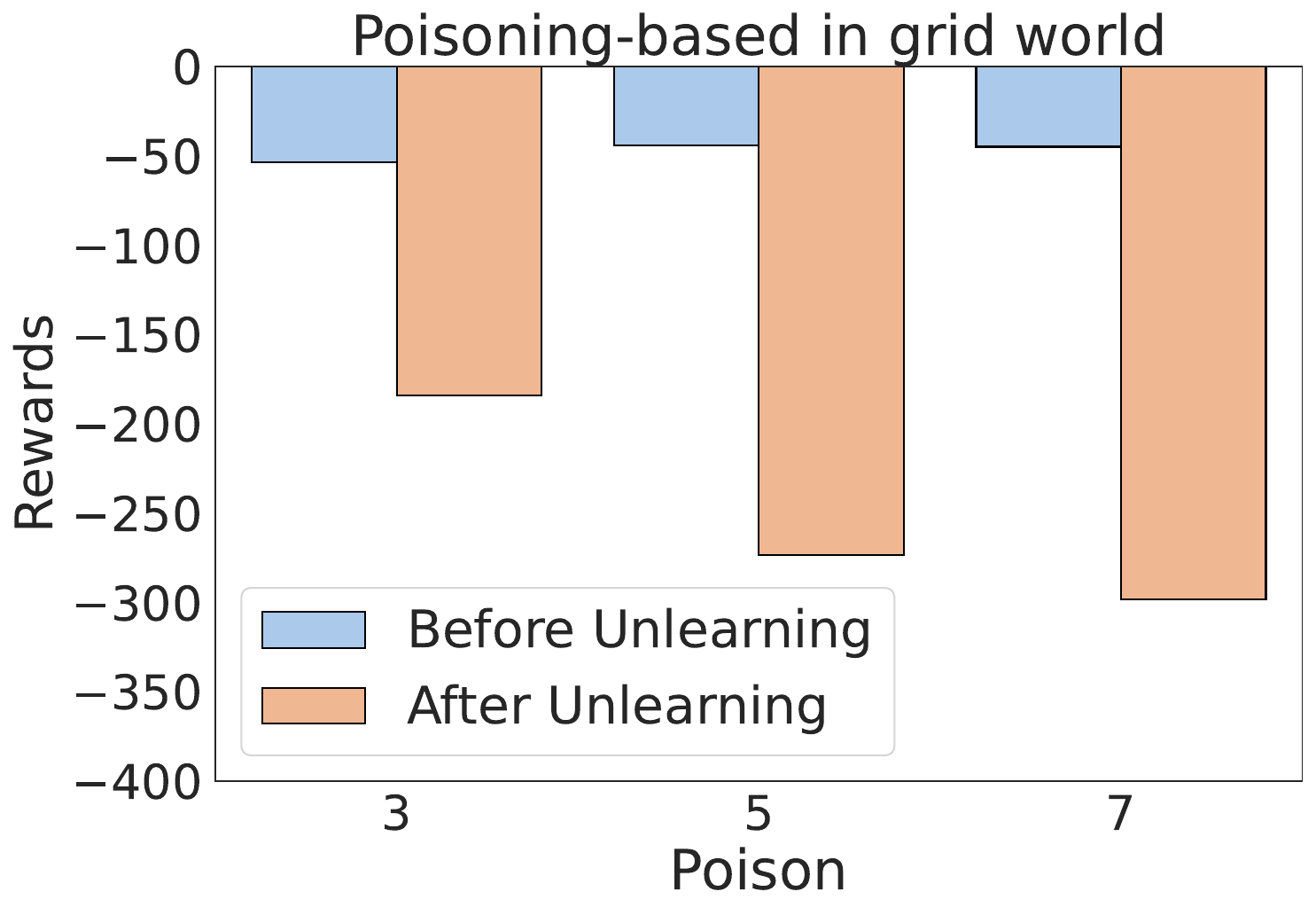}
			\label{fig:GridPoisonRewards}}\\[2ex]
    \end{minipage}
	\caption{Poisoning-based method in Grid World, different poison levels}
	\label{fig:GrdiPoison}
\end{figure}

Figure \ref{fig:GrdiPoison} illustrates the impact of changing the poisoning level on the evaluation metrics in the grid world setting. As the poisoning level increases, the values of all the evaluation metrics demonstrate a consistent downward trend. This trend indicates an improvement in the unlearning results with higher poisoning amounts.
The reason behind this promising phenomenon lies in the nature of the poisoning-based method and its strategic use of targeted perturbations. As the poisoning level escalates, the method introduces a more substantial amount of deceptive information into the agent's policy, causing a stronger deviation from the optimal path.

This increase in poisoning intensity effectively compels the agent to unlearn its previous behaviors more forcefully, encouraging it to abandon suboptimal policies. Thus, the agent's learned policy becomes more adaptable and resilient, leading to enhanced performance in unlearning unwanted knowledge.
Moreover, higher poisoning amounts facilitate a more efficient exploration of the policy space, allowing the agent to escape local optima and discover better solutions. Thus, the unlearning process becomes more effective in refining the agent's behavior and enhancing its performance. 

\begin{figure}[ht]
\centering
	\begin{minipage}{1\textwidth}
       \subfigure[\scriptsize{Steps of poisoning-based}]{
    \includegraphics[scale=0.16]{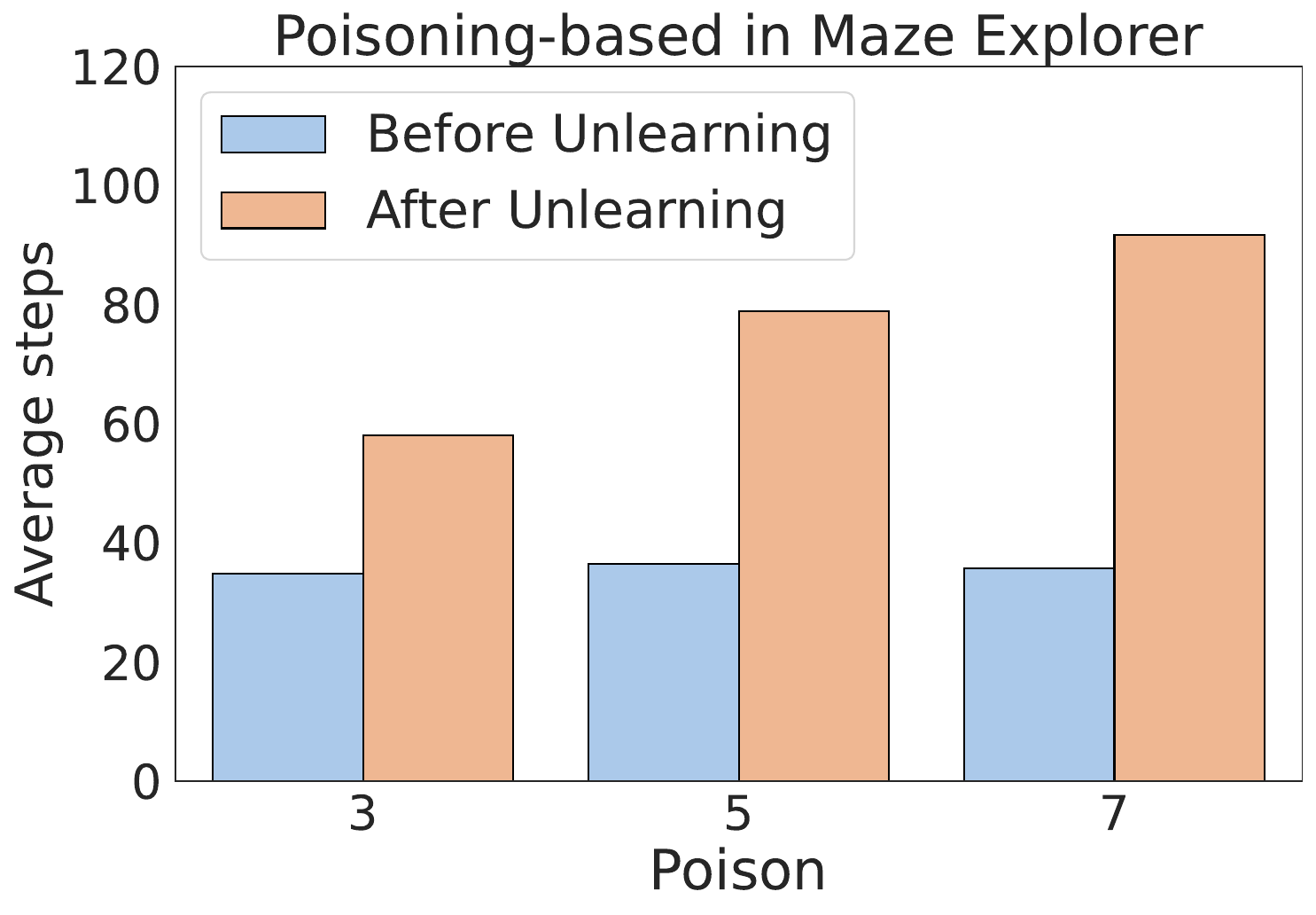}
			\label{fig:MazePoisonSteps}}
	\subfigure[\scriptsize{Rewards of poisoning-based}]{
    \includegraphics[scale=0.16]{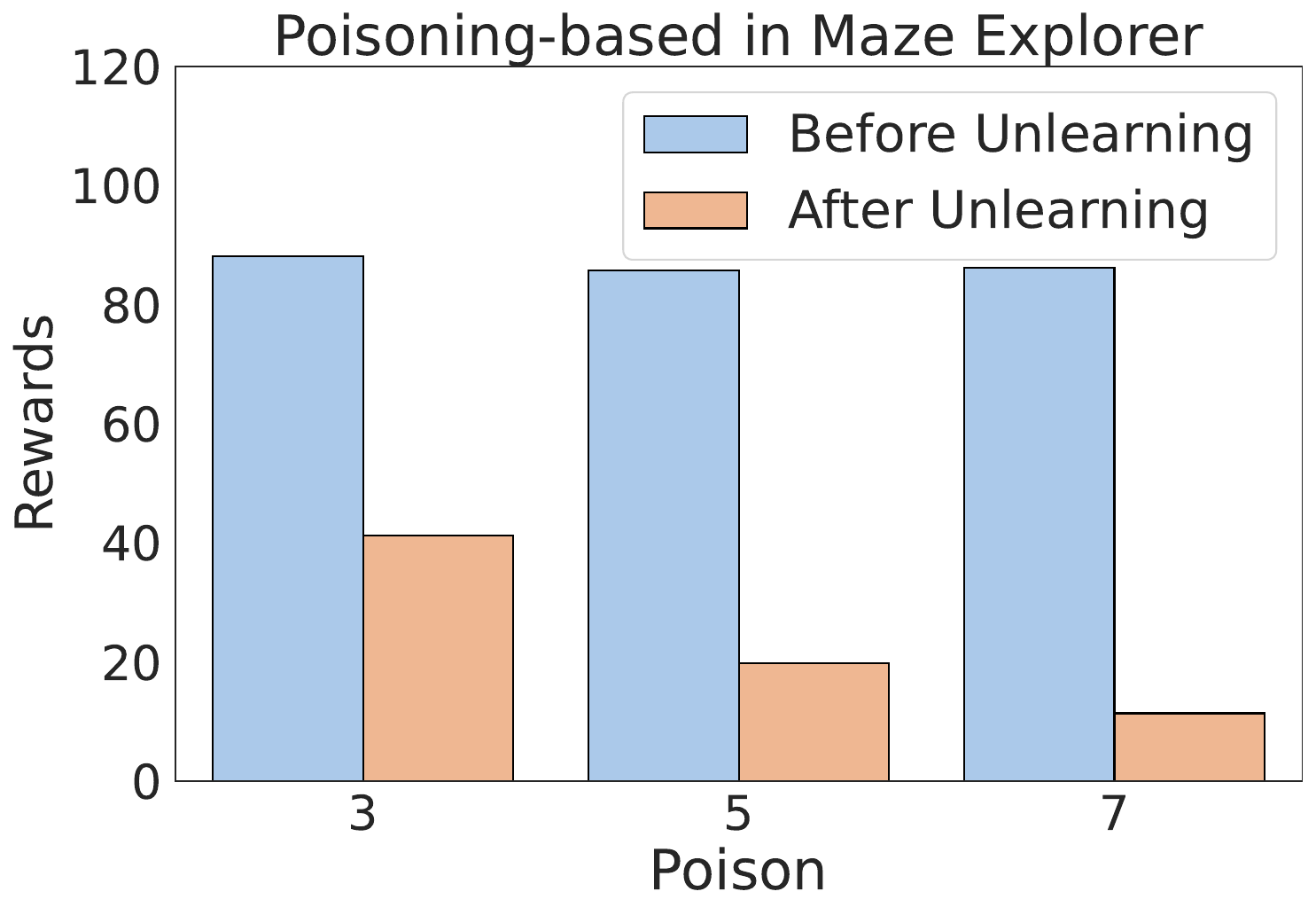}
			\label{fig:MazePoisonRewards}}\\[2ex]
    \end{minipage}
	\caption{Poisoning-based method in Virtual Home, different poison levels}
	\label{fig:VHomePoison}
\end{figure}

\begin{figure}[ht]
\centering
	\begin{minipage}{1\textwidth}
       \subfigure[\scriptsize{Steps of poisoning-based}]{
    \includegraphics[scale=0.16]{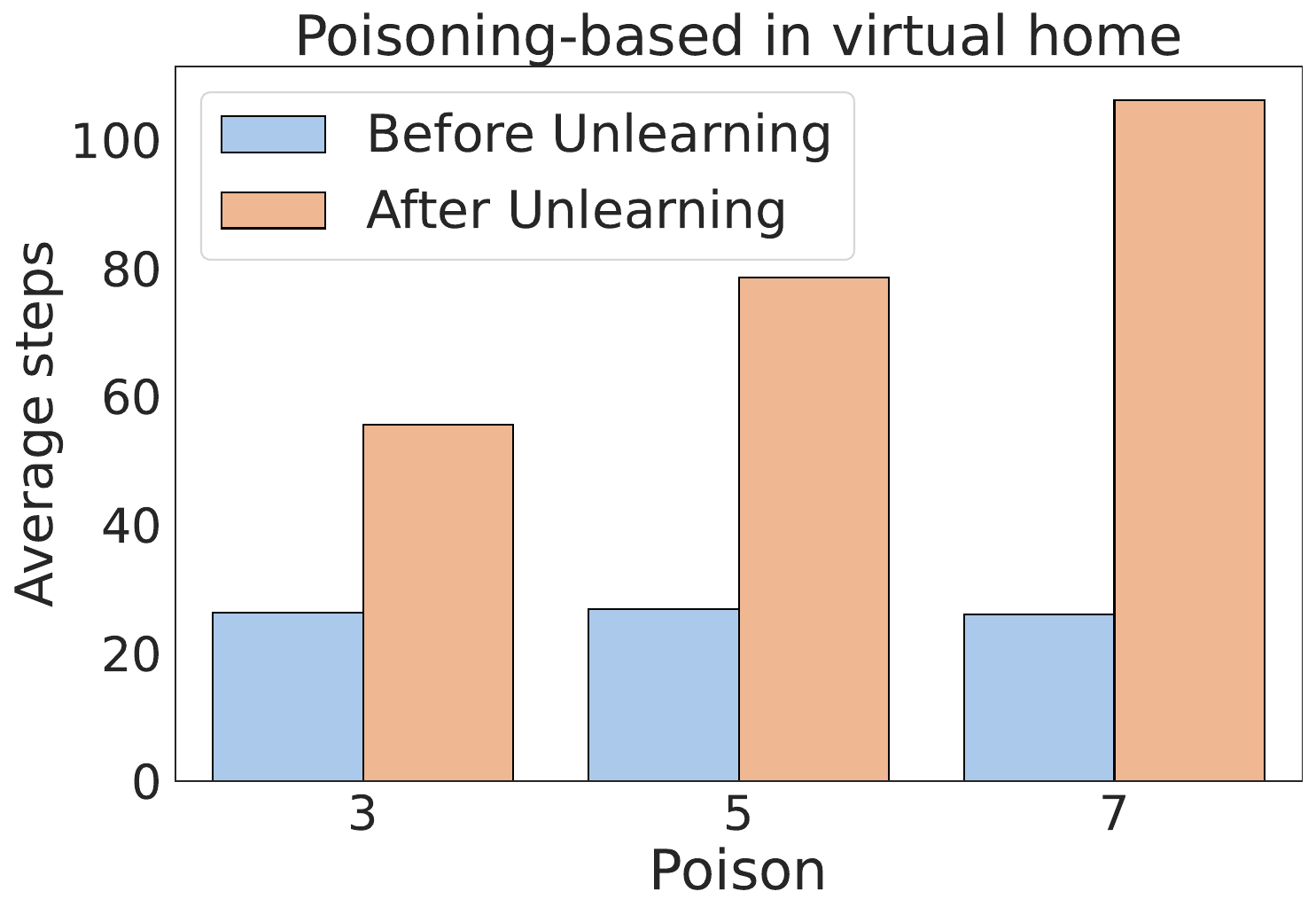}
			\label{fig:VHomePoisonSteps}}
	\subfigure[\scriptsize{Rewards of poisoning-based}]{
    \includegraphics[scale=0.16]{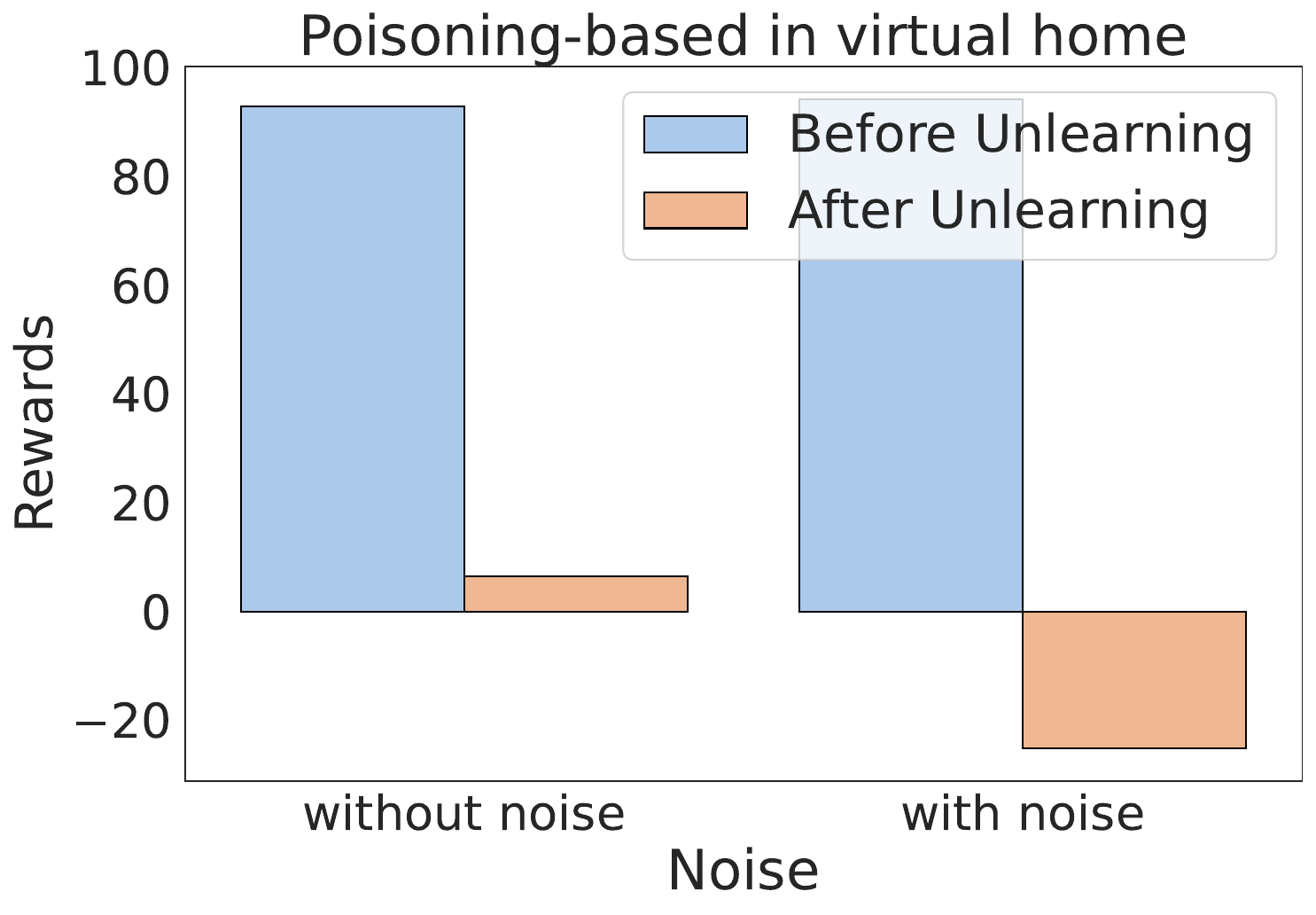}
			\label{fig:VHomePoisonRewards}}\\[2ex]
    \end{minipage}
	\caption{Poisoning-based method in Maze Explorer, different poison levels}
	\label{fig:MazePoison}
\end{figure}

Consistent results across the remaining scenarios, shown in Figure \ref{fig:VHomePoison} (virtual home) and Figure \ref{fig:MazePoison} (maze explorer), further strengthen the observations in grid world. The increased poison during unlearning correlates with improved results. 

\subsection{Adaptability Study}

\noindent\textbf{Dynamic Environments.~} To simulate dynamic environments, where the features and layouts of environments can change during agent training, we introduce a slight modification to the unlearning problem. Specifically, considering an unlearning environment $\mathcal{M}_u$, we employ time steps to represent changes in the environment. As time progresses in $t$ steps, the evolution of the unlearning environment can be represented as $\mathcal{M}^1_u, \ldots, \mathcal{M}^t_u$. Thus, the problem of unlearning $\mathcal{M}_u$ transforms into the task of unlearning $\mathcal{M}^1_u, \ldots, \mathcal{M}^t_u$.

\begin{figure}[ht]
\centering
	\begin{minipage}{1\textwidth}
    \subfigure[\scriptsize{Steps of the four methods}]{
    \includegraphics[scale=0.12]{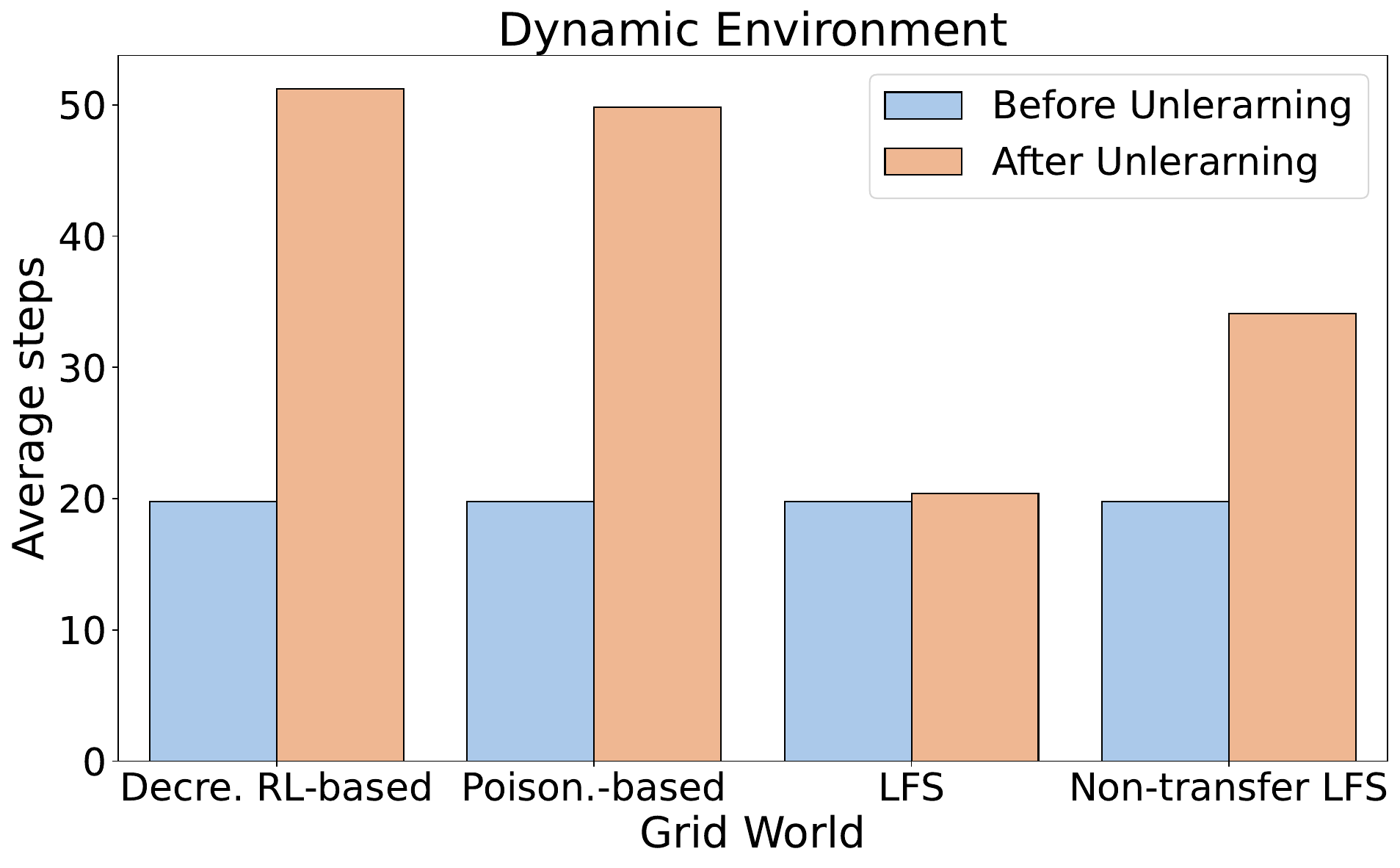}
			\label{fig:DynamicEnvironmentsSteps}}
	\subfigure[\scriptsize{Rewards of the four methods}]{
    \includegraphics[scale=0.12]{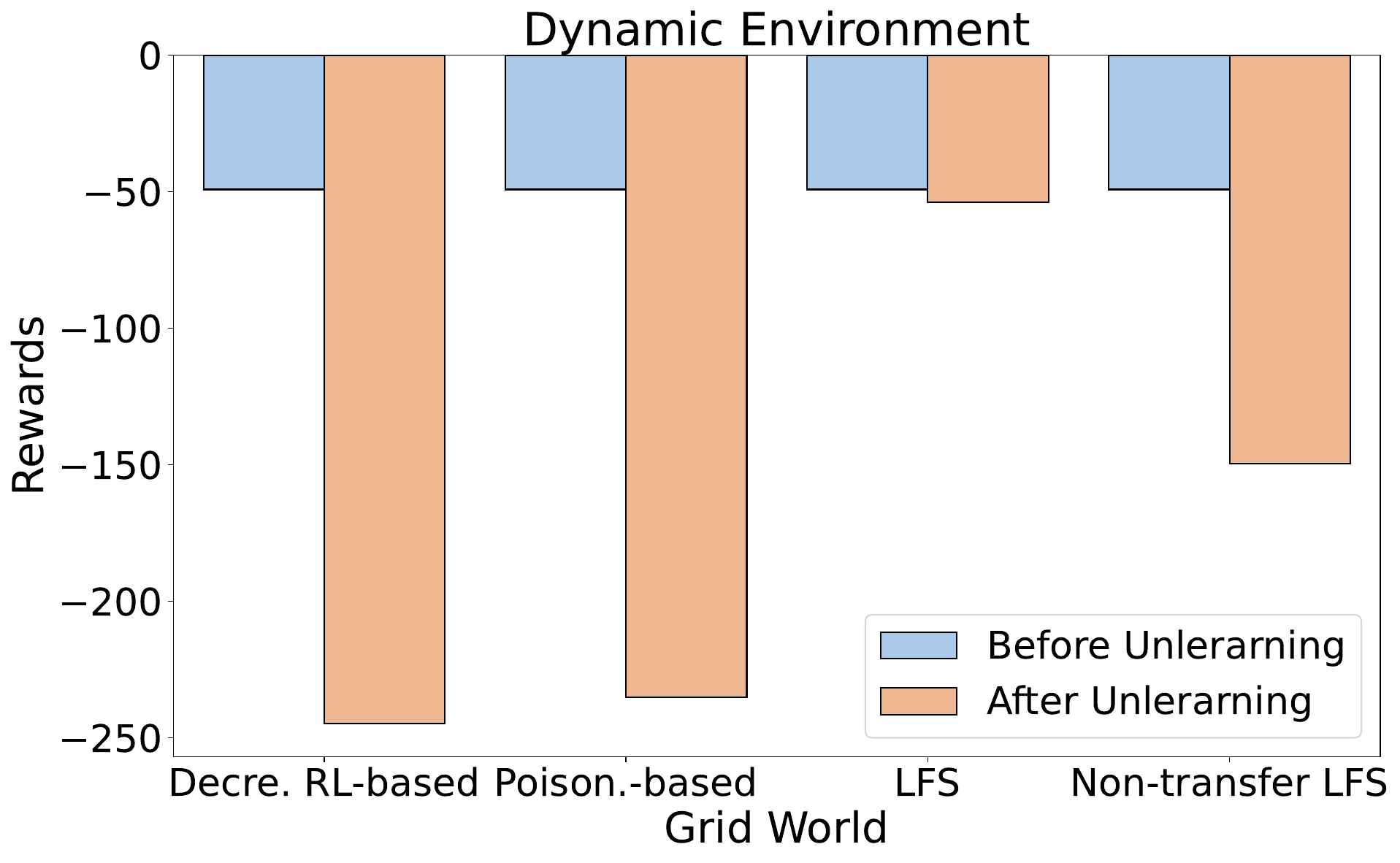}
			\label{fig:DynamicEnvironmentsRewards}}\\[2ex]
    \end{minipage}
	\caption{The four methods in dynamic environments in Grid World}
	\label{fig:DynamicEnvironments}
\end{figure}

The experimental results, shown in Figure~\ref{fig:DynamicEnvironments}, were derived by setting $t=5$ and averaging the outcomes across the five environments. Similar outcomes can be observed for other values of $t$, e.g., $3$ and $8$. The results indicate that even in this dynamic setting, the outcomes of post-unlearning remain notably favorable. The effectiveness is attributed to the carefully designed mechanisms inherent in our reinforcement unlearning methods. 
The decremental RL-based method dynamically adjusts the agent's knowledge, ensuring it remains effective even as the environment undergoes alterations. Similarly, the poisoning-based method introduces dynamism by modifying the unlearning environment, ensuring the agent to perform optimally in the evolving environment.

\vspace{2mm}
\noindent\textbf{Generalization.~}
To assess the generalization capability of our unlearning methods, we evaluated the unlearned models in unseen environments. We established the ratio between training environments and unseen environments as $4:1$, employing $20$ training environments and $5$ unseen environments. This configuration is analogous to the typical setting of the ratio between the size of the training set and the test set in conventional machine learning. 

\vspace{-3mm}
\begin{figure}[ht]
\centering
	\begin{minipage}{1\textwidth}
    \subfigure[\scriptsize{Steps of the four methods}]{
    \includegraphics[scale=0.2]{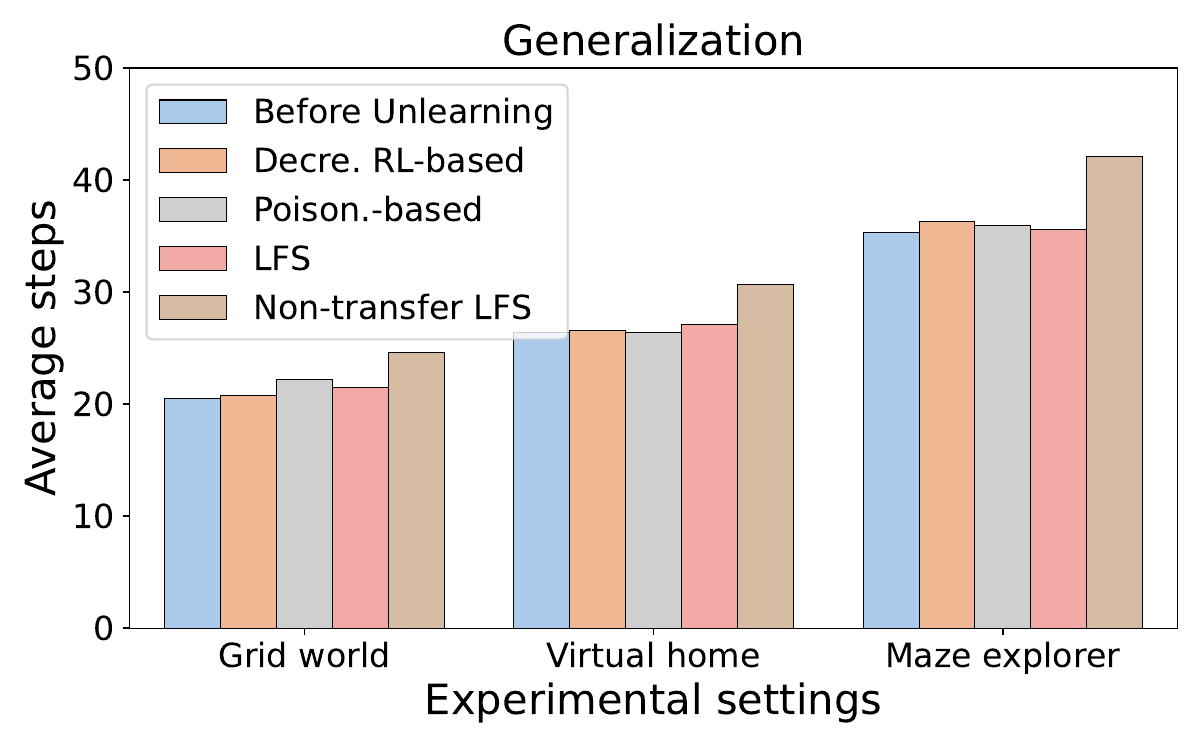}
			\label{fig:GeneralizationSteps}}
	\subfigure[\scriptsize{Rewards of the four methods}]{
    \includegraphics[scale=0.2]{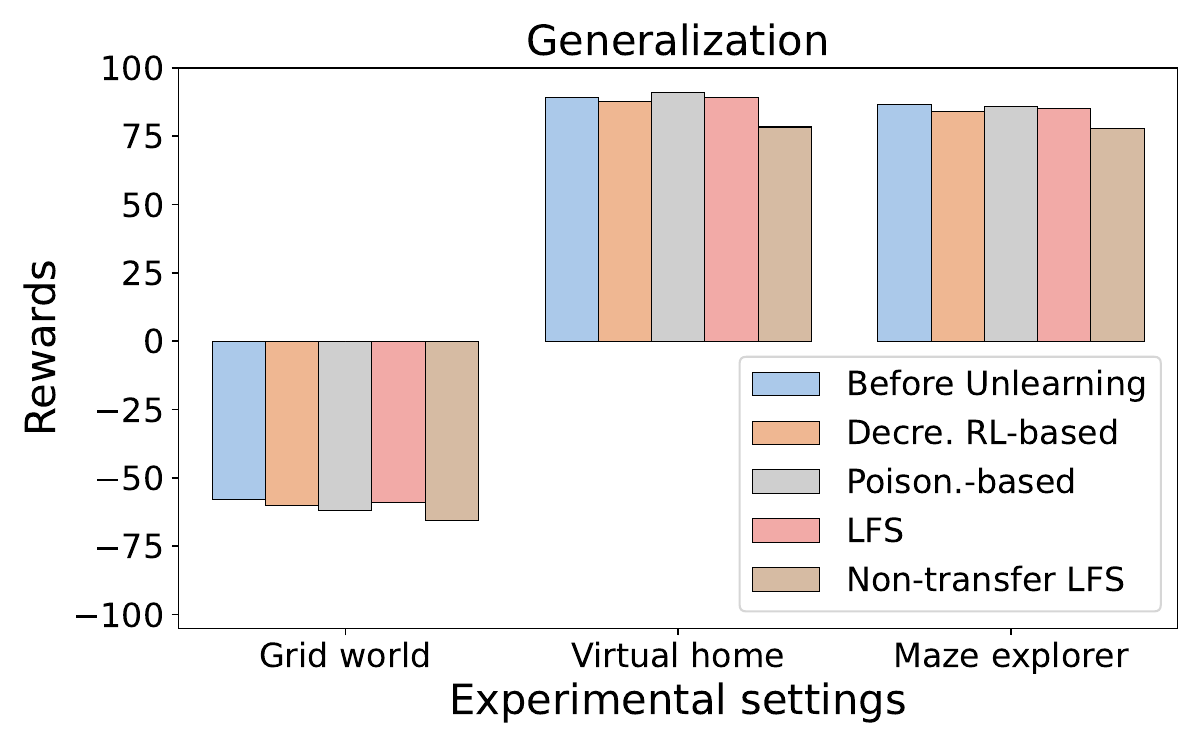}
			\label{fig:GeneralizationRewards}}\\[2ex]
    \end{minipage}
    \vspace{-4mm}
	\caption{Generalization of the four methods in Grid World.}
	\vspace{-1mm}
	\label{fig:Generalization}
\end{figure}

The outcomes, shown in Figure \ref{fig:Generalization}, were derived by averaging the results across the five unseen environments. The results indicate that the performance is sustained in these unseen settings. This suggests that our methods do not compromise the models' generalization ability; instead, they selectively impact the models' performance in the unlearning environments. This success can be attributed to the precision of our unlearning methods, which erase only the features specific to each unlearning environment while preserving the underlying rules gained from training environments.

\vspace{2mm}
\noindent\textbf{Robustness.~}
Robustness gauges the strength and resilience of a method in the face of external perturbations. Evaluating robustness entails assessing how the unlearning methods perform when subjected to external perturbations. 
To conduct the evaluation, we introduce noise to the agent's actions during both training and unlearning.  
The introduction of noise is achieved by randomly perturbing the probability distribution over the agent’s actions. This perturbation involves adding a small randomly generated number, falling within the range of $[-0.1, 0.1]$, to a randomly selected probability in the distribution.
This noise represents random variations or disturbances that can occur in real-world scenarios. 

\vspace{-2mm}
\begin{figure}[ht]
\centering
	\begin{minipage}{1\textwidth}
    \subfigure[\scriptsize{Steps of decre. RL-based with noise}]{
    \includegraphics[scale=0.21]{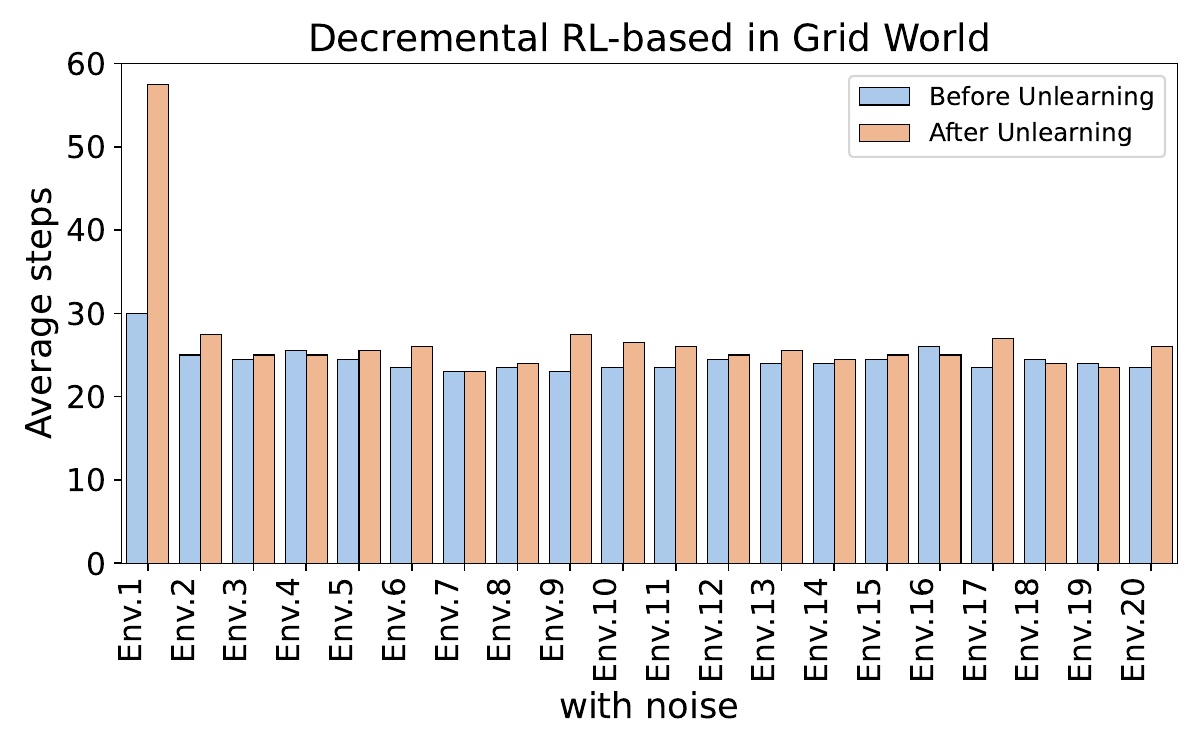}
			\label{fig:GridRobustStepsMethod1Noise}}\vspace{-4mm}
       \subfigure[\scriptsize{Steps of decre. RL-based without noise}]{
    \includegraphics[scale=0.21]{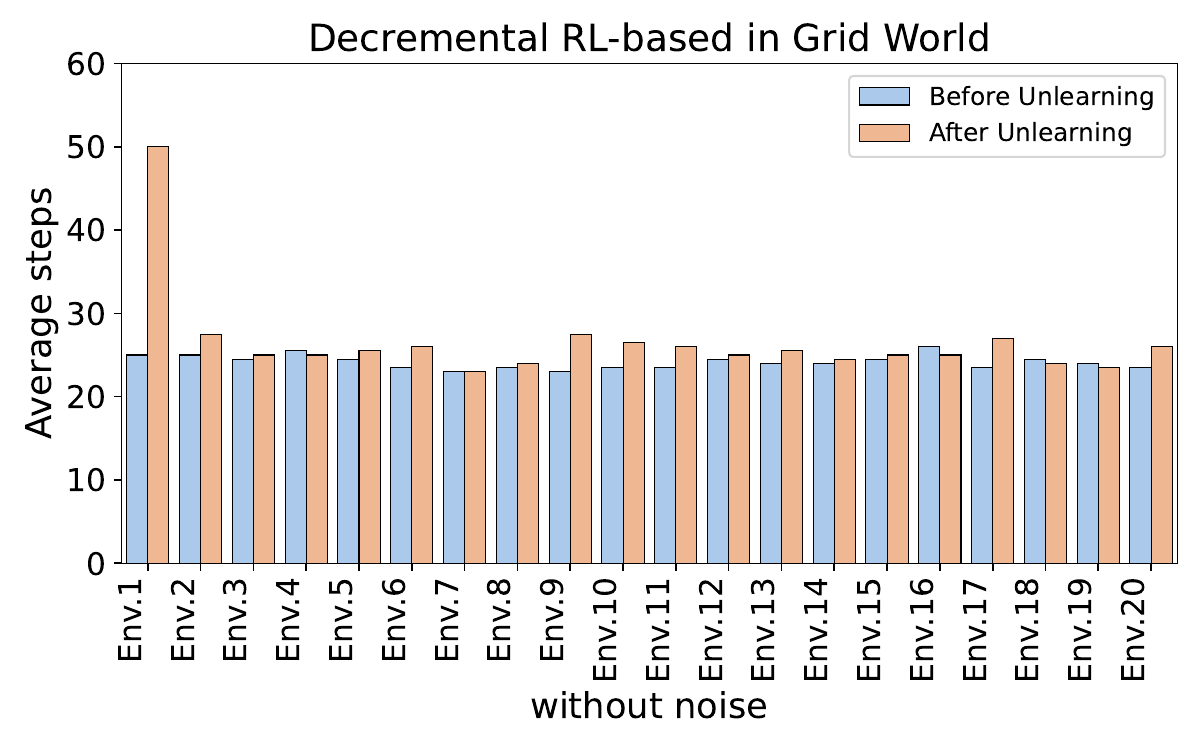}
			\label{fig:GridRobustStepsMethod1WONoise}}\\[2ex]\vspace{-4mm}
	\subfigure[\scriptsize{Rewards of decre. RL-based with noise}]{
    \includegraphics[scale=0.21]{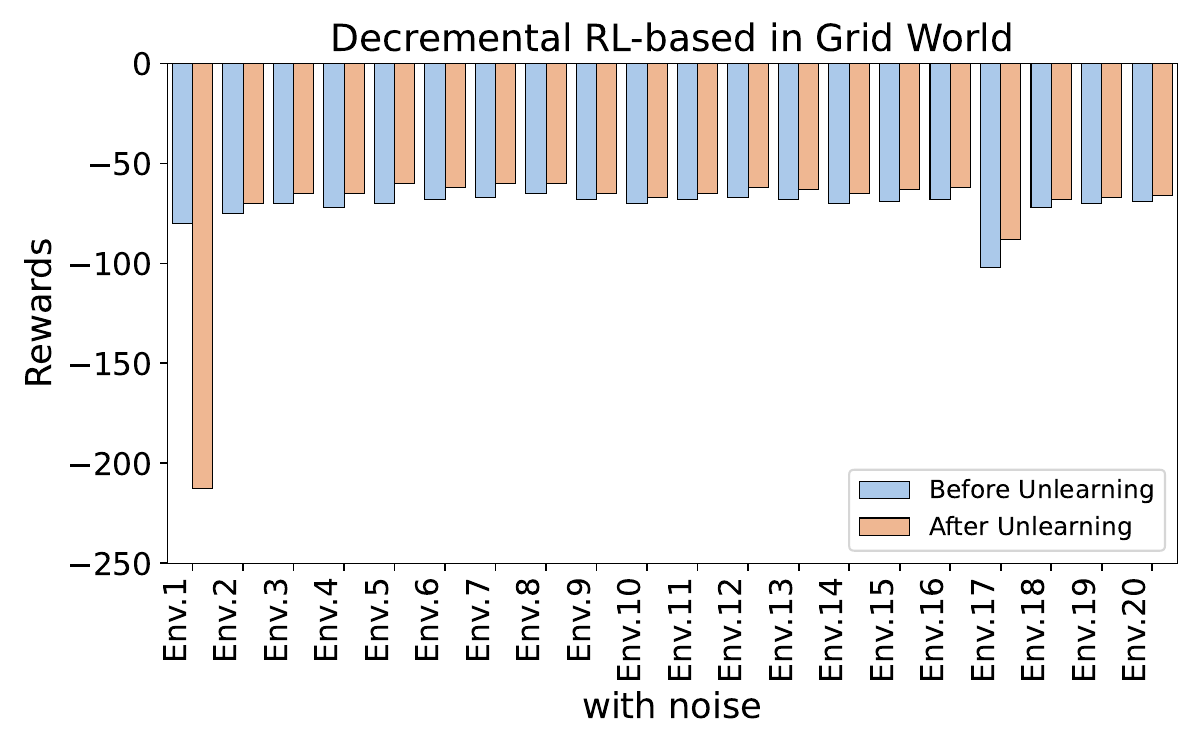}
			\label{fig:GridRobustRewardsMethod1Noise}}
   	\subfigure[\scriptsize{Rewards of decre. RL-based without noise}]{
    \includegraphics[scale=0.21]{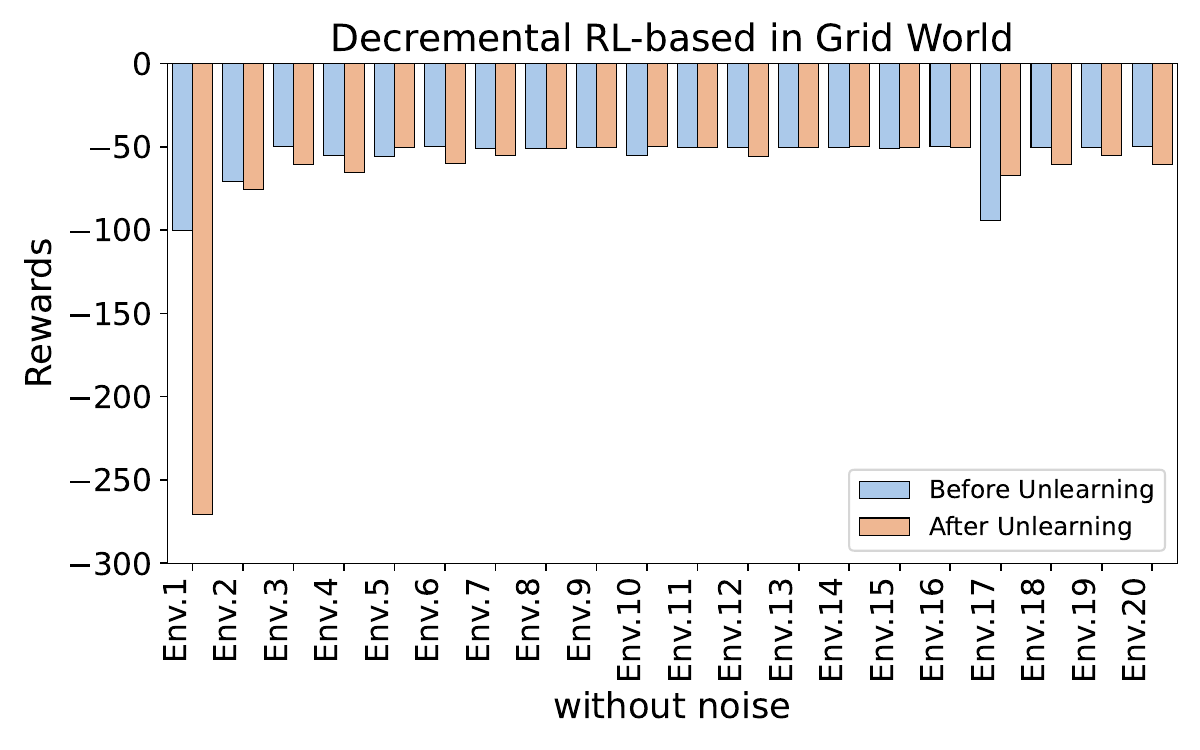}
			\label{fig:GridRobustRewardsMethod1WONoise}}\\[2ex]\vspace{-4mm}
       \subfigure[\scriptsize{Steps of poisoning-based with noise}]{
    \includegraphics[scale=0.21]{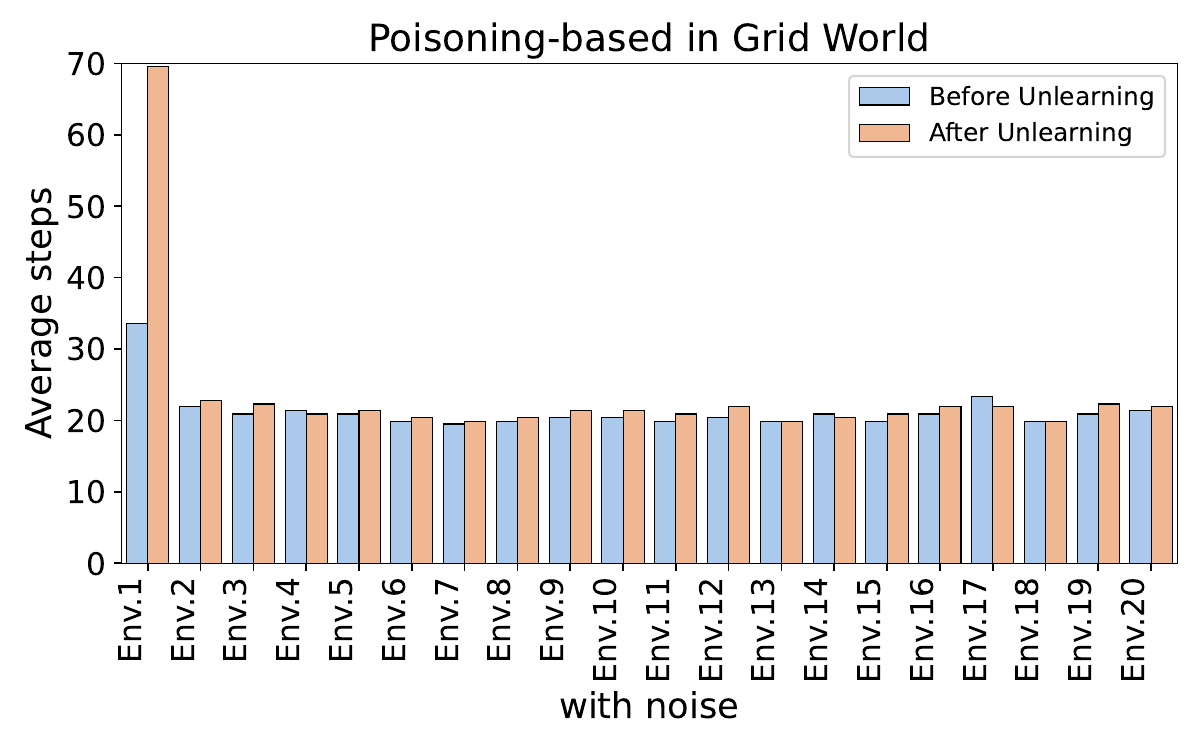}
			\label{fig:GridRobustStepsMethod2Noise}}
          \subfigure[\scriptsize{Steps of poisoning-based without noise}]{
    \includegraphics[scale=0.21]{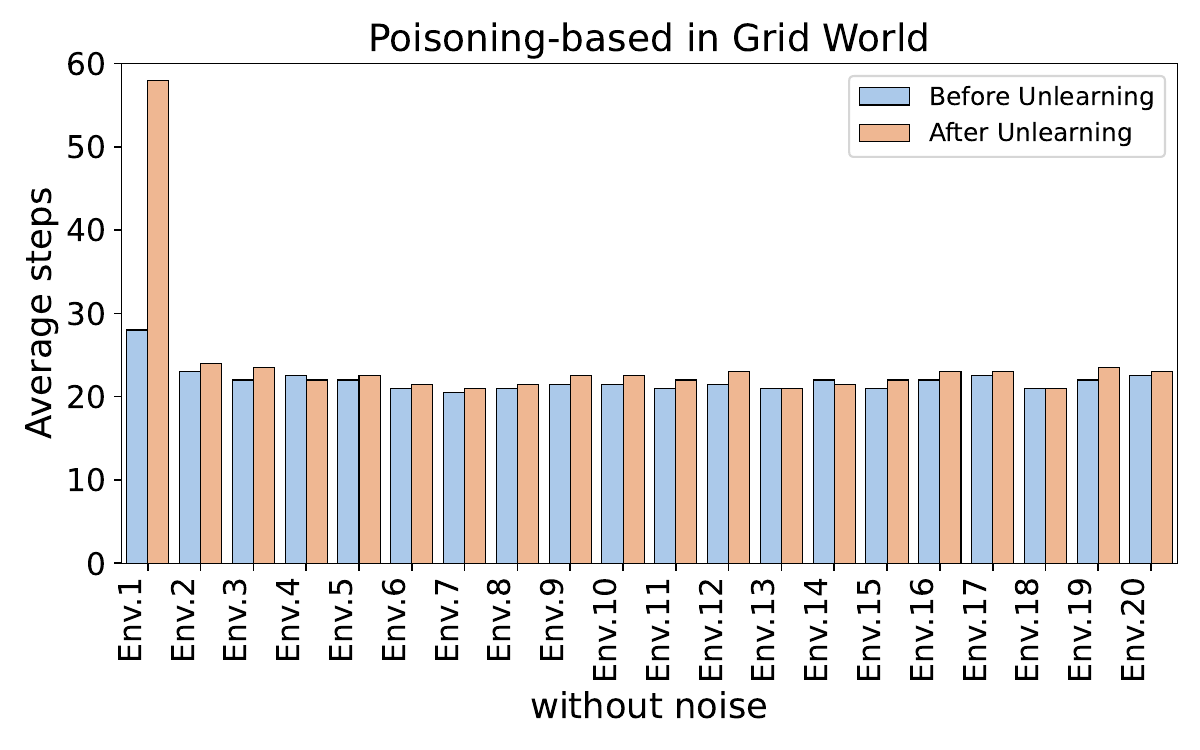}
			\label{fig:GridRobustStepsMethod2WONoise}}\\[2ex]
	\subfigure[\scriptsize{Rewards of poisoning-based with noise}]{
    \includegraphics[scale=0.21]{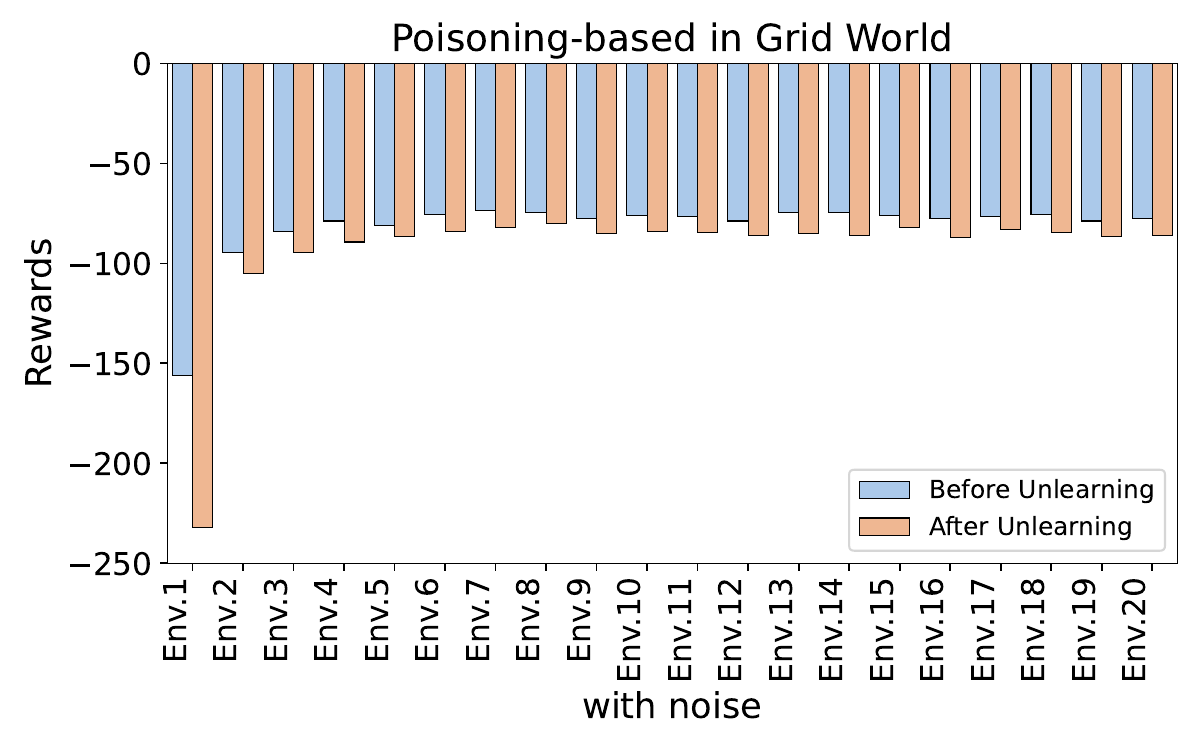}
			\label{fig:GridRobustRewardsMethod2Noise}}
   	\subfigure[\scriptsize{Rewards of poisoning-based without noise}]{
    \includegraphics[scale=0.21]{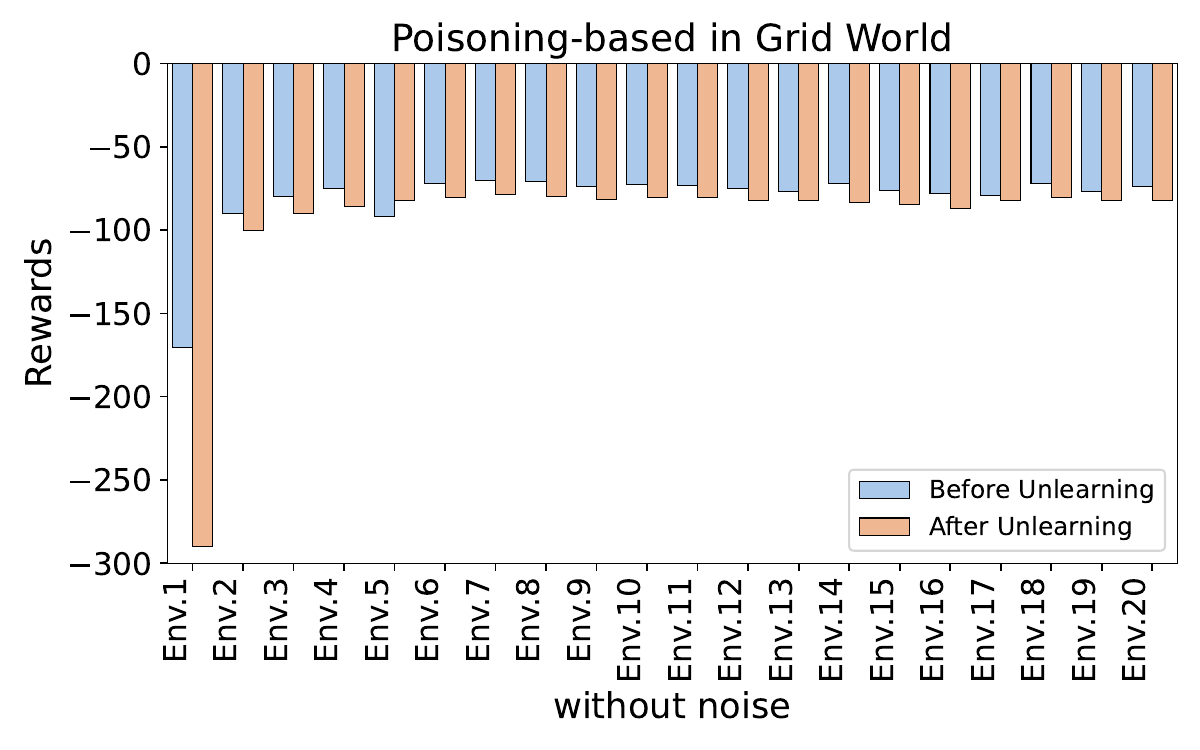}
			\label{fig:GridRobustRewardsMethod2WONoise}}
    \end{minipage}
    \vspace{-3mm}
	\caption{The decremental RL-based and poisoning-based methods in Grid World with and without noise}
	\vspace{-1mm}
	\label{fig:GridRobust}
\end{figure}

The corresponding results of the grid world setting are presented in Figure \ref{fig:GridRobust}. 
Upon analyzing the outcomes, a notable observation emerges: both the decremental RL-based and poisoning-based methods exhibit remarkable robustness against external noise. Despite the introduction of noise, the agent's performance in both the unlearning and remaining environments remains consistently stable across both methods, showing no significant difference compared to conditions without noise. 
This robustness can be attributed to the inherent adaptability and resilience of the unlearning methods. In the decremental RL-based method, the gradual modification of the agent's policy allows it to withstand minor variations in observations, ensuring that its learned behavior remains stable despite external noise.
Similarly, the poisoning-based method's strategic use of targeted perturbations enables the agent to develop a more adaptive policy. Thus, the agent's behavior proves to be less affected by the noise, maintaining its consistency in unlearning undesired knowledge.

\begin{figure}[ht]
\centering
	\begin{minipage}{1\textwidth}
    \subfigure[\scriptsize{Average steps of the decremental RL-based method with noise}]{
    \includegraphics[scale=0.092]{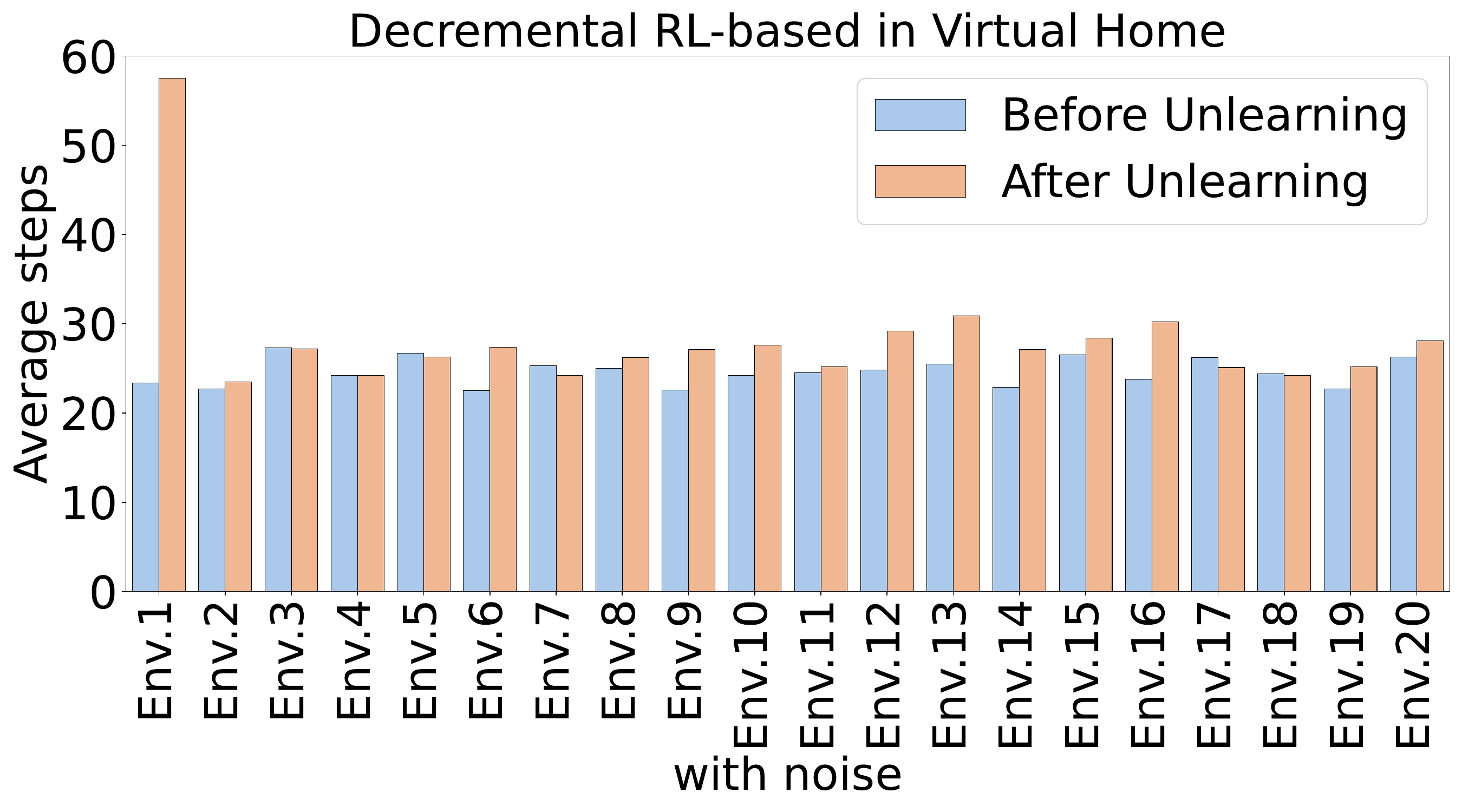}
			\label{fig:VHomeRobustStepsMethod1Noise}}
   \subfigure[\scriptsize{Average steps of the decremental RL-based method without noise}]{
    \includegraphics[scale=0.092]{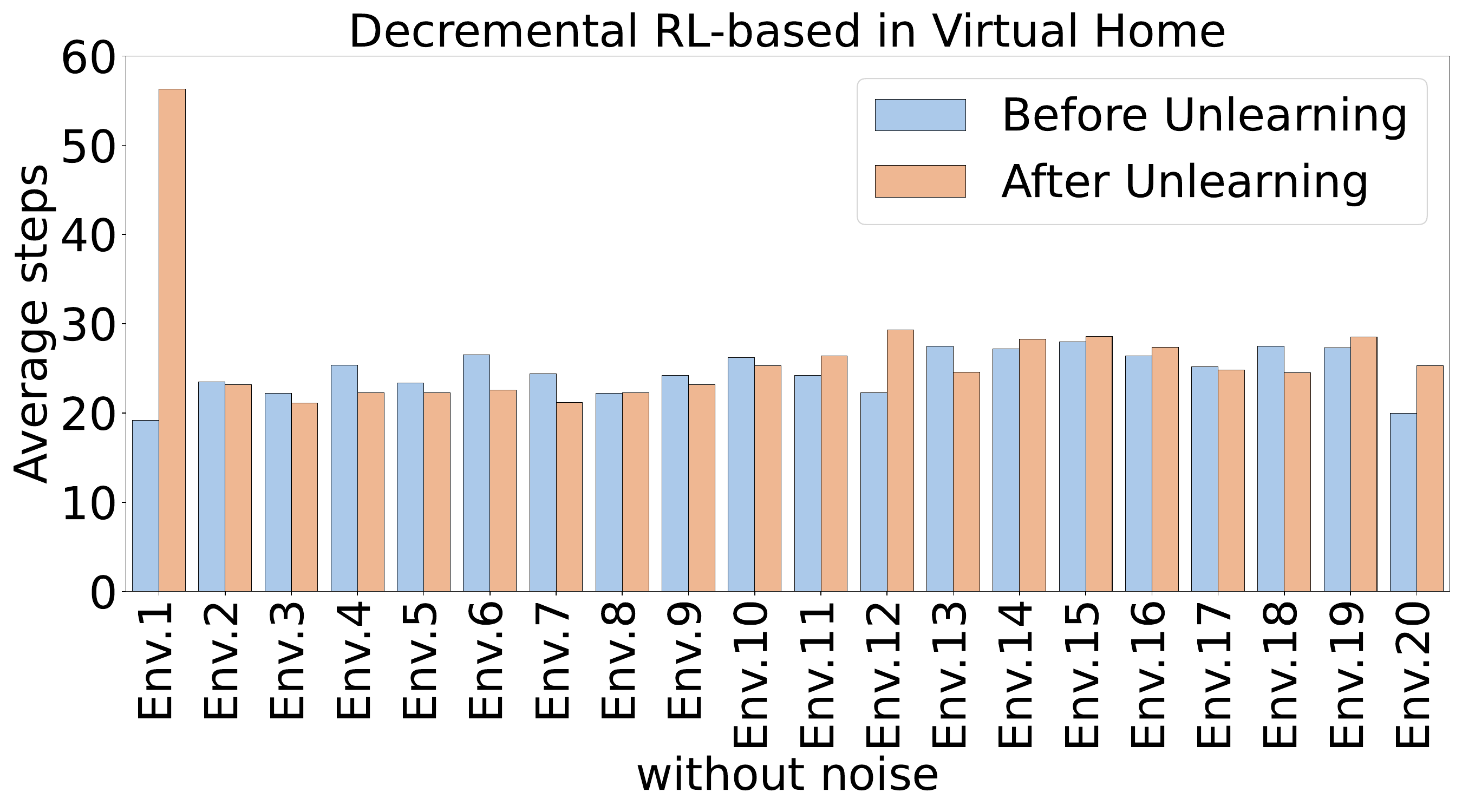}
			\label{fig:VHomeRobustStepsMethod1WONoise}}\\[2ex]
	\subfigure[\scriptsize{Rewards of the decremental RL-based method with noise}]{
    \includegraphics[scale=0.092]{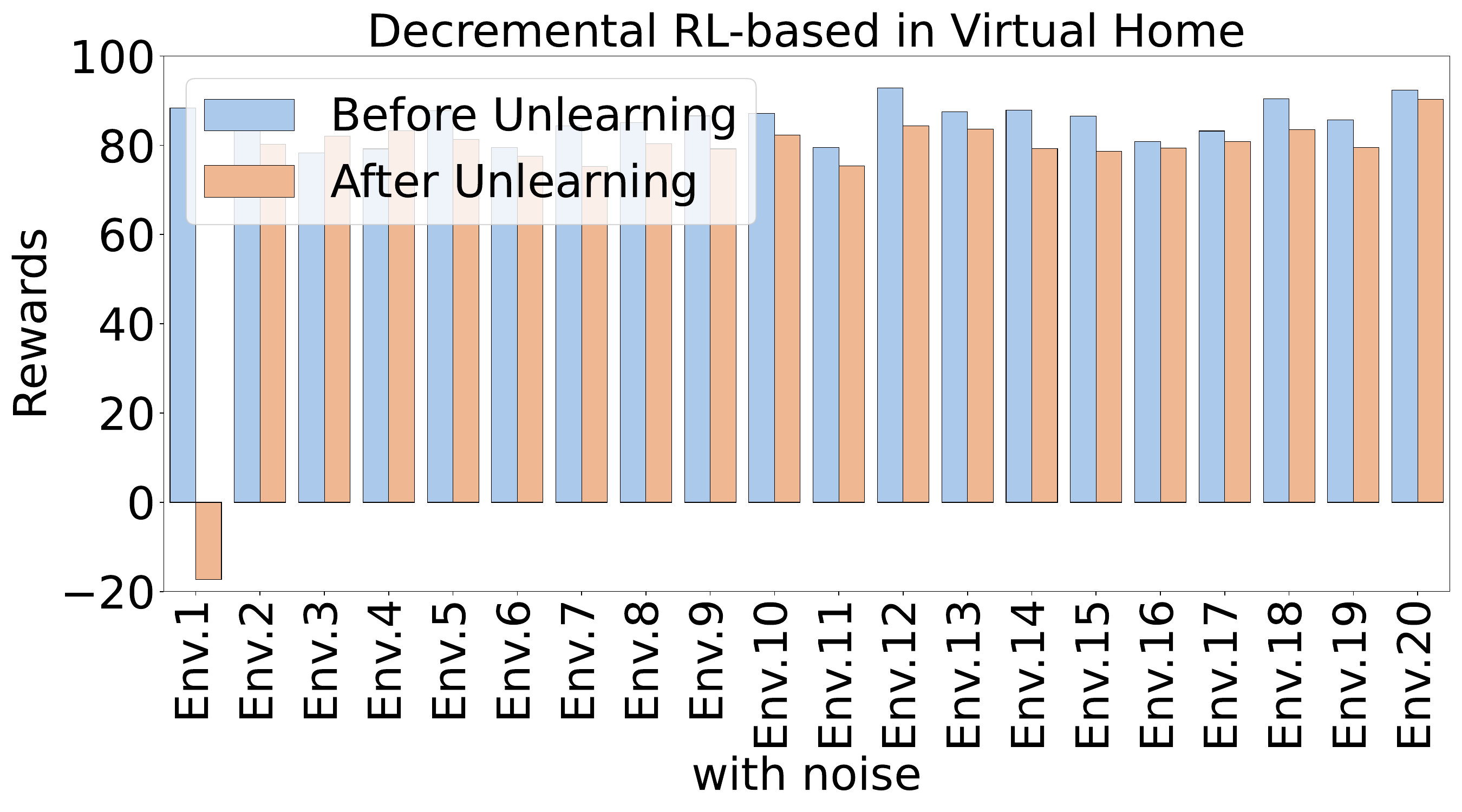}
			\label{fig:VHomeRobustRewardsMethod1Noise}}
   \subfigure[\scriptsize{Rewards of the decremental RL-based method without noise}]{
    \includegraphics[scale=0.092]{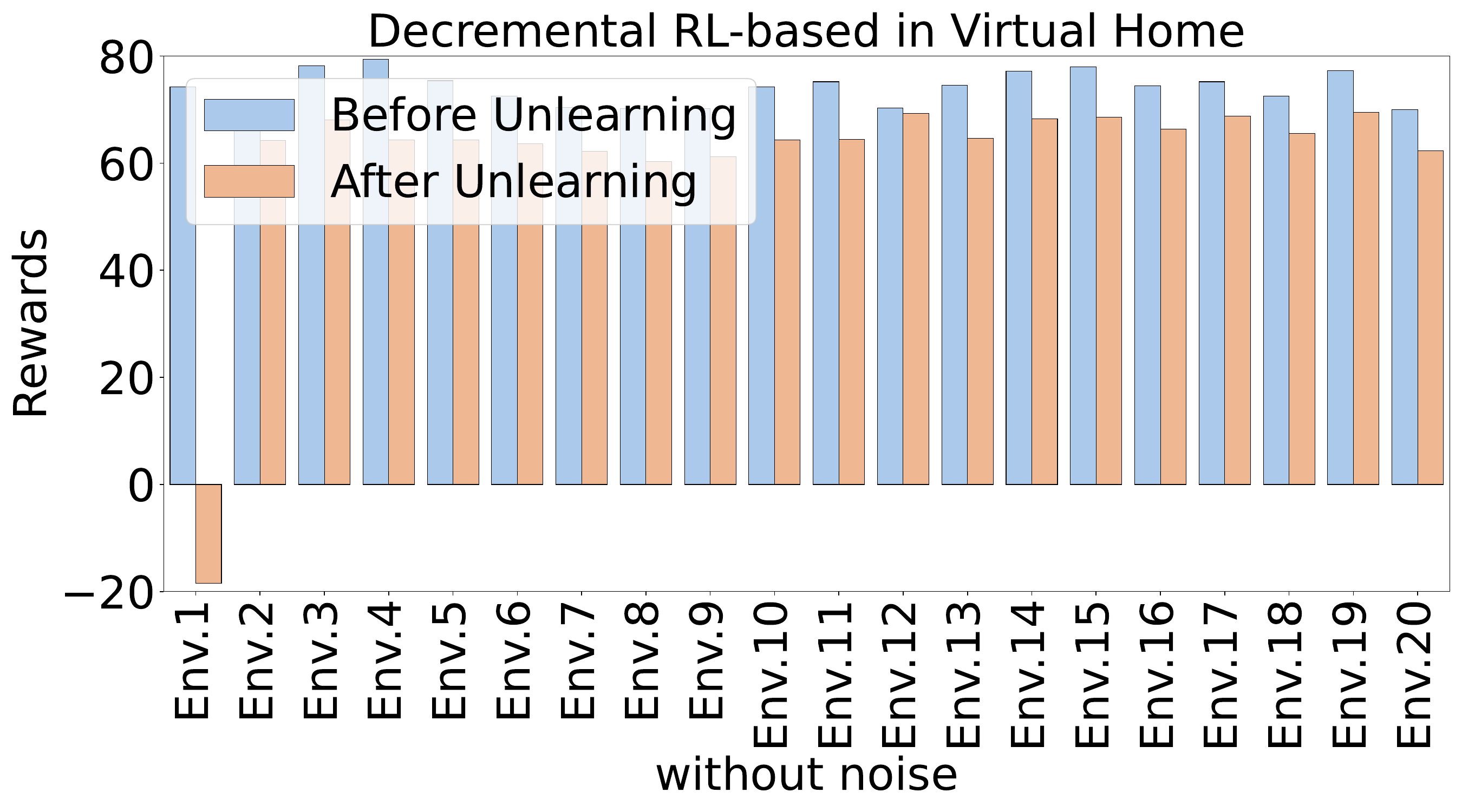}
			\label{fig:VHomeRobustRewardsMethod1WONoise}}\\[2ex]
       \subfigure[\scriptsize{Average steps of the poisoning-based method with noise}]{
    \includegraphics[scale=0.092]{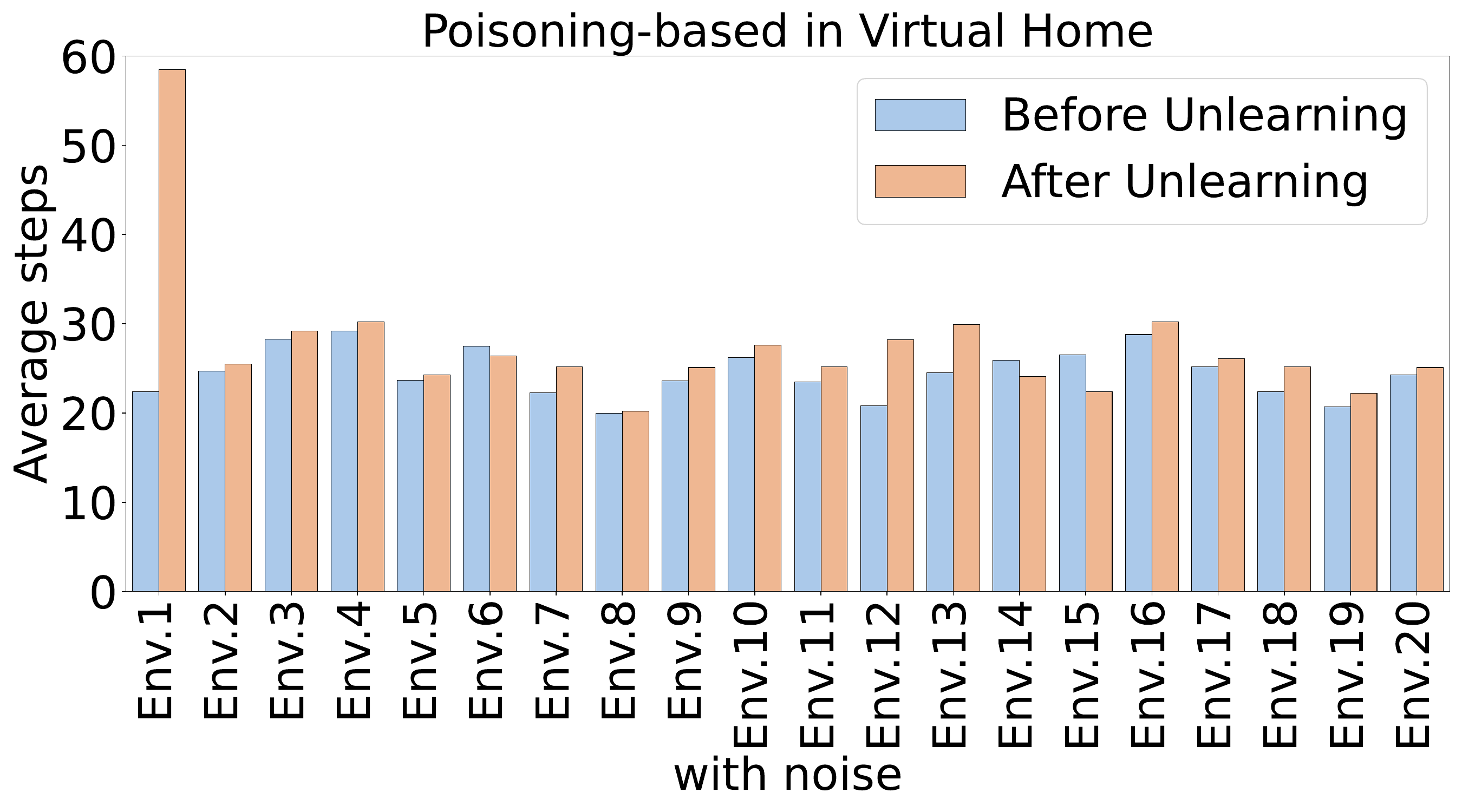}
			\label{fig:VHomeRobustStepsMethod2Noise}}
    \subfigure[\scriptsize{Average steps of the poisoning-based method without noise}]{
    \includegraphics[scale=0.092]{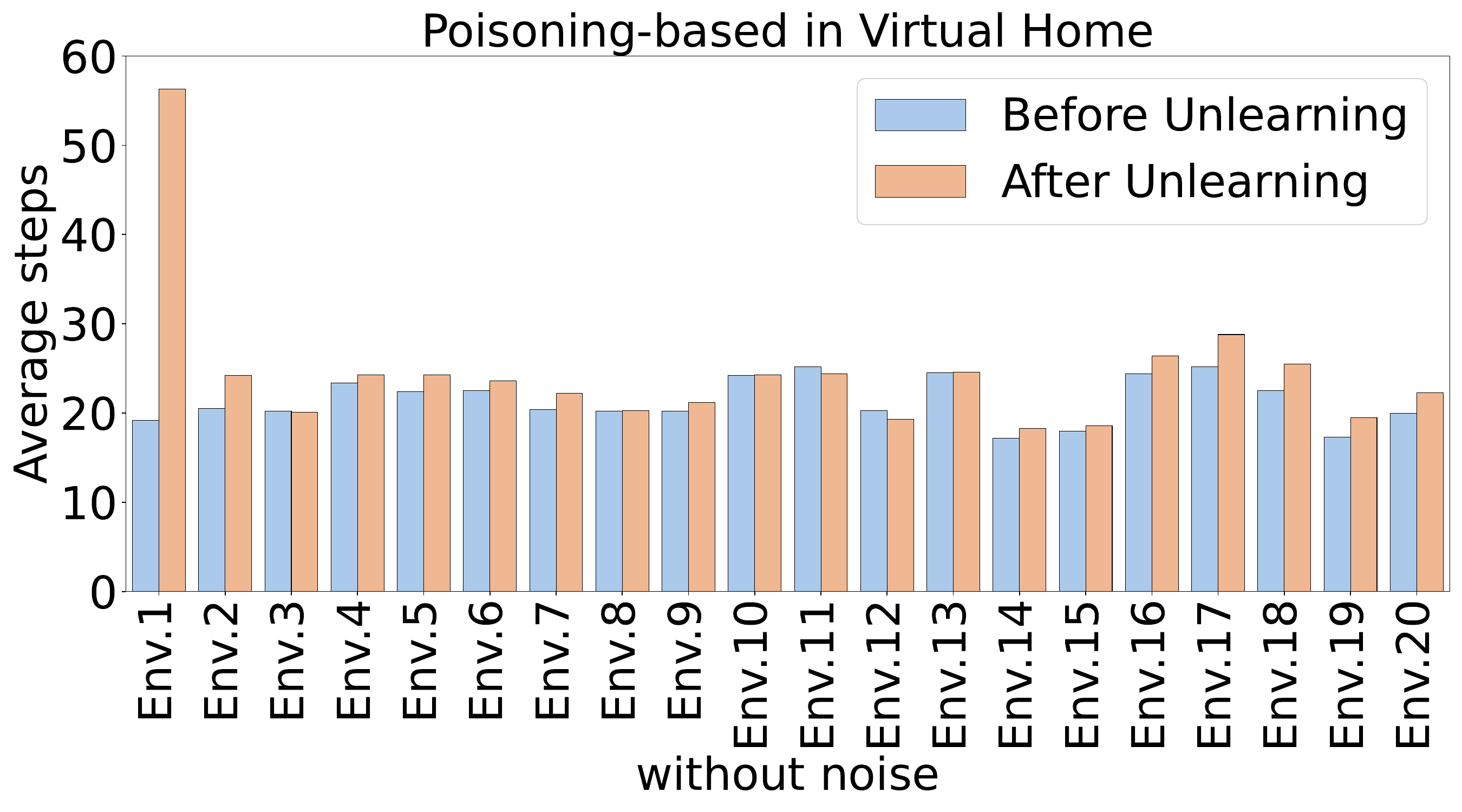}
			\label{fig:VHomeRobustStepsMethod2WONoise}}\\[2ex]
	\subfigure[\scriptsize{Rewards of the poisoning-based method with noise}]{
    \includegraphics[scale=0.092]{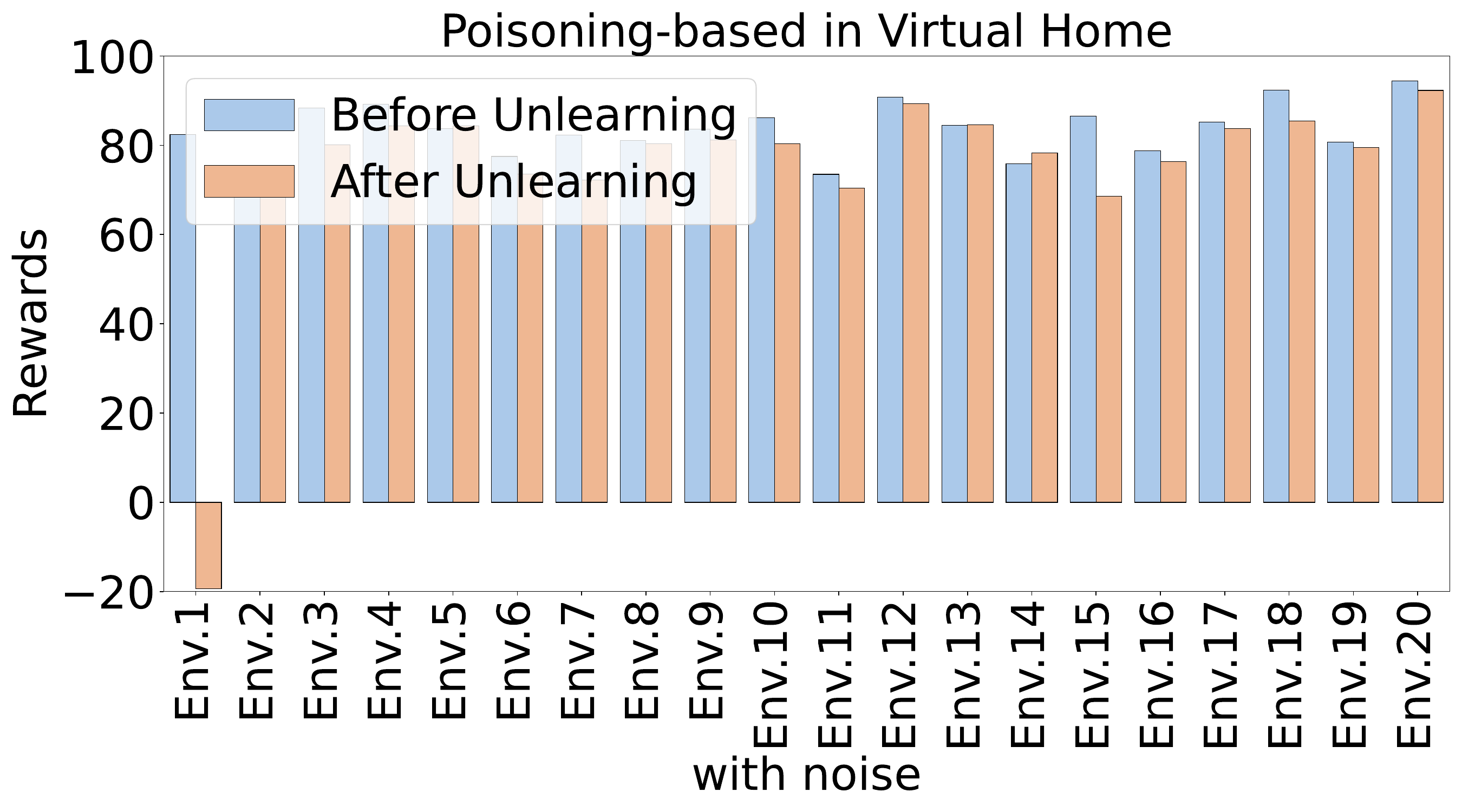}
			\label{fig:VHomeRobustRewardsMethod2Noise}}
   \subfigure[\scriptsize{Rewards of the poisoning-based method without noise}]{
    \includegraphics[scale=0.092]{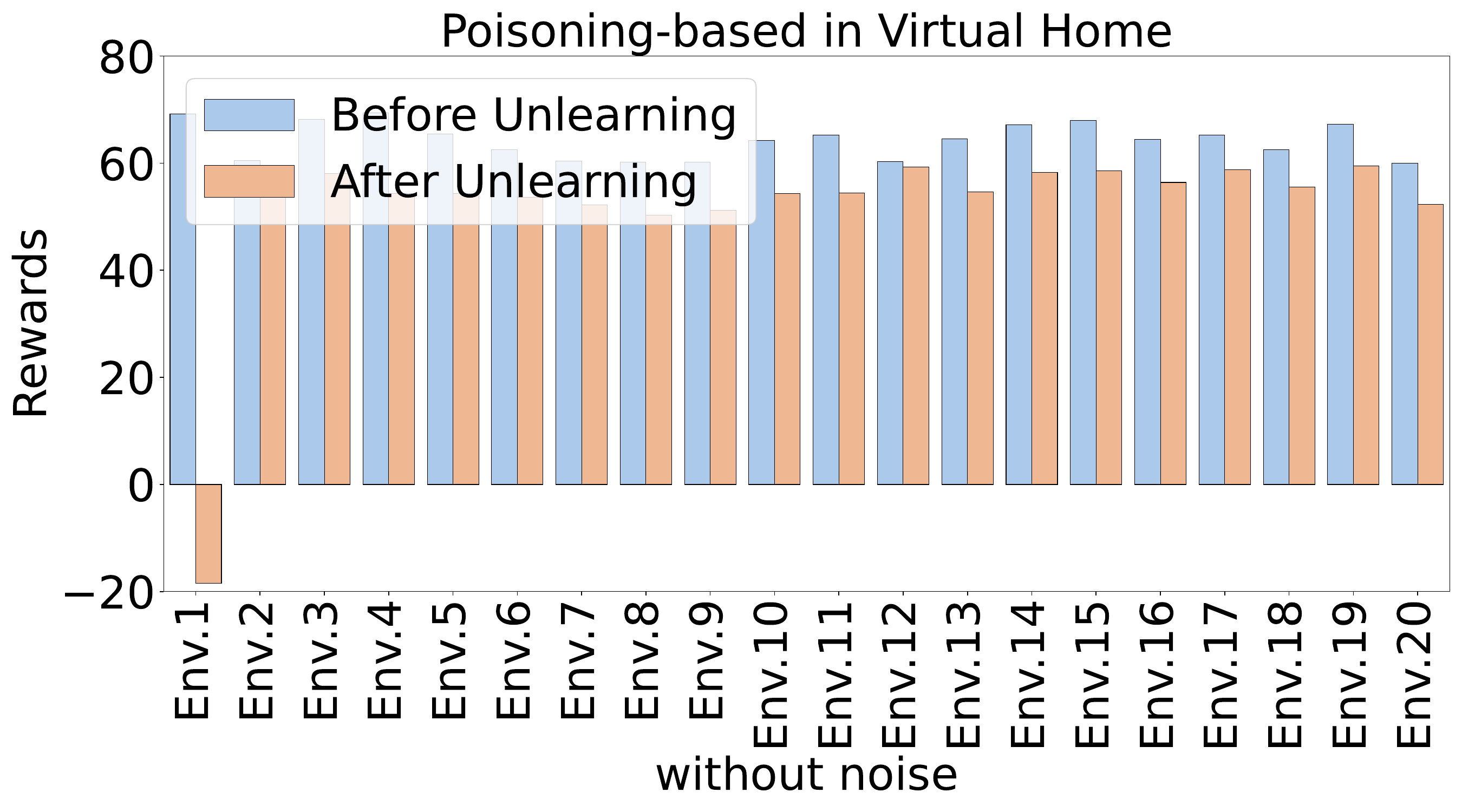}
			\label{fig:VHomeRobustRewardsMethod2WONoise}}
    \end{minipage}
	\caption{The decremental RL-based and poisoning-based methods in Virtual Home with and without noise}
	\label{fig:VHomeRobust}
\end{figure}

\begin{figure}[ht]
\centering
	\begin{minipage}{1\textwidth}
    \subfigure[\scriptsize{Average steps of the decremental RL-based method with noise}]{
    \includegraphics[scale=0.092]{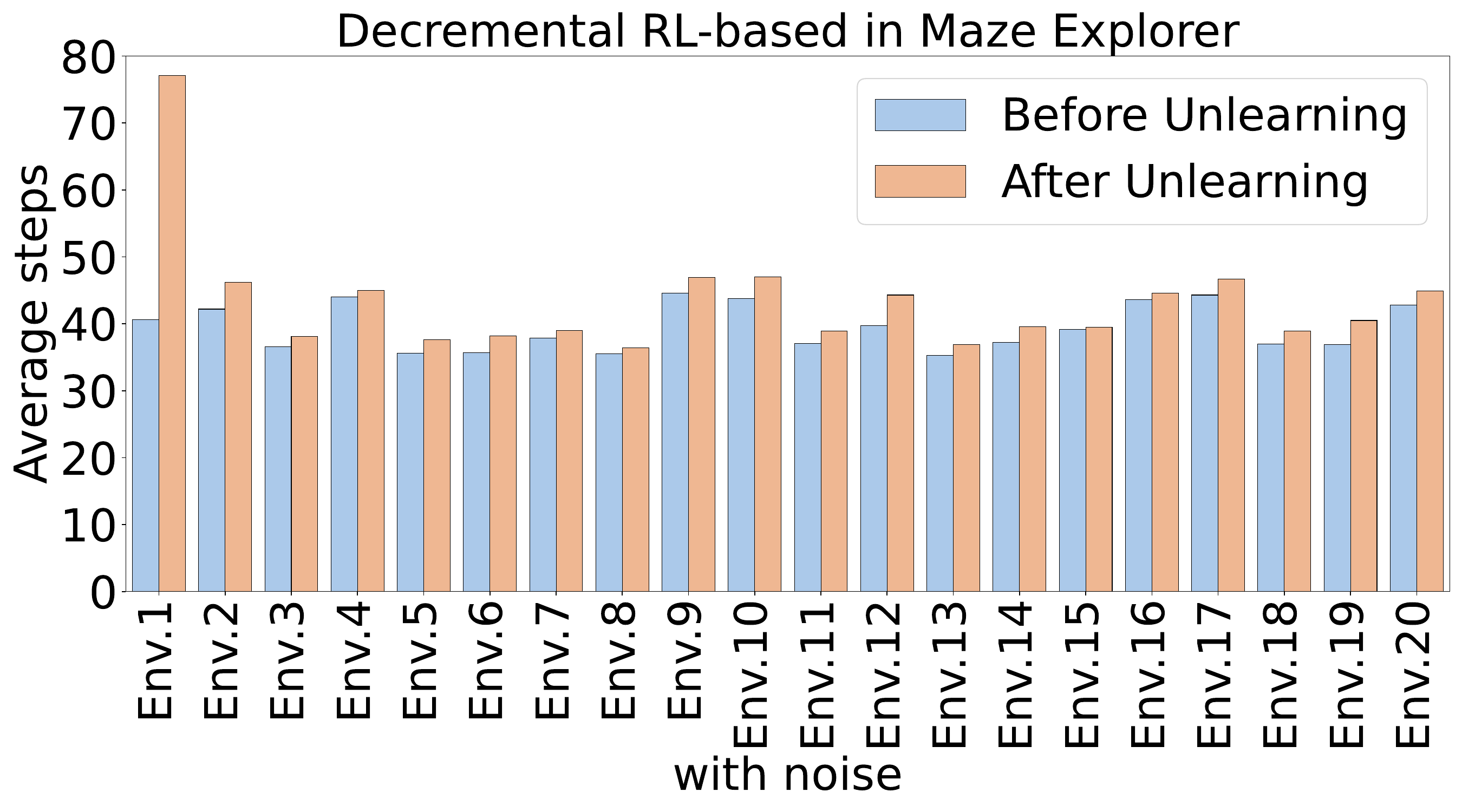}
			\label{fig:MazeRobustStepsMethod1Noise}}
    \subfigure[\scriptsize{Average steps of the decremental RL-based method without noise}]{
    \includegraphics[scale=0.092]{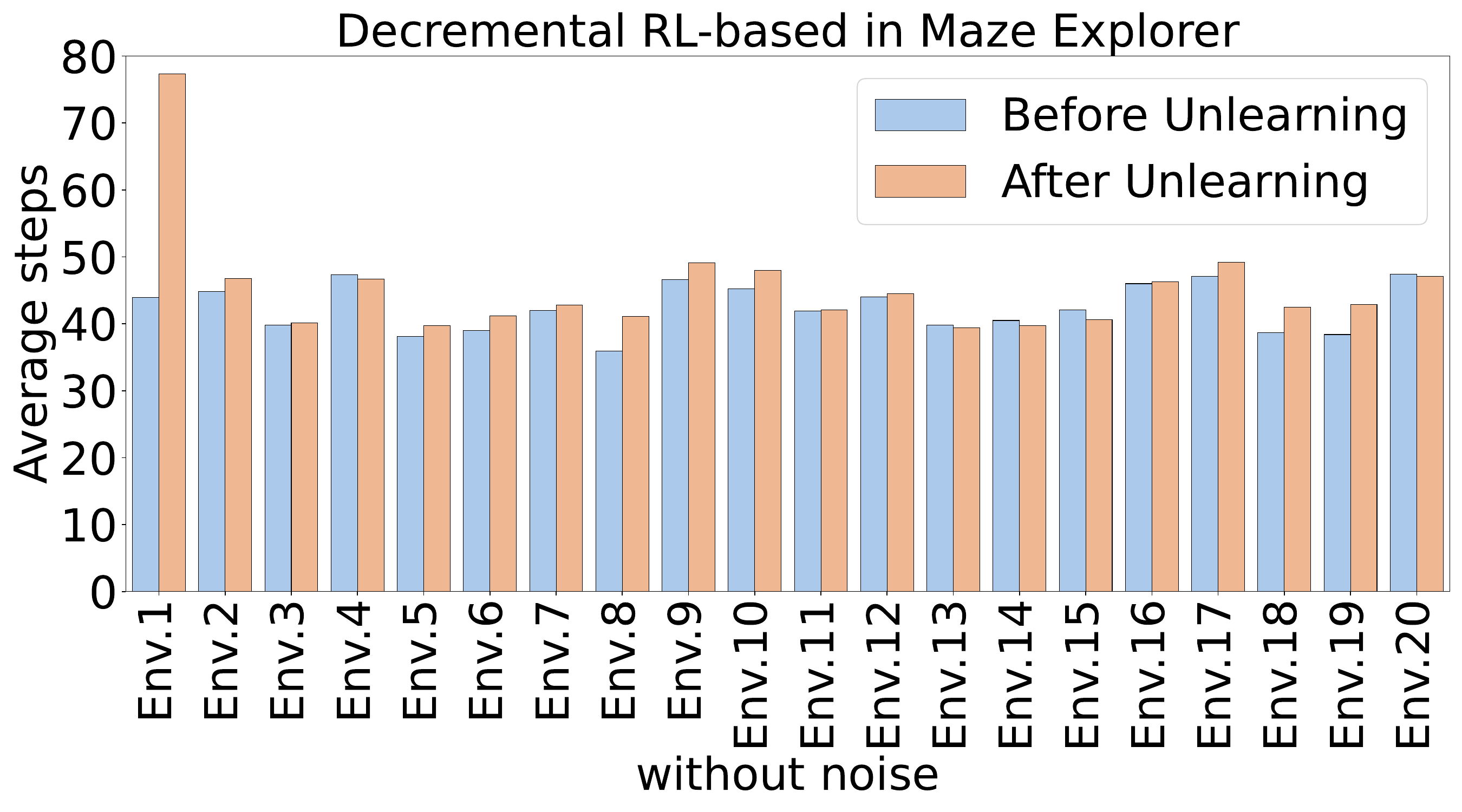}
			\label{fig:MazeRobustStepsMethod1WONoise}}\\[2ex]
	\subfigure[\scriptsize{Rewards of the decremental RL-based method with noise}]{
    \includegraphics[scale=0.092]{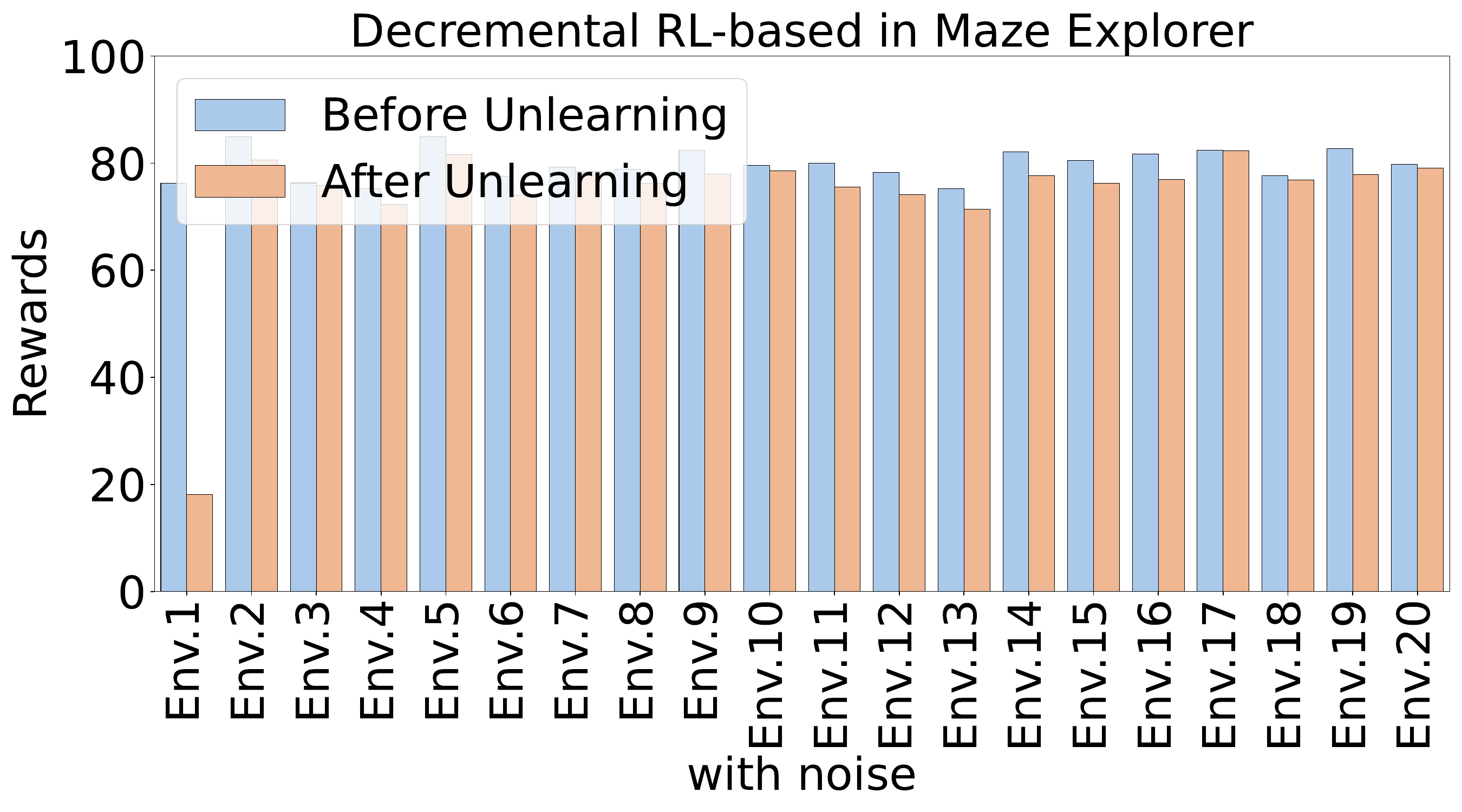}
			\label{fig:MazeRobustRewardsMethod1Noise}}
   	\subfigure[\scriptsize{Rewards of the decremental RL-based method without noise}]{
    \includegraphics[scale=0.092]{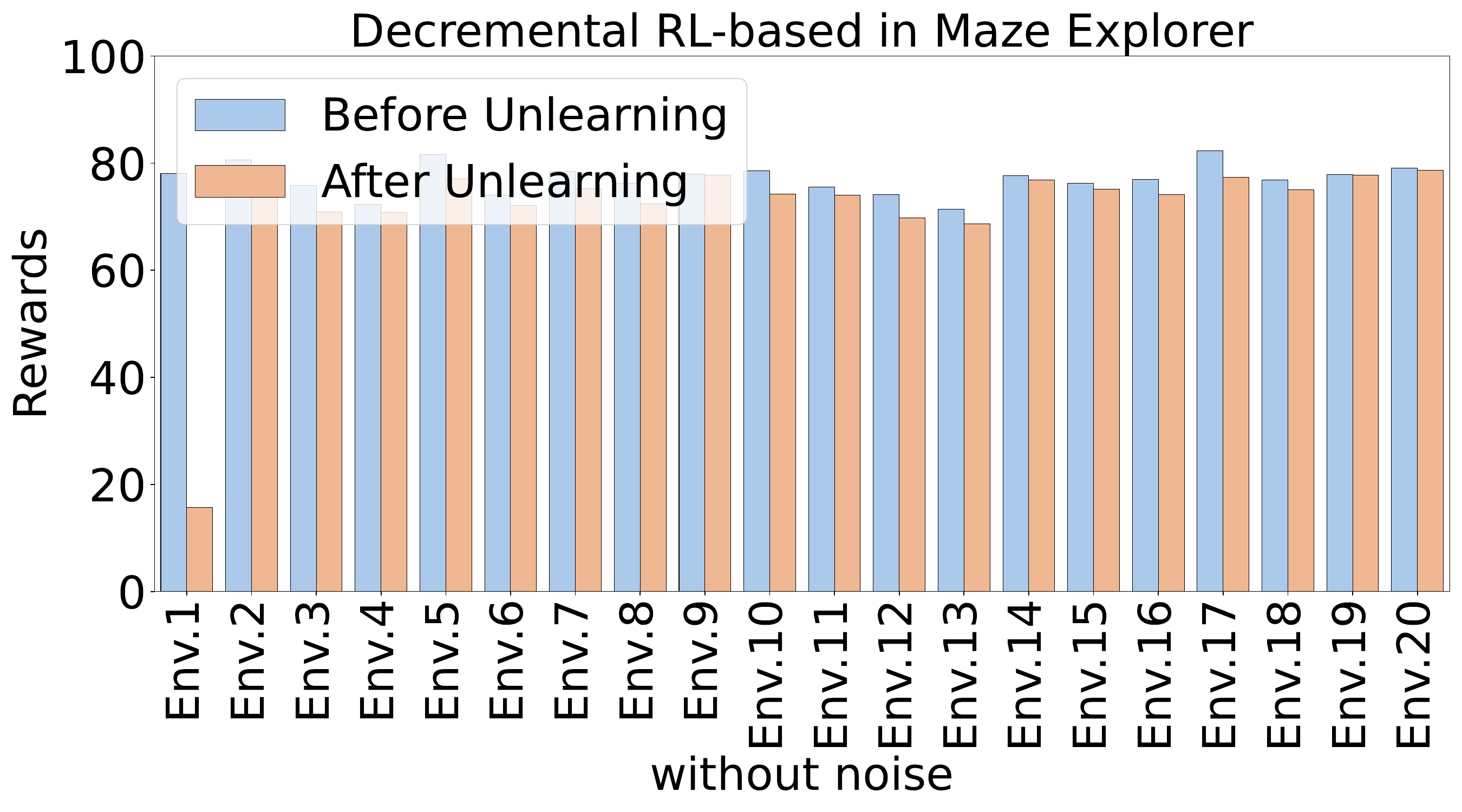}
			\label{fig:MazeRobustRewardsMethod1WONoise}}\\[2ex]
       \subfigure[\scriptsize{Average steps of the poisoning-based method with noise}]{
    \includegraphics[scale=0.092]{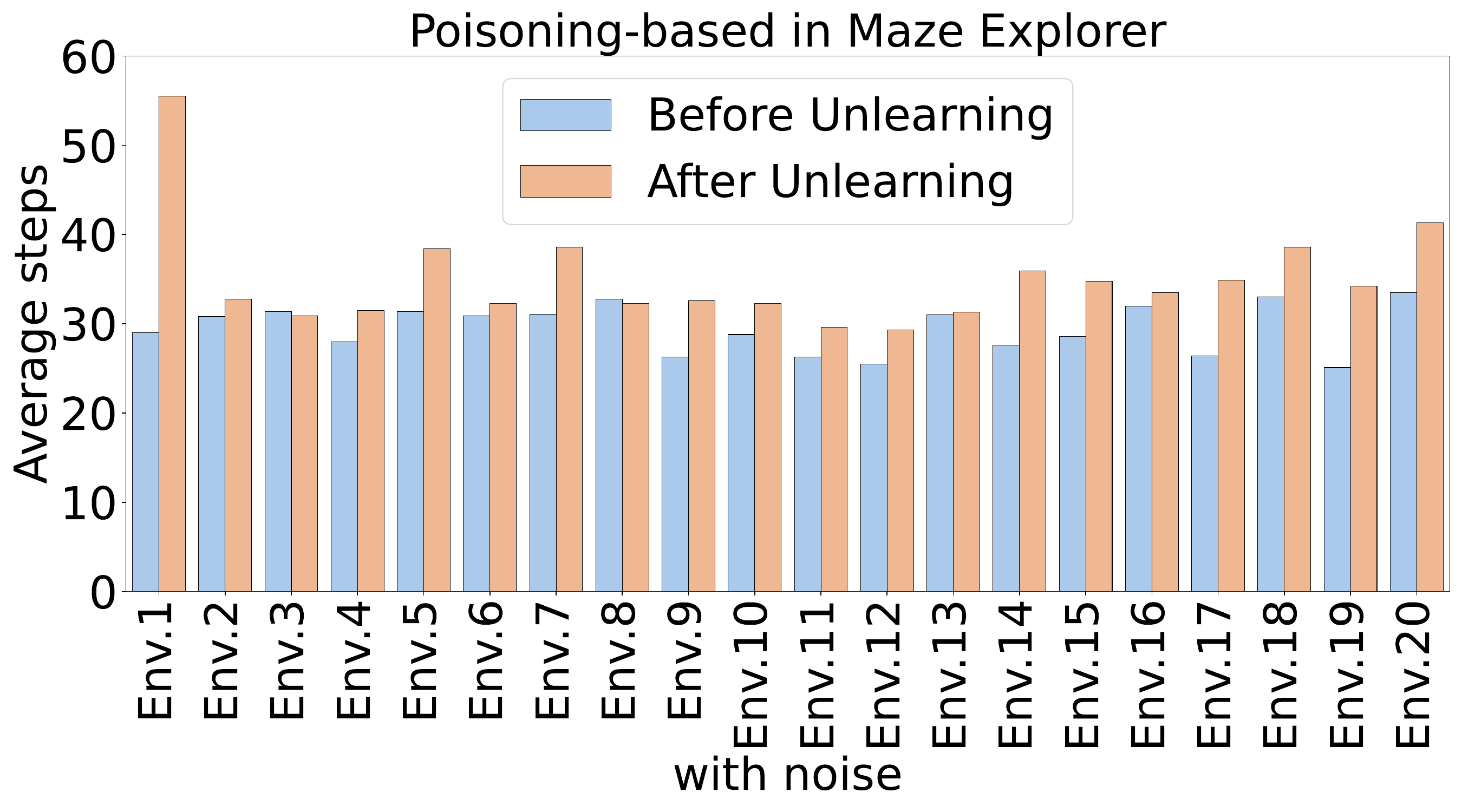}
			\label{fig:MazeRobustStepsMethod2Noise}}
          \subfigure[\scriptsize{Average steps of the poisoning-based method without noise}]{
    \includegraphics[scale=0.092]{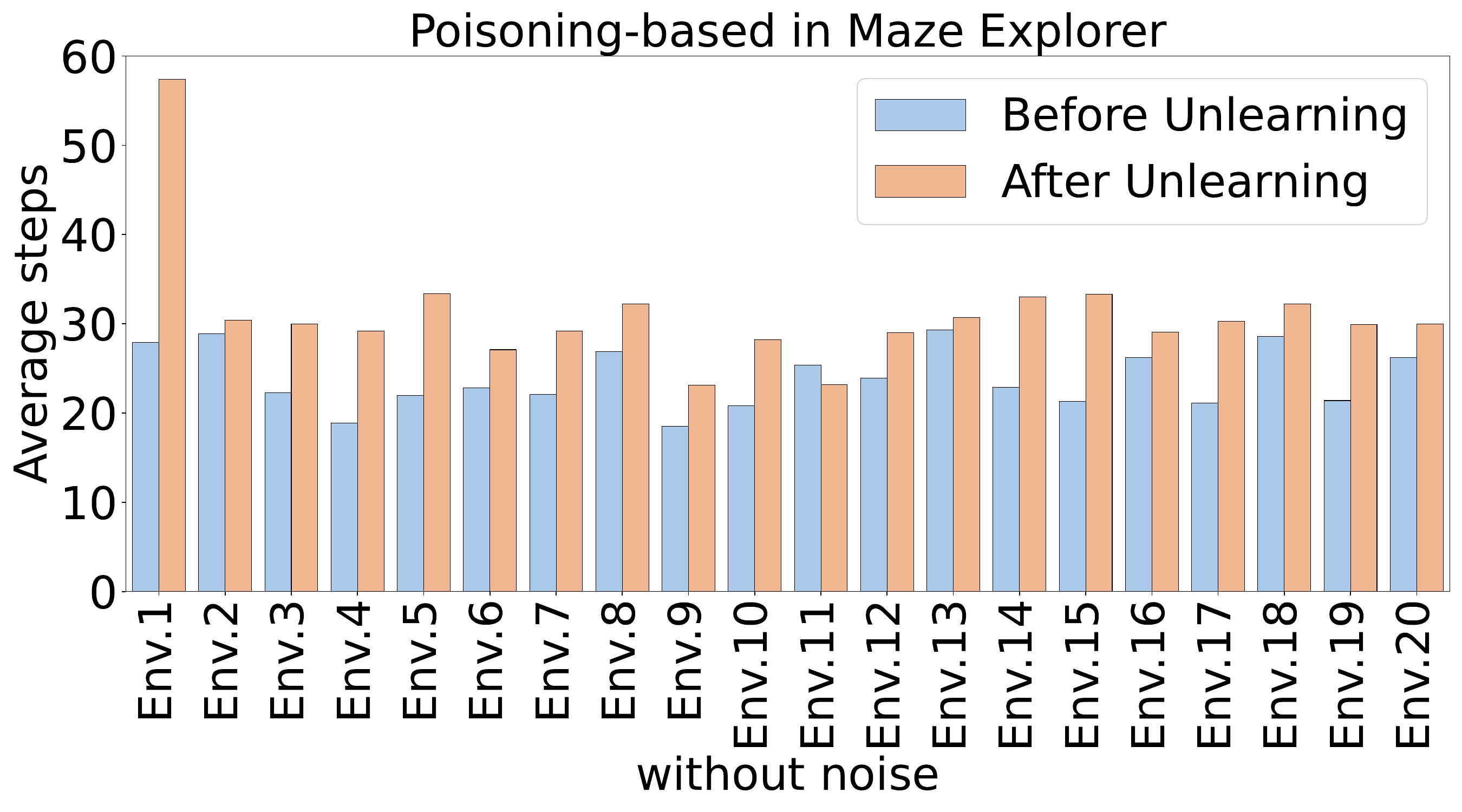}
			\label{fig:MazeRobustStepsMethod2WONoise}}\\[2ex]
	\subfigure[\scriptsize{Rewards of the poisoning-based method with noise}]{
    \includegraphics[scale=0.092]{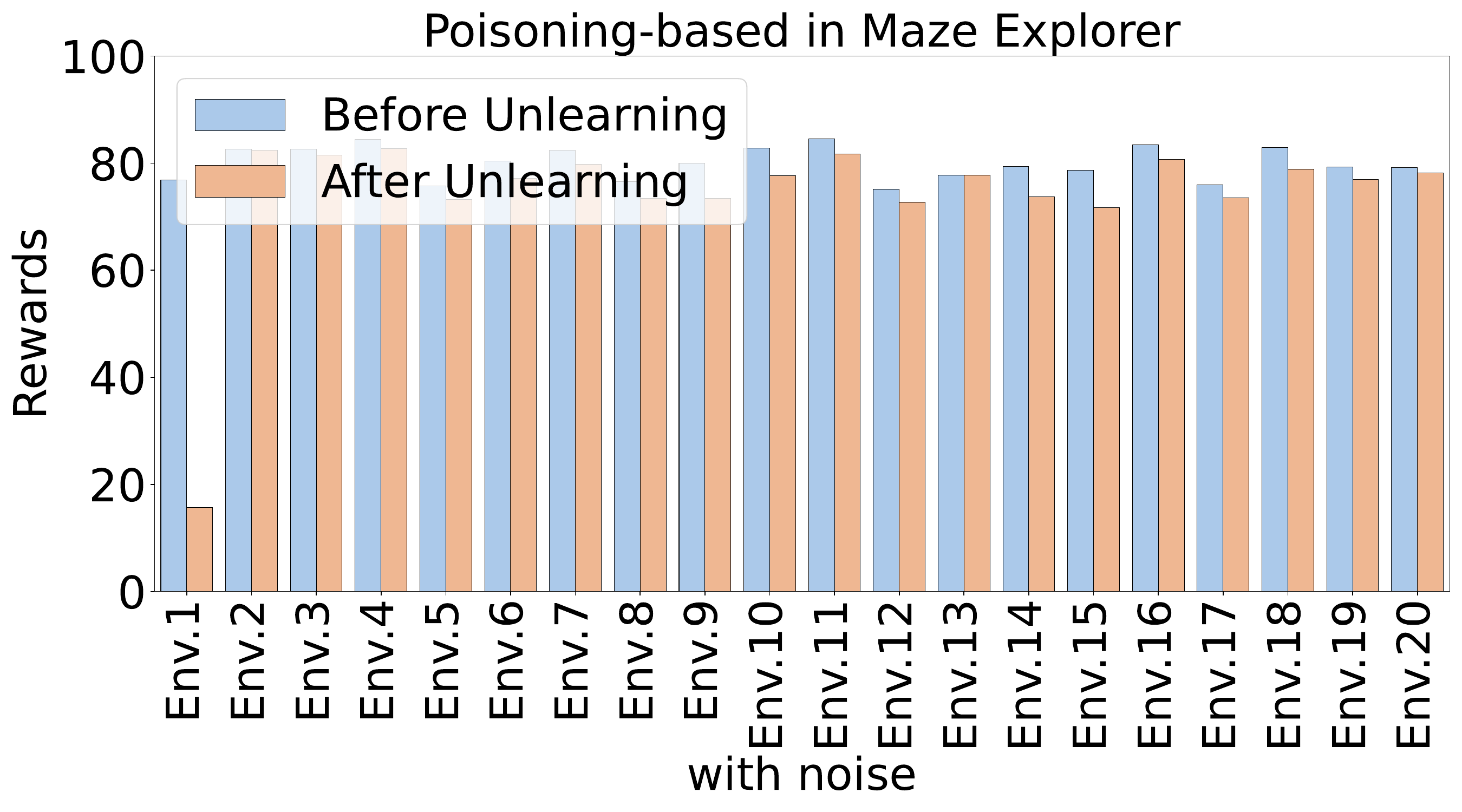}
			\label{fig:MazeRobustRewardsMethod2Noise}}
   	\subfigure[\scriptsize{Rewards of the poisoning-based method without noise}]{
    \includegraphics[scale=0.092]{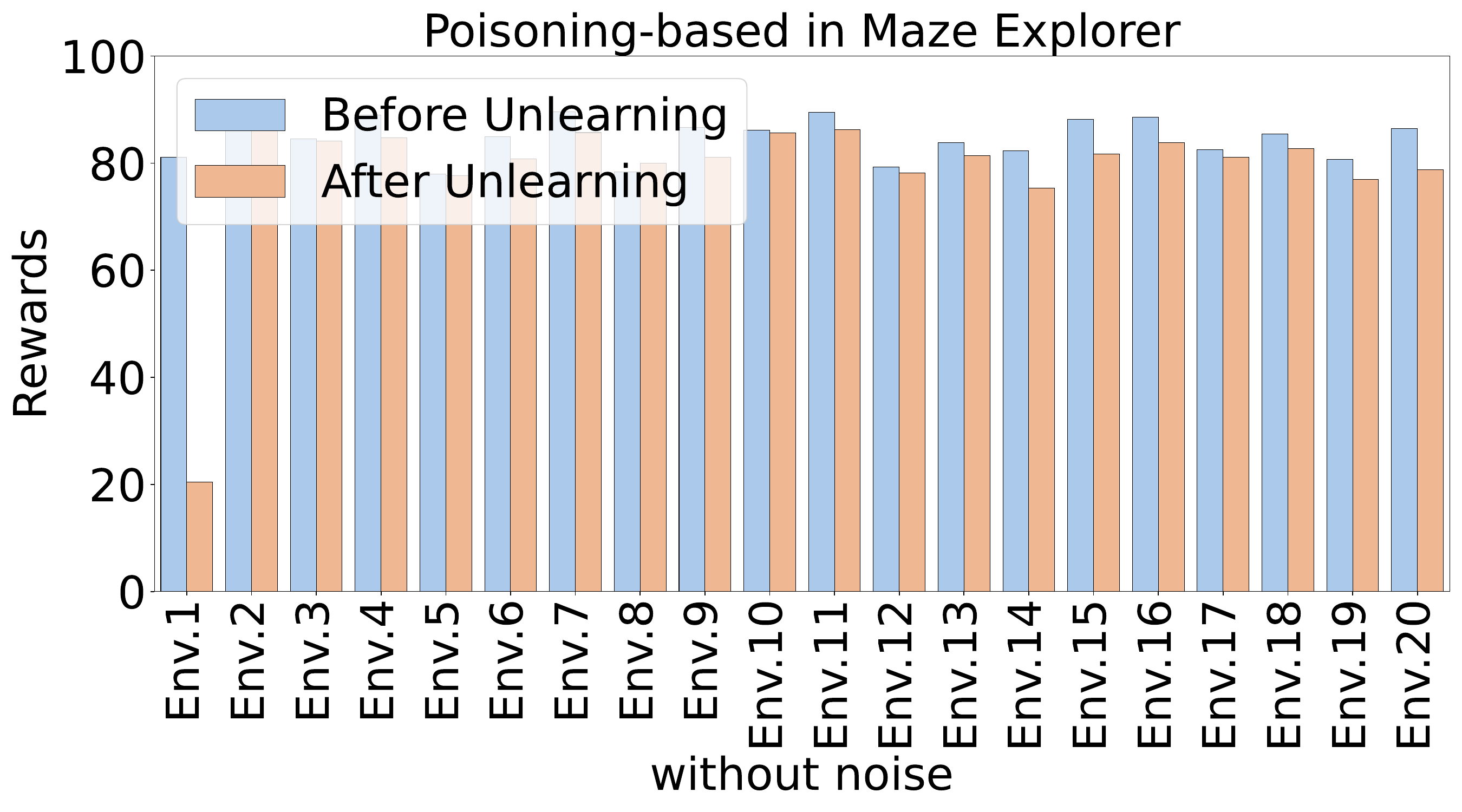}
			\label{fig:MazeRobustRewardsMethod2WONoise}}
    \end{minipage}
	\caption{The decremental RL-based and poisoning-based methods in Maze Explorer with and without noise}
	\label{fig:MazeRobust}
\end{figure}

The results observed in the virtual home (Figure \ref{fig:VHomeRobust}) and maze explorer (Figure \ref{fig:MazeRobust}) scenarios closely align with those in the grid world setting, showcasing the robustness of both the decremental RL-based and poisoning-based methods against external noise. In these two scenarios, the agent's behavior continues to exhibit consistent patterns even when external noise is introduced during both the training and unlearning process. This consistency is crucial for practical real-world applications, where agents must maintain their adaptability and performance despite uncertainties and disturbances. 


\vspace{2mm}
\noindent\textbf{Computational Adaptability.} To assess the computational adaptability of the four methods, we initially examined their computational overhead in scenarios where one environment is unlearned from a set of twenty. This evaluation was then expanded to the scenario of unlearning ten environments from a hundred. The outcomes are detailed in Table \ref{tab:computation}.

\begin{table}[!ht]\scriptsize
	\centering
 	\caption{Computation Overhead of the Four Methods in Grid World (seconds)}
\begin{tabular} {ccc}
\toprule
Methods & \makecell{Computation time of \\unlearning 1 environment} & \makecell{Computation time of \\unlearning 10 environments} \\
\midrule
Decremental RL & $18.80s$ & $64.83s$ \\
Poisoning & $20.17s$ & $73.56s$ \\
LFS & $198.62s$ & $746.25s$ \\
Non-transfer LFS & $196.97s$ & $742.51s$ \\
\bottomrule
\end{tabular}
	\label{tab:computation}
\end{table}

In comparison with the two baseline methods, the proposed decremental-RL and poisoning approaches show greater computational efficiency due to their less exhaustive exploration of the environments. Specifically, the decremental RL-based method is more efficient than the poisoning-based method. For instance, when the number of unlearning environments increases to $10$, the poisoning-based method requires $73.56$ seconds, whereas the decremental RL-based method only takes $64.83$ seconds, approximately $12\%$ faster. This efficiency stems from the decremental RL-based method's straightforward policy adjustments impacting rewards directly without the extensive need to recompute environmental dynamics, unlike the poisoning-based method, which involves computationally intensive alterations to the transition dynamics.


\subsection{Privacy Study}\label{sub:privacy study}
The privacy study is conducted using recommendation systems, where the key indicator is recommendation accuracy. Higher accuracy indicates a deep understanding of users' preferences and habits, reflecting potential privacy risks. Therefore, after a user revokes their data, the system's unlearning effectiveness is demonstrated by a decrease in recommendation accuracy for that user, suggesting the recommender has effectively forgotten the user’s preferences and is now generating recommendations randomly.

\begin{table}[!ht]\scriptsize
	\centering
 	\caption{Comparison of Recommendation Accuracy and Rewards Before and After Unlearning}
\begin{tabular} {ccccc}
\toprule
\multirow{2}*{Methods} & \multicolumn{2}{c}{\makecell{Performance for \\ the unlearned user}} & \multicolumn{2}{c}{\makecell{Average performance for \\ the remaining users}} \\\cline{2-5}
~ & Accuracy & Reward & Accuracy & Reward\\
\midrule
Before Unlearning & $92.07\%$ & $41.42$ & $91.52\%$ & $39.9$ \\
Decremental RL & $68.63\%$ & $20.03$ & $90.89\%$ & $38.82$ \\
Poisoning & $64.41\%$ & $18.25$ & $91.43\%$ & $37.17$ \\
\bottomrule
\end{tabular}
\vspace{-0mm}
	\label{tab:recommendation}
\end{table}

Table \ref{tab:recommendation} presents the recommendation results before and after unlearning. Initially, the recommendation accuracy for the unlearned user is high at $92.07\%$, indicating a strong understanding of the user's preferences. However, after unlearning, accuracy significantly decreases to $68.63\%$ and $64.41\%$ for the decremental RL-based and poisoning-based approaches respectively, suggesting effective unlearning and enhanced user privacy protection. This shift towards randomness in recommendations is further evidenced by the dramatic drop in rewards from $41.42$ to $20.03$ and $18.25$, indicating dissatisfaction with the recommendations after unlearning. Importantly, recommendations to other users remain consistent, highlighting the targeted nature of the proposed unlearning methods.

\subsection{Safety Study}\label{sub:safety study}
Since conventional machine unlearning can introduce security issues, such as reducing the adversarial robustness of unlearned models \cite{Zhao24ICML}, we aim to investigate whether reinforcement unlearning might lead to similar vulnerabilities. Our study focuses on a safety-critical scenario: aircraft landing. In this scenario, we measure safety performance by the number of collisions the agent encounters with obstacles. An increase in collisions after unlearning would indicate a higher level of safety concerns, suggesting that reinforcement unlearning could potentially compromise the safety of the system in ways analogous to the security issues observed in conventional machine unlearning.

\begin{figure}[ht]
\centering
	\begin{minipage}{1\textwidth}
    \subfigure[\scriptsize{The number of collisions under the decremental RL-based method}]{
    \includegraphics[scale=0.092]{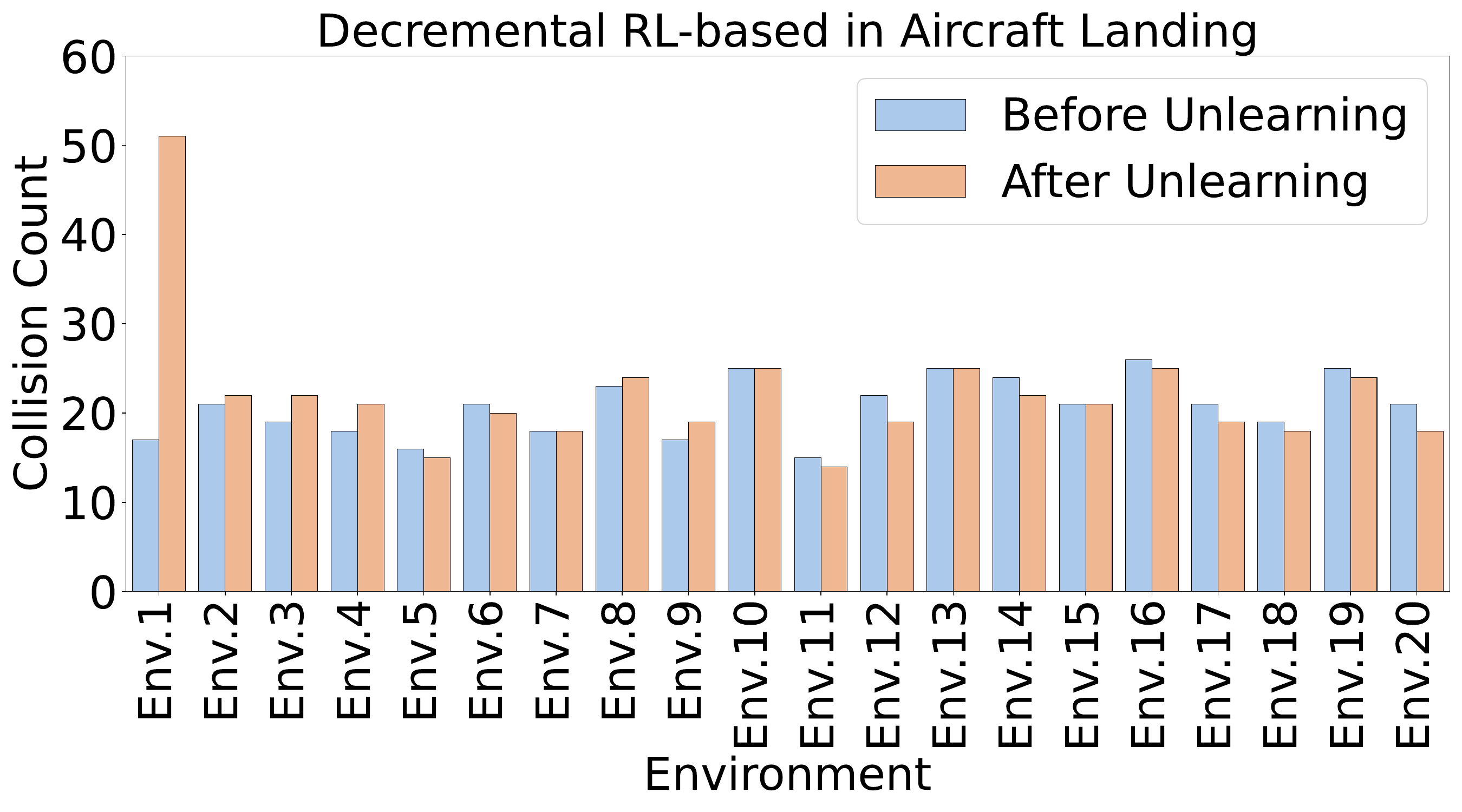}
			\label{fig:collisionDecremental}}
	\subfigure[\scriptsize{The number of collisions under the poisoning-based method}]{
    \includegraphics[scale=0.092]{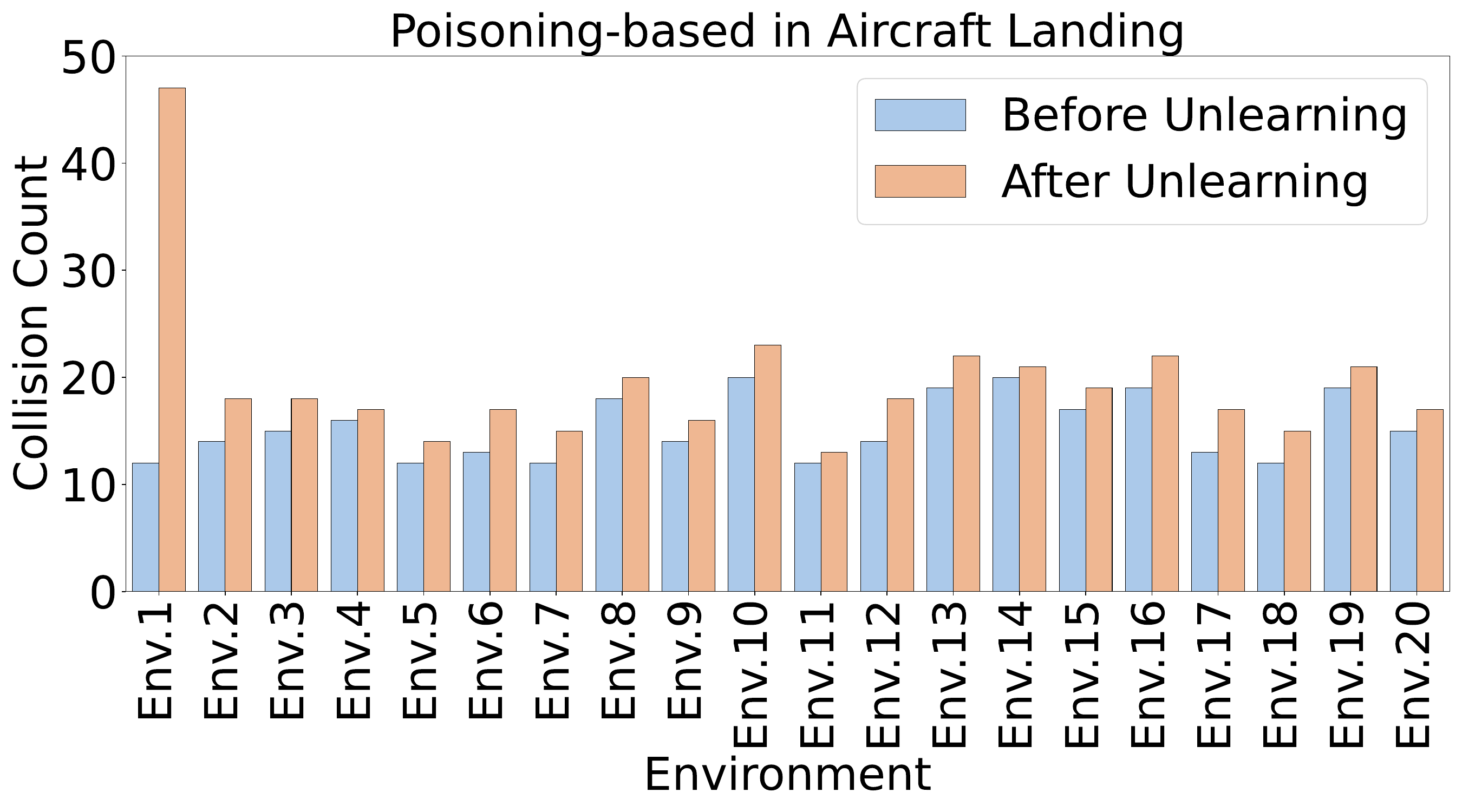}
			\label{fig:collisionPoison}}\\[2ex]
    \end{minipage}
	\caption{Collision count before and after unlearning in Aircraft Landing.}
	\label{fig:collision}
\end{figure}

The results shown in Figure \ref{fig:collision} reveal that after unlearning, the number of collisions under both methods has tripled compared to before unlearning. This substantial increase highlights the safety risks introduced when the agent forgets critical features of the environment. In safety-critical scenarios like aircraft landing, this observation underscores the need for extreme caution when implementing unlearning strategies. The trade-off between ensuring privacy and maintaining safety must be judiciously managed, as compromising safety can have serious consequences. Therefore, in contexts where safety is paramount, the approach to unlearning must prioritize minimal impact on operational safety.

\subsection{Environment Inference Testing}\label{sec:robustness}
This inference enables us to infer the environment that the agent needs to forget, allowing for a comparison of the inference outcomes before and after unlearning. 
We incorporate the $l_0$ distance as a quantitative measure to assess the inference results. Specifically, consider the unlearning environment as $\mathcal{M}_u$, and the inferred environments before and after unlearning as $\mathcal{M}_u^{before}$ and $\mathcal{M}_u^{after}$, respectively. The $l_0$ distance between two environments is defined as the number of differing dimensions between them.

\begin{figure}[ht]
\centering
    \includegraphics[scale=0.3]{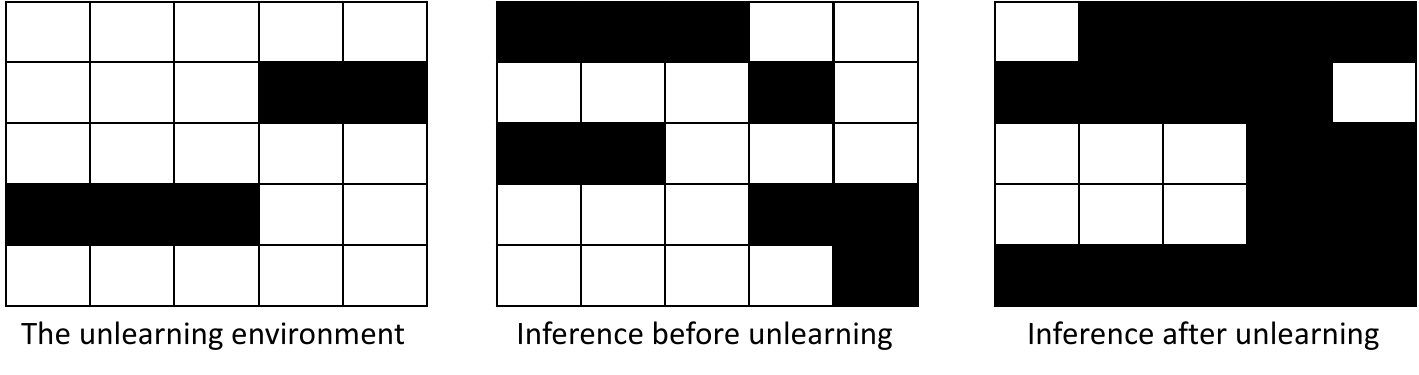}
	\caption{Inference results in Grid World}
	\label{fig:GridReconstruction}
\end{figure}


In Figure \ref{fig:GridReconstruction}, we present the results of the grid world setting for the decremental RL-based method.
The inference successfully recreates $50\%$ of the unlearning environment before unlearning, resulting in $l_0(\mathcal{M}_u,\mathcal{M}_u^{before})=12$ for this $5\times 5$ environment. However, after unlearning, this inference result significantly reduces to only $20\%$, yielding $l_0(\mathcal{M}_u,\mathcal{M}_u^{before})=20$. This result provides clear evidence of a successful unlearning process.
Note that the inference results before unlearning may not appear fully accurate due to the specific inference approach used, which is only intended to verify the effectiveness of the unlearning process, rather than to demonstrate an adversary's capabilities. The primary objective here is to assess how well the unlearning method obscures the agent’s prior knowledge of the environment.

The reason behind this success lies in the unlearning methods' capability to modify the agent's learned policy effectively. Both of the proposed methods adapt the agent's behavior to forget specific aspects of the environment while preserving essential knowledge. Thus, the environment inference becomes less effective in recreating the forgotten parts after unlearning.
Also, the visual comparison highlights the unlearning methods' efficiency in refining the agent's policy to eliminate unwanted behaviors. The process of inferring the forgotten environment confirms the success of our unlearning methods in reducing the agent's reliance on previously learned information and adapting to changes in the environment.

\section{Threats to Validity}
\noindent\textbf{Definition Validity.} The study defines unlearning in a specific way (minimizing performance in the unlearning environment). Alternative definitions of unlearning might yield different results. Moreover, the defined reward function might not capture all aspects of the agent's behavior and performance, leading to incomplete assessments of unlearning efficacy.

\vspace{2mm}
\noindent\textbf{Methodology Validity.} The effectiveness of the proposed unlearning methods may be dependent on the specific RL algorithms used in the study. Additionally, the unlearning process might be influenced by factors not accounted for in the study. 
For example, in recommendation systems, user preferences can change over time, making it challenging for the agent to provide accurate recommendations. While we have tested various RL algorithms and incorporated environmental dynamism into our experiments, our evaluations are not exhaustive in capturing all potential variables.

\vspace{2mm}
\noindent\textbf{Generalization Validity.} The experiments were conducted in simulated environments, which may not fully represent the complexities of real-world scenarios. Furthermore, the metrics used to evaluate unlearning performance might not capture all aspects of the agent's behavior in real-life applications. 

\vspace{-1mm}
\section{Related Work}

\noindent\textbf{Machine Unlearning.~} 
The concept of machine unlearning was initially introduced in \cite{Cao15}. They employed statistical query learning and decomposed the model into a summation form, enabling efficient removal of a sample by subtracting the corresponding summand. 
Later, Bourtoule et al. \cite{Bourtoule21} proposed SISA training, which involves randomly partitioning the training set into multiple shards and training a constituent model for each shard. 
In the event of an unlearning request, the model provider only needs to retrain the corresponding shard model.
Warnecke et al. \cite{Warnecke23} shifted the focus of unlearning research from removing samples to removing features and labels. Their approach is based on the concept of influence functions, which allows for estimating the influence of data on learning models. 
Machine unlearning has also been explored from a theoretical perspective. 
Ginart et al. \cite{Ginart19} introduced the concept of $(\epsilon,\delta)$-approximate unlearning, drawing inspiration from differential privacy (DP) \cite{Dwork06,Ruan23SP,Feng23SP}. 
Then, Guo et al. \cite{Guo20} formulated unlearning as certified removal and provided theoretical guarantees. They achieved certified removal by employing convex optimization followed by Gaussian perturbation on the loss function.
Gupta et al. \cite{Gupta21} considered update sequences based on a function of the published model. 
They leveraged differential privacy and its connection to max information to develop a data deletion algorithm. 
Thudi et al. \cite{Thudi23} argued that unlearning cannot be proven solely by training the model on the unlearned data. 
Instead, unlearning can only be defined at the level of the algorithms used for learning and unlearning. 

\vspace{1mm}
\noindent\textbf{Reinforcement Learning Security.~} 
While reinforcement unlearning remains an underexplored area, considerable research has been devoted to reinforcement learning security  \cite{Wu21,Yu23}.  
For instance, Chen et al. \cite{Chen23NIPS} proposed a backdoor detection and removal approach for DRL via unlearning, involving the re-initialization of the top $L$ neurons' weights to erase the most vulnerable shortcuts. However, these studies differ significantly from reinforcement unlearning for three reasons. Firstly, the focus of these studies revolves around identifying and addressing vulnerabilities within the learning process. In contrast, our work centers on the task of effectively forgetting previously acquired knowledge. Secondly, these studies commonly aim to train robust agents capable of withstanding diverse adversarial activities. Our goal, however, is to develop techniques for unlearning knowledge in RL agents, allowing them to adapt their policies based on unlearning requests. Lastly, these studies endeavor to explain learned policies, particularly within security applications. In contrast, our research extends beyond security concerns to address broader issues related to policy adaptation and forgetting in reinforcement learning. 





\section{Conclusion}\label{sec:conclusion}
This paper presents a pioneering research area, termed \emph{reinforcement unlearning}, which addresses the crucial need to protect the privacy of environment owners by enabling an agent to unlearn entire environments.
We propose two distinct reinforcement unlearning methods: decremental RL-based and environment poisoning-based approaches. These methods are designed to be adaptable to different situations and provide effective mechanisms for unlearning.
Also, we introduce a novel concept termed ``environment inference'' to evaluate the outcomes of the unlearning process. 

Our future work focuses on developing a unified framework for reinforcement unlearning that integrates various unlearning methods and supports easy application to different RL algorithms and environments. This includes designing a modular architecture that allows for interchangeable unlearning components tailored to specific needs of different RL scenarios.

\section*{Acknowledgments}
Tianqing Zhu is the corresponding author. This research is supported by the Australian Research Council (ARC) Discovery Grant DP230100246. Moreover, Minhui Xue's work is partially supported by ARC Grant DP240103068 and by the CSIRO–National Science Foundation (US) AI Research Collaboration Program.




\bibliographystyle{IEEEtranS}
\bibliography{references}

\section*{Appendix}
\setcounter{section}{0}
\renewcommand{\appendixname}{Appendix~\Alph{section}}

\section{Unlearning in a Single Environment}\label{appendix: single environment}
\noindent\textbf{Problem Definition.} Unlearning within a single environment primarily focuses on selectively forgetting specific knowledge from that environment. We formally define the problem as follows. Given an environment $\mathcal{M}=\{\mathcal{S},\mathcal{A},\mathcal{T},r\}$, a set of $n$ trajectories $(\tau_1,...,\tau_n)$ occur within it. Each trajectory $\tau_i=((s^i_1, a^i_1), \ldots, (s^i_k, a^i_k))$ is a sequence of $k$ states and actions experienced by the agent as it interacts with the environment according to a policy. Assume the agent aims to unlearn a trajectory $\tau_u$, the goal is to update the policy $\pi$ to $\pi'$ such that the reward obtained along trajectory $\tau_u$ is minimized:
\begin{equation}\label{eq:aim sinEnv}
    \min_{\pi'}\mathbb{E}_{(s,a)\sim\tau_u}[Q_{\pi'}(s,a)],
\end{equation}
while unaffecting the reward in other trajectories: 
\begin{equation}\label{eq:constraint sinEnv}
    \min_{\pi'}\mathbb{E}_{(s,a)\not\sim\tau_u}|Q_{\pi'}(s,a)-Q_{\pi}(s,a)|. 
\end{equation} 
In Eq. \ref{eq:aim sinEnv}, minimizing the reward for specific state-action pairs teaches the agent these pairs are less valuable. Consequently, the agent may select alternative actions in those states to avoid the reduced rewards. This change in action selection leads to different subsequent states, effectively altering the unlearned trajectory. Essentially, this makes the trajectory unlikely to occur, achieving the effect of forgetting it. In comparison, Eq. \ref{eq:constraint sinEnv} focuses on preserving the integrity of the remaining trajectories, effectively ensuring they remain unaffected. This goal is to maintain stability and reliability in the agent's performance across these trajectories, emphasizing remembering them, while selectively forgetting the specified trajectory.

\vspace{2mm}
\noindent\textbf{Methods.} To adapt our methods to unlearning in a single-environment scenario, we have modified the loss and reward functions accordingly. For the decremental RL-based method, the revised loss function is:
\begin{equation}\label{eq:lossu sinEnv}\nonumber
    \mathcal{L}_u=\mathbb{E}_{(s,a)\sim\tau_u}[Q_{\pi'}(s,a)]+\mathbb{E}_{(s,a)\not\sim\tau_u}|Q_{\pi'}(s,a)-Q_{\pi}(s,a)|,
\end{equation}
where the first term reduces rewards along the trajectory $\tau_u$, and the second term maintains policy performance elsewhere. Similarly, the reward function for the poisoning-based method is reformulated as:
\begin{equation}\label{eq:reward sinEnv}\nonumber
    \mathcal{R}_i:=\lambda_1\Delta(\pi_i(s_i)||\pi_{i+1}(s_i))+\lambda_2\sum_{s\not\sim\tau_u}\sum_a\pi_i(s,a)r(s,a),
\end{equation}
targeting a higher divergence between successive policies $\pi_i$ and $\pi_{i+1}$ and ensuring good performance on remaining trajectories.

\begin{figure}[ht]
\centering
	\begin{minipage}{1\textwidth}
    \subfigure[\scriptsize{Rewards of decremental RL-based}]{
    \includegraphics[scale=0.21]{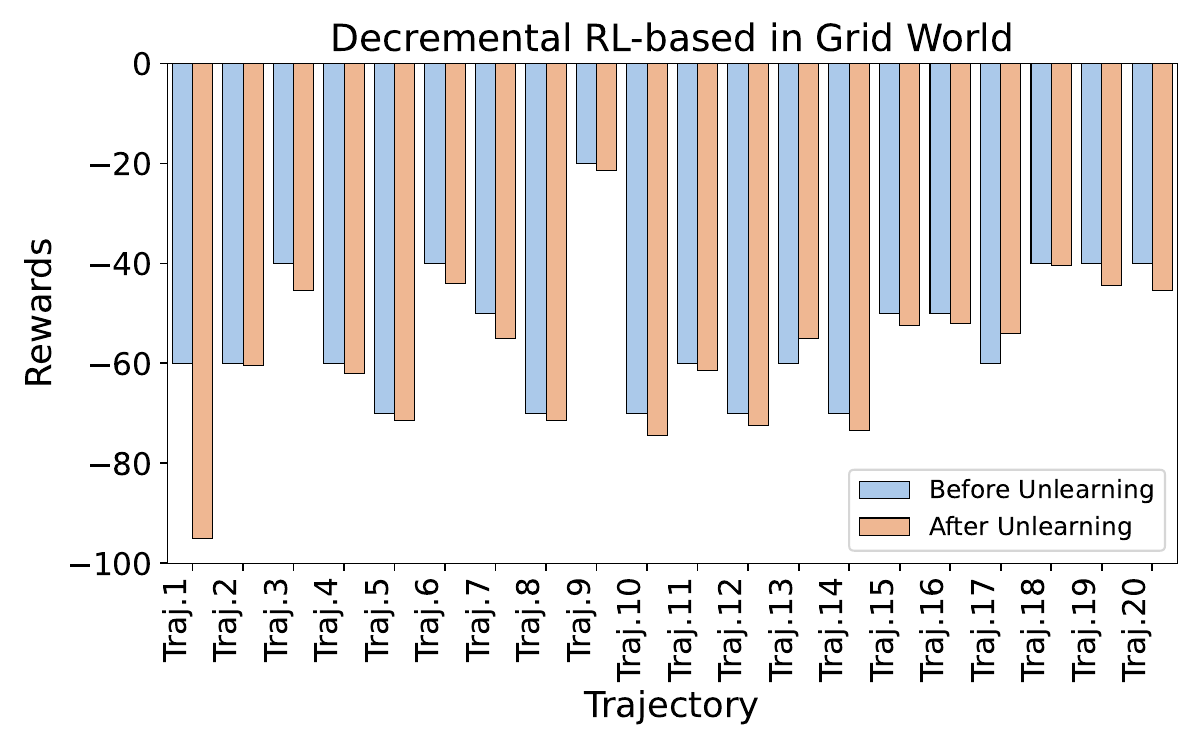}
			\label{fig:SinEnvMethod1Reward}}
	\subfigure[\scriptsize{Rewards of poisoning-based}]{
    \includegraphics[scale=0.21]{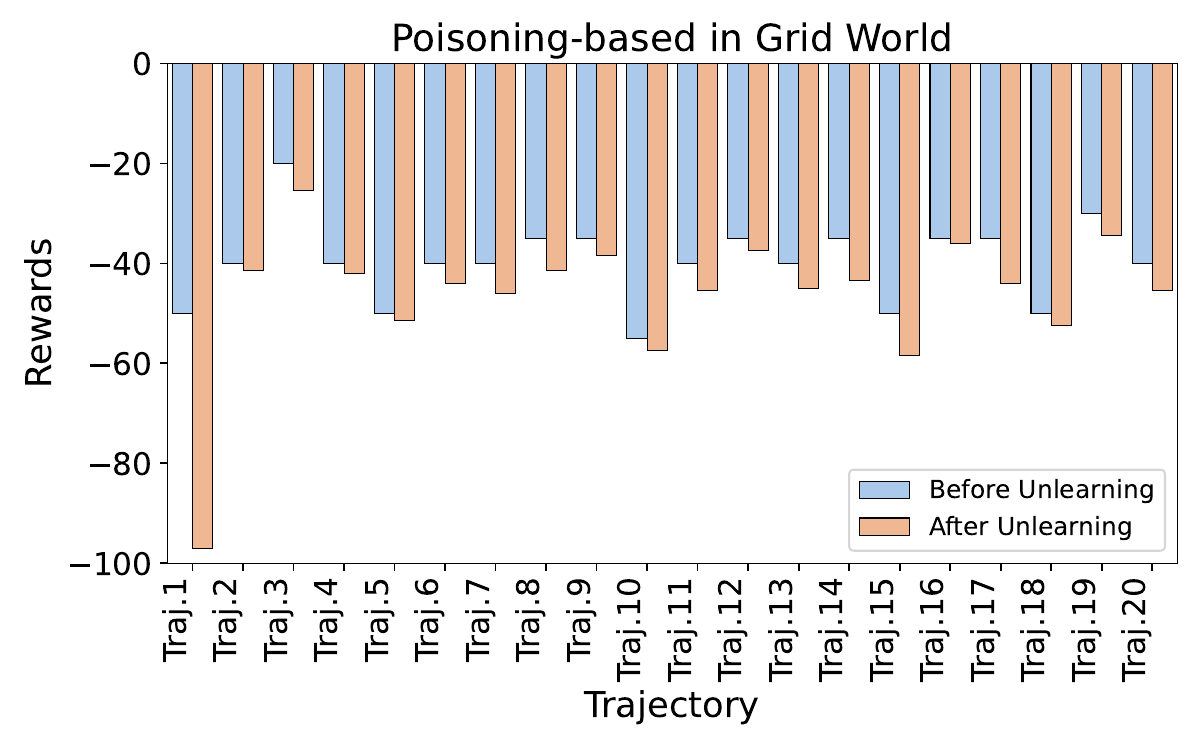}
			\label{fig:SinEnvMethod2Reward}}
    \end{minipage}
	\caption{The two methods in unlearning Trajectory 1 within a single environment using the grid world setting.}
	\label{fig:SingleEnvironment}
\end{figure}

\vspace{2mm}
\noindent\textbf{Experimental Results.} The results of unlearning within a single environment in the grid world scenario, depicted in Figure \ref{fig:SingleEnvironment}, align with those observed in multi-environment settings: the agent performs deterioratively in the unlearning trajectory, receiving lower rewards, while continuing to perform well in other trajectories. Note that counting steps is less relevant as an evaluation metric in the single-environment setting compared to multi-environment settings, because the focus is now on ensuring the agent effectively avoids specific states or actions, rather than on the speed or distance covered.

\section{Model Architecture} 
In the grid world and aircraft landing settings, we employ a fully-connected neural network as the model for our reinforcement learning agent. The neural network architecture consists of an input layer, two hidden layers, and an output layer.
The input layer takes a 10-dimensional vector as input, representing the relevant features of the environment. The output layer generates a 4-dimensional vector, representing the probability distribution over the four possible actions: up, down, left, and right. The first hidden layer comprises 64 neurons, while the second one consists of 32 neurons. 
Similarly, for recommendation systems, the architecture is slightly adjusted with an input layer, two hidden layers, and an output layer with neuron counts of 4, 128, 128, and 2 respectively. The input layer manages four movie attributes: ID, release year, genre, and rating, while the output layer determines whether to recommend the movie.

In the virtual home and maze explorer settings, we employ a Convolutional Neural Network (CNN) comprising three CNN blocks and one hidden layer with 512 neurons. This network receives visual information as input with a size of $140\times 120$ and produces a 4-dimensional vector, indicating the probability distribution across the four possible actions: up, down, left, and right. The weights of these neural networks are randomly initialized.

\section{Comparison of Various RL Algorithms}\label{app:overall performance}
\begin{figure}[ht]
\centering
    \subfigure[c][\scriptsize{The average number of steps with different RL algorithms}]{
    \centering\includegraphics[scale=0.21]{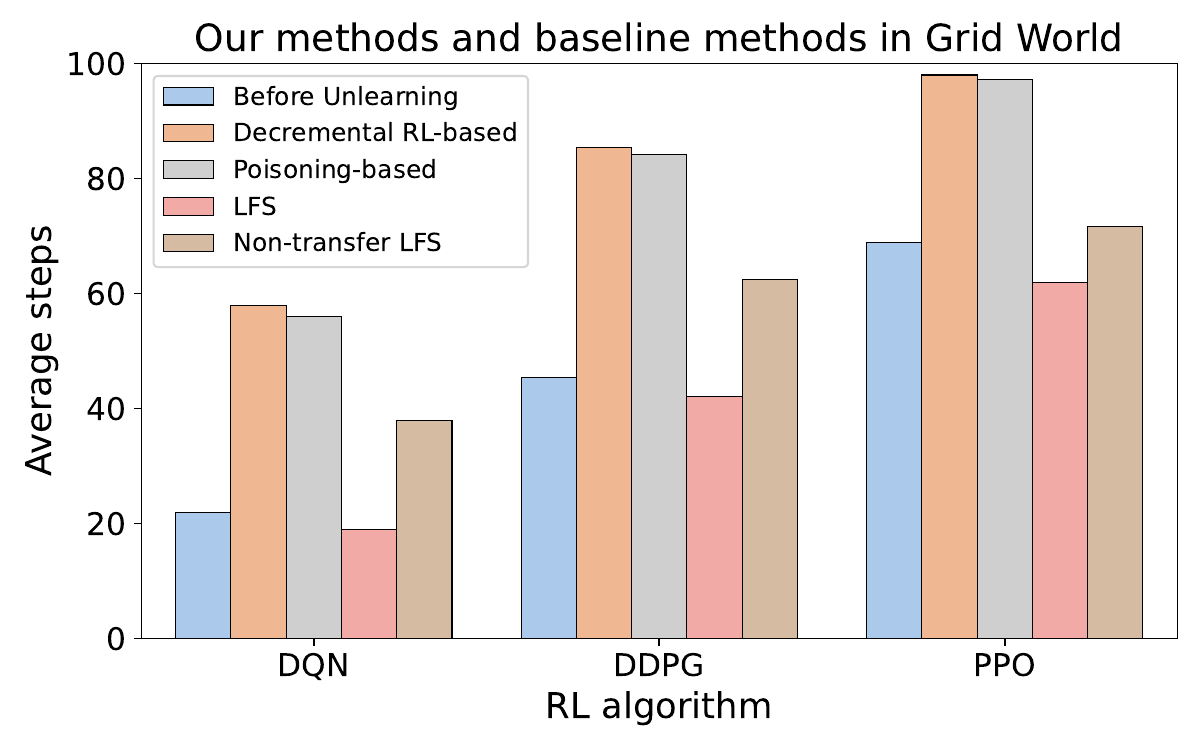}
			\label{fig:Steps}}
	\subfigure[c][\scriptsize{The average rewards received by the agent with different RL algorithms}]{
    \centering\includegraphics[scale=0.21]{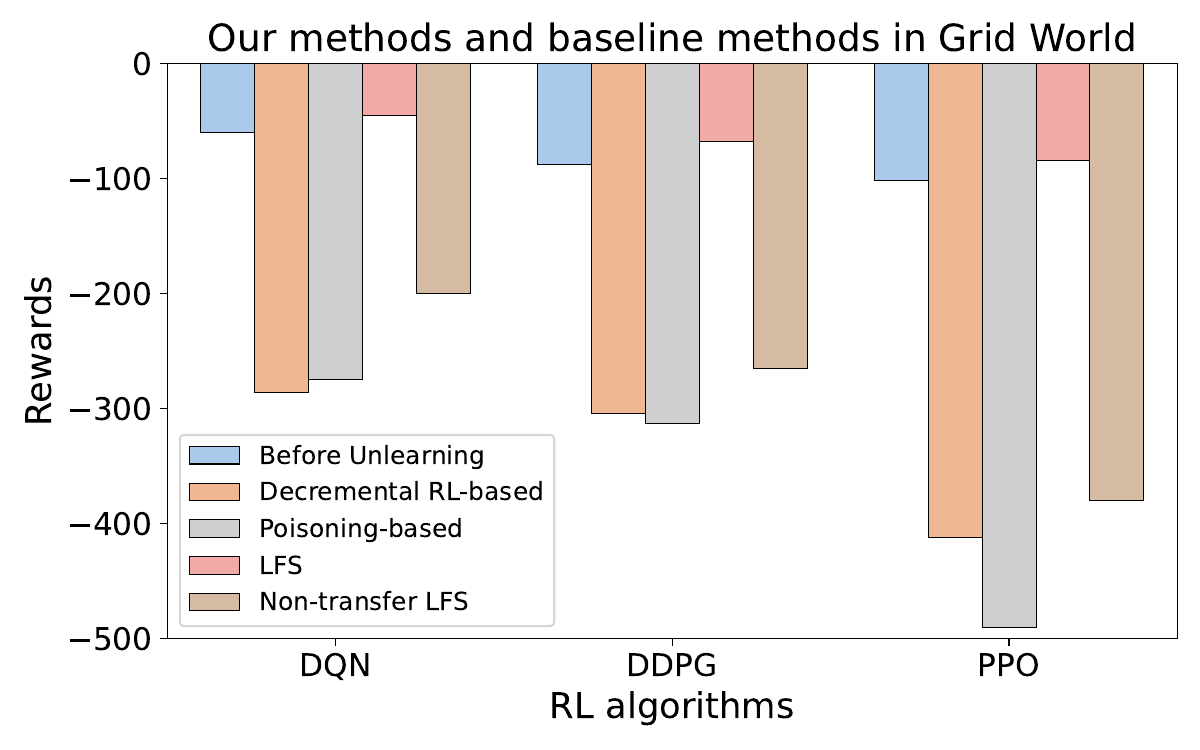}
			\label{fig:Rewards}}
	\caption{Performance of the three RL algorithms in grid world.}
	\label{fig:3RLGridWorld}
\end{figure}

To assess the effectiveness of different RL algorithms in our setting, we evaluate the performance of PPO and DDPG and compare them with the adopted DQN algorithm.
As illustrated in Figure \ref{fig:3RLGridWorld}, PPO and DDPG closely mirrors that of the DQN algorithm during the unlearning process, albeit with inferior results. Specifically, the agent achieves lower rewards and requires more steps to complete tasks. Several factors may contribute to this disparity. Firstly, DQN's exploration-exploitation trade-off may be better suited for efficiently navigating complex environments. Additionally, its experience replay mechanism enhances sample efficiency and learning convergence, further bolstering its performance. Thus, PPO and DDPG are not extensively utilized in our experiments.

	\label{tab:MazeExplorer}

\end{document}